\documentclass[square,numbers,twocolumn,superscriptaddress]{revtex4-2}
\usepackage[utf8]{inputenc}
\usepackage[margin=1in]{geometry}
\usepackage{parskip}

\usepackage{amsmath,amsfonts,amssymb,amsthm,bbold,mathtools,braket,physics,bm}

\usepackage{afterpage}

\usepackage{qcircuit}

\usepackage[algoruled,ruled,vlined,noline,linesnumbered]{algorithm2e}

\usepackage[usenames,dvipsnames]{xcolor}

\usepackage{tikz}
\usetikzlibrary{matrix,decorations.pathreplacing,calligraphy}

\usepackage[customcolors]{hf-tikz}

\usepackage[percent]{overpic}

\renewcommand{\dim}{\texttt{\textbf{len}}}

\newcommand{\AlgTagLocal}{[\textcolor{gray}{local}]}
\newcommand{\AlgTagDistributed}{[\textcolor{gray}{distributed}]}
\newcommand{\AlgTagStateVector}{[\textcolor{gray}{statevector}]}
\newcommand{\AlgTagDensityMatrix}{[\textcolor{gray}{density matrix}]}

\newcommand{\AlgTagBops}[1]{\vspace{6pt}%
[{\color[rgb]{
    0.471412, 0.108766, 0.527016
}#1 bops}]}
\newcommand{\AlgTagFlops}[1]{[{\color[rgb]{
    0.266122, 0.486664, 0.802529
}#1 flops}]}
\newcommand{\AlgTagNumSerialRounds}[1]{[{\color[rgb]{
    0.513417, 0.72992, 0.440682
}#1 exchanges}]}
\newcommand{\AlgTagAmpsTransferred}[1]{[{\color[rgb]{
    0.863512, 0.670771, 0.236564
}#1 exchanged}]}
\newcommand{\AlgTagMemOverhead}[1]{[{\color[rgb]{
    1, .5, .5
}#1 memory}]}
\newcommand{\AlgTagMemWrites}[1]{[{\color[rgb]{
    0.857359, 0.131106, 0.132128
}#1 writes}]}
\newcommand{\AlgTagNumSendRounds}[1]{[{\color[rgb]{
    0.513417, 0.72992, 0.440682
}#1}]}

\def\AlgThreadCommentColor{
    \color[rgb]{.25,.46,.89}
}
\let\oldnl\nl
\newcommand{\nonl}{\renewcommand{\nl}{\let\nl\oldnl}}
\newcommand{\AlgThreadComment}[1]{{\AlgThreadCommentColor{}\nonl\# #1}}

\newcommand{\codegap}{\vspace{.2cm}}

\newcommand{\iu}{\mathrm{i}}

\DeclarePairedDelimiter\floor{\lfloor}{\rfloor}

\renewcommand{\vec}[1]{\underline{#1}}
\newcommand{\matr}[1]{\underline{\underline{#1}}}

\newcommand{\choivec}[1]{\ket{\!\ket{#1}\hspace{-2.7pt}}}

\newcommand{\numsub}[1]{_{\color{lightgray}#1}}

\newcommand*\BitAnd{\mathbin{\texttt{\&}}}
\newcommand*\BitOr{\mathbin{\texttt{|}}}
\newcommand*\BitXor{\mathbin{\texttt{\^{}}}}
\newcommand*\BitShiftLeft{\mathbin{\texttt{<<}}}
\newcommand*\BitShiftRight{\mathbin{\texttt{>>}}}
\newcommand*\BitMod{\mathbin{\texttt{\%}}}
\newcommand*\BitComp{\ensuremath{\mathbin{\sim}}}
\newcommand*\LogicalNot{\mathbin{\texttt{!}}}

\newcommand*\PlusEq{\texttt{+=}}
\newcommand*\MinusEq{\texttt{-=}}
\newcommand*\TimesEq{\texttt{*=}}
\newcommand*\EqEq{\texttt{==}}
\newcommand*\NotEqEq{\texttt{!=}}

\newcommand{\intrange}[2]{[#1..#2)}

\def\subketradius{1pt}

\definecolor{subketcolor1}{rgb}{1,.9,.9}
\definecolor{subketborder1}{rgb}{1,.5,.5}
\tikzset{subketstyle1/.style={
    draw=subketborder1,
    fill=subketcolor1,
    thin,
    rounded corners=\subketradius
}}

\definecolor{subketcolor2}{rgb}{.85,.89,.98}
\definecolor{subketborder2}{rgb}{.25,.46,.89}
\tikzset{subketstyle2/.style={
    draw=subketborder2,
    fill=subketcolor2,
    thin,
    rounded corners=\subketradius
}}

\newcommand*\mystrut[1]{\vrule width0pt height0pt depth#1\relax}

\usepackage{lmodern}
\SetAlFnt{\ttfamily}

\newcommand\mycommfont[1]{\small\ttfamily\textcolor{cyan}{#1}}
\SetCommentSty{mycommfont}

\SetAlgoCaptionLayout{raggedright}

\SetNlSty{ttfamily}{\color{lightgray}}{}

\newtagform{equnumcol}{\color{lightgray}(}{)}
\usetagform{equnumcol}

\definecolor{customlinkcol}{HTML}{7B50B5}
\definecolor{customcitecol}{HTML}{E6439F}
\usepackage[
colorlinks=true,
linkcolor=customlinkcol,
citecolor=customcitecol
]{hyperref}

\renewcommand{\arraystretch}{1.5}

\usepackage[titles]{tocloft}
\cftsetindents{section}{0em}{2em}
\cftsetindents{subsection}{1em}{2em}
\cftsetindents{subsubsection}{2em}{2em}

\usepackage{enumitem}
\setlist[enumerate]{font=\color{red}\bfseries,leftmargin=15pt}

\setlist[itemize]{font=\color{red}\bfseries,leftmargin=15pt}

\let\oldput\put
\def\put(#1,#2)#3{%
  \oldput(#1,#2){\scriptsize #3}%
}

\begin{document}

\title{Distributed Simulation of Statevectors and Density Matrices}
\author{Tyson Jones}
\email{tyson.jones.input@gmail.com}
\author{B\'alint Koczor}
\author{Simon C. Benjamin}
\affiliation{Department of Materials, University of Oxford, Parks Road, Oxford OX1 3PH, United Kingdom}
\affiliation{Quantum Motion Technologies, Pearl House, 5 Market Road, London N7 9PL, United Kingdom}

\begin{abstract}

Classical simulation of quantum computers is an irreplaceable step in the design of quantum algorithms. 
Exponential simulation costs demand the use of high-performance computing techniques, and in particular \textit{distribution}, whereby the quantum state description is partitioned between a network of cooperating computers -- necessary for the exact simulation of more than approximately 30 qubits.
Distributed computing is notoriously difficult, requiring bespoke algorithms dissimilar to their serial counterparts with different resource considerations, and which appear to restrict the utilities of a quantum simulator.
This manuscript presents a plethora of novel algorithms for distributed full-state simulation of gates, operators, noise channels and other calculations in digital quantum computers. We show how a simple, common but seemingly restrictive distribution model actually permits a rich set of advanced facilities including Pauli gadgets, many-controlled many-target general unitaries, density matrices, general decoherence channels, and partial traces.
These algorithms include asymptotically, polynomially improved simulations of exotic gates,
and thorough motivations for high-performance computing techniques
which will be useful for even non-distributed simulators. 
Our results are derived in language familiar to a quantum information theory audience, and our algorithms formalised for the scientific simulation community.
We have implemented all algorithms herein presented into an isolated, minimalist \texttt{C++} project, hosted open-source on 
Github\footnote{\label{fn:link_to_repo}\href{https://github.com/TysonRayJones/Distributed-Full-State-Algorithms}{\texttt{github.com/TysonRayJones/Distributed-Full-State-Algorithms}}} with a permissive MIT license, and extensive testing.
This manuscript aims both to significantly improve the high-performance quantum simulation tools available, and offer a thorough introduction to, and derivation of, full-state simulation techniques.

\vspace{1.1cm}

\end{abstract}

\maketitle

\setcounter{tocdepth}{2}
\tableofcontents

\clearpage

\section{Introduction}
\label{sec:introduction}

Quantum computers are anticipated to revolutionise high-performance computing and scientific simulation. In the meantime, the converse is true; high-performance computing and simulation play an essential role in the development of quantum computers. 
Naturally such methods are necessary for \textit{emulating} quantum devices in the interim to their experimental realisation, in order to study the behaviour of analytically intractable quantum algorithms like variational schemes~\cite{yuan2019theory} or those upon variably noisy devices~\cite{vankov2019methods}.
But the utility of classical simulation is far grander.
It is central in
verifying experimental prototypes~\cite{villalonga2019flexible}, motivating supremacy~\cite{arute2019quantum} and error correction thresholds~\cite{knill1998resilient}, discovering~\cite{koczor2020variational} and recompiling~\cite{jones2022robust} quantum circuits, and even proves useful for benchmarking supercomputers themselves~\cite{willsch2019benchmarking,jones2019quest}.
Even after fault-tolerant and insimulably large quantum computers have been achieved, there will remain questions answerable only through strong simulation.

A common method of strong simulation is the ``full-state" technique, also referred to as ``direct evolution", ``statevector" and ``brute force" simulation~\cite{huang2020explicit,nes2010classical}. These evolve a precise representation of the quantum state - a dense statevector or density matrix - and are a prudent first choice of simulator when it is unknown which (if any) specialised simulation techniques to employ for a given circuit or calculation.
Alas, full-state simulation of a generic quantum circuit is exponentially costly in time or memory, and sometimes both. A floating-point (with double precision) statevector description of the idealised $53$ qubit Sycamore processor~\cite{arute2019quantum} would require approximately $144$ petabytes, and a density matrix description to precisely model its true, noisy behaviour would require $1.3\times10^{18}$ petabytes; or the combined memory of one quintillion ARCHER2 supercomputers~\cite{archer2_hardware}. Serially simulating even very simple quantum operations upon such large states is prohibitively slow.

Fortunately, methods of statevector simulation parallelise extremely well. The modification of a numerical quantum state representation under the action of gates and channels can very
effectively utilise classical hardware acceleration techniques like multithreading and general-purpose GPU parallelisation. Alas, the limitations of finite memory remain. A $15$-qubit dense mixed state, for example, has already become too large to process by a $12\,$GB GPU despite a similar $14$-qubit state being rapidly simulable. 

This makes the use of distribution essential, wherein multiple machines cooperating over a network each store a tractable subset of the full state description.
The memory aggregate between all machines enables the study of larger quantum systems with greater parallelisation. Under ideal weak scaling, $1024$ compute nodes of a supercomputer could together simulate a $40$ qubit quantum state in a similar time it takes a single node to simulate a $30$ qubit state.
Indeed, distributing large data structures between machines and modifying each through local parallelisation strategies like multi-threading is a typical application of supercomputing platforms.

Developing simulations to run on distributed and parallel systems often proves an advanced exercise in programming and algorithm design. Many of the facilities leveraged by serial programs, like globally accessible memory and synchronised program clocks, are not available.
The metrics by which we measure and predict the runtime performance of a distributed program are also different. Accessing data stored on another node requires synchronised inter-node communication at a relatively significant network penalty.
Despite supercomputing facilities boasting powerful network architectures, exchanging information over the network remains orders of magnitude slower than local operations like local memory access~\cite{motlagh1998memory}. And since statevector simulation is typically `memory bandwidth bound'~\cite{haner2016high}, communication becomes the most significant resource in distributed runtime accounting, and the most pertinent metric to optimise.
Yet, devising even an \textit{unoptimised} distributed algorithm to simulate certain quantum operators can be an immense challenge.

In this manuscript, we derive nineteen novel, distributed algorithms to accelerate the simulation of statevectors and density matrices, achieving provable constant to polynomial speedups.
%
All algorithms and their costs are summarised in Table~\ref{tab:alg_summary_table}.

The manuscript structure is as follows.
The remainder of Sec.~\ref{sec:introduction} defines our notation.
Sec.~\ref{sec:local_statevector_algorithms} presents the performance measures of \textit{local} (that is, \textit{non}-distributed) simulation, and derives three conventional local algorithms.
Sec.~\ref{sec:distribution} describes how the quantum state is partitioned between multiple compute nodes, how these nodes can communicate, and the performance measures of distributed simulation.
Sec.~\ref{sec:distributed_statevector_algorithms} reviews the simple one-target gate, then
derives 6 novel distributed algorithms for simulating unitary and Hermitian operators upon statevectors.
Sec.~\ref{sec:distributed_dens_matr_algorithms} reviews an existing technique for distributed representation of \textit{density matrices} and simulation of unitary gates (described in a previous work~\cite{jones2019quest}), then derives 13 novel distributed algorithms for simulating decoherence channels, and evaluating expectation values and partial traces.
Sec.~\ref{sec:final_summary} summarises the core mechanisms of our algorithms.

\subsection{Notation}
\label{sec:notation}


Let underlined $\vec{v} = \{v_0, \, v_1, \, \dots \}$ notate an ordered set with $i$-th element $v_i$ or $\vec{v}[i]$, indexing from $i \ge 0$. 
Index variable $i\in\mathbb{N}$ is italicised to distinguish it from the imaginary unit $\iu=\sqrt{-1}$.
Let $\matr{M}$ denote a matrix with elements $\matr{M}[i,j]$ (uppercase) or $m_{ij} \equiv m_{i,j}$ (lowercase), also indexing from zero. $\hat{M}$ indicates an operator, and $\hat{M}_t$ indicates one explicitly targeting the $t$-th qubit ($t \ge 0$), where the $0$-th qubit is the rightmost in a ket (similarly of bits forming a binary integer).
$\hat{M}_{\vec{t}}$ is a many-qubit operator targeting all qubits within array $\vec{t}$ of length $\dim(\vec{t})$. The operator's corresponding matrix representation is $\matr{M}\in\mathbb{C}^{2^{n} \times 2^{n}}$, where $n=\dim(\vec{t})$. Symbol $\hat{\sigma}$ is reserved for Pauli operators $\hat{\mathbb{1}},\hat{X}$,$\hat{Y}$,$\hat{Z}$, while $\hat{U}$ is reserved for unitary operators, $\hat{H}$ for Hermitian operators and $\hat{K}$ for Kraus operators. $\hat{\mathbb{1}}^{\otimes n}$ notates an $n$-qubit identity operator, and $\matr{\mathbb{1}}^{\otimes n}$ its
$2^n\times 2^n$ diagonal matrix. Unless specified otherwise, all objects are presented in the $\hat{Z}$-basis.

A ket $\ket{\psi}$ (with a Greek symbol) indicates a general pure quantum state, and $\ket{\psi}\numsub{N}$ (grey subscript) explicitly indicates one of $N$ qubits; its statevector is a ($2^N$)-length complex array $\vec{\psi}$. Meanwhile, $\ket{i}\numsub{N}$ (with a Latin symbol) indicates the $i$-th of $2^N$ computational basis states, indexing from zero and where the rightmost qubit is least significant. Ergo
\begin{align}
    \ket{9}\numsub{5} \equiv \ket{0}\numsub{1}\ket{1}\numsub{1}\ket{0}\numsub{1}\ket{0}\numsub{1}\ket{1}\numsub{1},
\end{align}
where $\ket{a}\ket{b} \equiv \ket{a}\otimes\ket{b}$ and $\otimes$ is the Kronecker product. 
Let $i_{[t]} \in \{0,1\}$ denote the $t$-th bit of $i \in \mathbb{N}$, so that $9_{[0]} = 1$ and $9_{[4]} = 0$. 
When a sum does not explicitly specify the lower bound, it is assumed to begin from \textit{\textbf{zero}} and the upper bound becomes \textit{\textbf{exclusive}}. 
Our notations so far imply
\begin{gather}
    i \equiv \sum\limits_{t=0}^{n-1} i_{[t]} \, 2^t
    \;\;\; \equiv \;\;\;
    \sum\limits_{t}^{n}
    i_{[t]} \, 2^t
\\
    \ket{i}\numsub{N} \equiv \ket{i_{[N-1]}} \ket{i_{[N-2]}} \dots \ket{i_{[1]}}\ket{i_{[0]}},
\\
    \ket{\psi}\numsub{N} \equiv \sum\limits_{i}^{2^N} \alpha_i \ket{i}\numsub{N}, 
    \;\;\;\;
    \alpha_i \in \mathbb{C},
\\
    \vec{\psi} = \{ \alpha_0, \; \dots, \; \alpha_{2^N-1} \},
\\
    \hat{U}_t \ket{\psi}\numsub{N}
    \simeq
    \left( 
    \matr{\mathbb{1}}^{\otimes (N-t-1)} \otimes
    \matr{U} \otimes 
    \matr{\mathbb{1}}^{\otimes t}
    \right)
    \vec{\psi}.
\end{gather}
We also use $i_{\lnot t}$ to indicate the natural number produced by flipping the $t$-th bit of $i\in\mathbb{N}$, and $i_{\lnot \vec{t}}$ is that resulting from flipping all bits with indices $t \in \vec{t}$.

A bold $\bm{\rho}$ notates a density operator with density matrix $\matr{\rho}$ and elements $\matr{\rho}[i,j] = \rho_{ij}$, which is the amplitude of basis state $\ketbra{i}{j}$, indexing from zero. 
In Sec.~\ref{sec:distributed_dens_matr_algorithms}, we will introduce a so-called ``Choi-vector" representation of $\bm{\rho}$ as an unnormalised statevector notated $\choivec{\bm{\rho}}$, instantiated with ordered set $\vec{\rho}$.

%
%
%
%

We present algorithms in pseudocode reminiscent of Python programming code, but also invoke the below symbols, and make regular use of the bit-twiddling functions in Alg.~\ref{alg:bit_twiddles}.
\begin{center}
\resizebox{1.1\columnwidth}{!}{%
    \begin{tabular}{r|lr|l}
         $\BitMod$ & modulo operator 
            &
         $\BitShiftLeft$ & bitwise left-shift 
                \\
         $\BitShiftRight$ & bitwise right-shit 
            &
         $\BitOr$ & bitwise \textit{or} 
                \\ 
         $\BitXor$ & bitwise \textit{exclusive or} 
            &
         $\BitAnd$ & bitwise \textit{and} 
                \\
         $\BitComp$ & bitwise complement
            &
         $\LogicalNot$ & logical \textit{not}
                \\
         $\cup$ & union \& list concatenation 
            &
         $x^*$ & complex conjugate of $x$
                \\
         $\matr{M}^T$ & transpose of matrix $\matr{M}$
            &
         $\hat{U}^\dagger$ & adjoint of operator 
                \\
         $\{ \, \}$ & empty list
            &
            &
                    (or matrix) $\hat{U}$ 
        \\
        {\mycommfont{//}} & comment
            &
        \AlgThreadComment{} & multithreading pragma
    \end{tabular} %
} 
\end{center}

%
\begin{algorithm}[b]
\DontPrintSemicolon
\caption{
    Bit-twiddling functions of unsigned integers, most adapted from Ref.~\cite{anderson2005bit}.
}
\label{alg:bit_twiddles}

\textbf{getBit}($n$, $t$):

\Indp 

    \textbf{return} $(n \BitShiftRight t) \BitAnd 1$

\Indm

\codegap 

\textbf{flipBit}($n$, $t$): 

\Indp
    
    \textbf{return} $n \BitXor (1 \BitShiftLeft t)$
    
\Indm

\codegap

\textbf{flipBits}($n$, $\vec{t}$): 

\Indp

    \textbf{for} $q$ \textbf{in} $\vec{t}$:

    \Indp 

        $n$ = \textbf{flipBit}($n$, $q$)

    \Indm 
    
    \textbf{return} $n$
    
\Indm

\codegap

\textbf{insertBit}($n$, $t$, $b$): 

\Indp
    
    $l = (n \BitShiftRight t) \BitShiftLeft (t+1)$

    $m = b \BitShiftLeft t$
    
    $r = n \BitAnd ((1 \BitShiftLeft t) - 1)$
    
    \textbf{return} $l \BitOr m \BitOr r$
    
\Indm

\codegap 


    
    
    
    




    
    


\textbf{insertBits}($n$, $\vec{t}$, $b$): %
\tcp*{$\vec{t}$ must be sorted}

\Indp
                        
    \textbf{for} $q$ \textbf{in} $\vec{t}$:
                        
    \Indp 
                        
        $n = \textbf{insertBit}(n,\,q,\,b)$
                            
    \Indm
    
    \textbf{return} $n$
    
\Indm 

\codegap

\textbf{setBits}($n$, $\vec{t}$, $v$):

\Indp 

    \textbf{for} $q$ \textbf{in} \textbf{range($0$, $\dim(\vec{t}))$}:
    
    \Indp 
    
        $b$ = \textbf{getBit($v$, $q$)} $\BitShiftLeft$ $\vec{t}[q]$
        
        $n = (n \BitAnd (\BitComp b)) \BitOr b$
    
    \Indm

    \textbf{return} $n$

\Indm

\codegap 

\textbf{allBitsAreOne}($n$, $\vec{t}$): \tcp*{can be made\hphantom{$\mathcal{O}(1)$}}

\Indp 

    $v = 1$
    \tcp*{$\mathcal{O}(1)$ using mask}

    \textbf{for} $q$ \textbf{in} $\vec{t}$:
    
    \Indp 
    
        $v = v \BitAnd \textbf{getBit}$($n$, $q$)
        
    \Indm
    
    \textbf{return} $v$

\Indm

\codegap 

\textbf{getBitMask}($\vec{t}$):

\Indp 

    \textbf{return} \textbf{flipBits}($0$, $\vec{t}$)

\Indm 






\end{algorithm}

\clearpage

\section{Local statevector algorithms}
\label{sec:local_statevector_algorithms}

Before we can discuss the challenges of \textit{distributed} simulation, we must first understand the simpler problem of \textit{local} (i.e. non-distributed), shared-memory simulation. This paradigm already invites a myriad of high-performance computing considerations such as memory striding~\cite{stridecite}, hierarchical caching~\cite{bottomley2004understanding}, branch prediction~\cite{smith1998study}, vectorisation~\cite{larsen2000exploiting,shin2005superword}, type-aware  arithmetic~\cite{anderson1967ibm} and inlining~\cite{chen1993effect}. And on multiprocessor architectures, non-uniform memory access~\cite{manchanda2010non} and local parallelisation paradigms like multithreading~\cite{nemirovsky2013multithreading} with incurred nuances such as cache-misses and false sharing~\cite{bolosky1993false}.
We here forego an introduction to these topics, referring the interested reader to a discussion in quantum simulation settings in Sec~6.3 of Ref.~\cite{jones2022thesis}. Still, this manuscript's algorithms will be optimised under these considerations in order to establish a salient performance threshold.
In this section we review and derive the basic principles and algorithms of a local full-state simulator. 

Such simulators maintain a dense statevector $\ket{\psi}$, typically instantiated as an array $\vec{\psi}$ of scalars $\alpha_i \in \mathbb{C}$ related by
\begin{align}
    \ket{\psi}\numsub{N} &= \sum\limits_{i}^{2^N} \alpha_i \ket{i}\numsub{N}
    \tag*{}
    \;\;\;\;
    \leftrightarrow
    \\
    \vec{\psi} &= \{\alpha_0, \; \dots, \; \alpha_{2^N-1} \}.
    \label{eq:local_statevec_model}
\end{align}
The scalars can adopt any complex numerical implementation such as Cartesian or polar, or a reportedly well-performing adaptive polar encoding~\cite{de2019massively}, to which the algorithms and algebra in this manuscript are agnostic.

Naturally, a classical simulator can store an unnormalised array of complex scalars and ergo represent non-physical states satisfying
\begin{align} 
\sum\limits_{i}^{2^N} |\alpha_i|^2 \ne 1 \, .
\end{align}
This proves a useful facility in the simulation of variational quantum algorithms~\cite{jones2020grad,jones2020natgrad} and the representation of density matrices presented later in this manuscript.

\begin{figure}[t]
    \centering
    \includegraphics[width=\columnwidth]{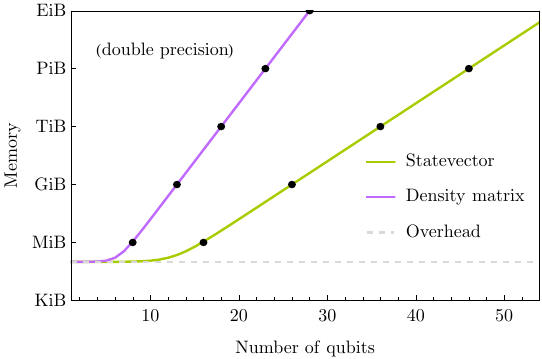}
    \caption{
        The memory costs to represent pure states (via statevectors) and mixed states (via density matrices), assuming that a single complex amplitude requires 16\,B (like a \texttt{C++} complex at double precision) and a process overhead of 100\,KiB.
    }
    \label{fig:serial_total_mem_costs}
\end{figure}

By storing all of a statevector's amplitudes, a full-state simulator maintains a complete description of a quantum state and permits precise \textit{a posteriori} calculation of any state properties such as probabilities of measurement outcomes and observable expectation values. It also means that the memory and runtime costs of simulation are roughly homogeneous across circuits of equal size, independent of state properties like entanglement and unitarity, making resource prediction trivial. For these reasons, full-state simulators are a natural first choice in much of quantum computing research.
Their drawback is that representing an $N$-qubit pure state requires simultaneous storage of $2^N$ complex scalars (an exponentially growing memory cost), and simulating an $n$-qubit general operator acting upon the state requires $\mathcal{O}(2^{N+n})$ floating-point operations (an exponentially growing time cost).
We illustrate these memory costs (and the equivalent for a dense $N$-qubit mixed state) in Figure~\ref{fig:serial_total_mem_costs}.
Thankfully, operators modify the statevector in simple, regular ways, admitting algorithms which can incorporate many high-performance computing techniques, as this section presents.

%

\subsection{Local costs}

We will measure the performance of local algorithms in this manuscript via the following metrics:
\begin{itemize}
    \item 
    \AlgTagBops{}
    The number of prescribed basic operations or ``\textit{bops}", such as bitwise, 
    arithmetic, logical and indexing operations. Performing a bop in the arithmetic logic unit (ALU) of a modern CPU can be orders of magnitude faster than the other primitive operations listed below, but their accounting remains relevant to data-dominated HPC~\cite{wang2018bops} like quantun simulation. While bitwise operations (like those of Alg.~\ref{alg:bit_twiddles}) and integer arithmetic are both classified as bops, the former is significantly faster~\cite{furer2007faster} and is preferred in tight loops. Our example code will be optimised thusly.
    
    \item 
    \AlgTagFlops{}
    The number of floating-point operations or \textit{flops}. These are accounted separately from bops because they are significantly slower to perform (typically implemented as a sequence of bops) and done so within a separate CPU sub-processor; the floating-point unit (FPU)~\cite{anderson1967ibm}.
    
    \item 
    \AlgTagMemOverhead{}
    The {memory} overhead, i.e. the size of temporary data structures created during an algorithm. 
    This excludes persistent pre-allocated memory dedicated to the quantum state representation.
    For memory efficiency, all algorithms in this manuscript will modify a pre-allocated statevector (or density matrix) in-place, though some will require non-negligible non-distributed temporary structures, exponentially smaller than a state.
    
    \item 
    \AlgTagMemWrites{}
    The number of memory \textit{writes}, i.e. modification of data structures which reside in the application's heap memory and (when not cached) in the simulating hardware's main memory (like RAM). In this manuscript, modifying a statevector amplitude constitutes a memory write (and encounters all the caching and multithreading nuances) while all other data (like local scalar variables and gate descriptions) are assumed negligible stack items preloaded into registers or fast caches. Accessing heap memory is assumed significantly slower than stack primitives, especially in the tight memory-bandwidth-bound loops typical of quantum simulation. In serial simulation, this is our most important metric to minimise. 
    
\end{itemize}

We will succinctly summarise an algorithm's cost under these respective metrics through the below tags displayed within an algorithm's caption.
    \begin{center}
        \AlgTagBops{$a$}%
        \AlgTagFlops{$b$}%
        \AlgTagMemOverhead{$c$}%
        \AlgTagMemWrites{$d$}
    \end{center}

Note that these measures do not capture \textit{all} local performance considerations; our presented algorithms are optimised to avoid branching, enable auto-vectorisation, cache efficiently, and avoid false-sharing when possible in multithreaded settings. Some of these considerations are introduced by example in the one-target gate of the next section.

We now derive local algorithms to simulate three canonical pure-state operations; the one-target gate, the many-control one-target, and the many-target gate. These will later serve to demonstrate the challenges introduced by distribution and the main study of this manuscript.


\subsection{One-target gate} %
\label{sec:intro_sim_1targ_gate} %
\noindent 
\begin{equation*}
\Qcircuit @C=1em @R=.7em {
& \qw & \qw \\
& \gate{\hat{M}} & \qw \\
& \qw & \qw 
}
\end{equation*}

The one-target gate (or ``single-qubit gate") is the most frequently appearing class of operator in quantum circuits, describing 1-local non-entangling unitaries, qubit projectors, Pauli operators, rotations, and a wide family of named gates like the Hadmard. It is arguably the simplest operator in a full-state simulator, and an important motivator for the performance goals when simulating more complicated operators. We here derive a local algorithm to in-place simulate the one-target gate acting upon an $N$-qubit pure state in $\mathcal{O}(2^N)$ bops and flops, $2^N$ memory writes, and an $\mathcal{O}(1)$ memory overhead.

Let $\hat{M}_t$ be a general one-target gate upon qubit $t\ge 0$, described by a complex matrix 
\begin{align} 
\matr{M} = \begin{pmatrix} m_{00} & m_{01} \\ m_{10} & m_{11}
\end{pmatrix}.
\end{align}
While bespoke, optimised simulation is possible when $\matr{M}$ is diagonal (like a phase gate), anti-diagonal (like $\hat{X}$ and $\hat{Y}$) or unit (like projector $\ketbra{0}{0}$), we will here assume $\matr{M}$ is completely general and unconstrained; therefore operator $\hat{M}_t$ can be non-physical.
We seek to apply $\hat{M}_t$ upon an arbitrary $N$-qubit pure state $\ket{\psi}$ (stored as array $\vec{\psi}$) of amplitudes $\alpha_i \in \mathbb{C}$. This means modifying $\vec{\psi}$ in-place to describe the resulting state
\begin{align}
\hat{M}_t \ket{\psi} 
&=
\left( 
\hat{\mathbb{1}}^{\otimes (N-t-1)}
\otimes 
\hat{M}
\otimes 
\hat{\mathbb{1}}^{\otimes t}
\right)
\ket{\psi},
\end{align}
although naturally there is no need to instantiate such a prohibitively large $\mathbb{C}^{2^N\times 2^N}$ operator matrix.
While there exists a large HPC literature on efficient multiplication of generic Kronecker products upon distributed vectors~\cite{tadonki2001parallel},  
 the exponential size of our vector (here, the statevector) warrants a dedicated treatment.

Denote the $t$-th bit of binary-encoded natural number $i \in \mathbb{N}$ as $i_{[t]}$.
By expanding the general initial state into a basis of bit sequences,
\begin{align}
\ket{\psi}\numsub{N} &= \sum\limits_{i}^{2^N} \alpha_i \ket{i}\numsub{N}
\\
&=
\sum\limits_{i}^{2^N} \alpha_i \ket{i_{[N-1]}}\numsub{1} \dots \ket{i_{[0]}}\numsub{1},
\end{align}
we can express the action of the operator as
\begin{align}
\hat{M}_t \ket{\psi} 
&=
\sum\limits_{i}^{2^N} \alpha_i \ket{i_{[N-1]}}\numsub{1} \dots \left( \hat{M} \ket{i_{[t]}}\numsub{1} \right) \dots \ket{i_{[0]}}\numsub{1}
\label{eq:serial_1qb_gate_hithere}
\end{align}
where the single-qubit kets are mapped to
\begin{align}
    \hat{M} \ket{i_{[t]}}\numsub{1}
    &=
    \begin{cases}
    m_{00}\ket{0}\numsub{1} + m_{10}\ket{1}\numsub{1}
        &
            i_{[t]} = 0
        \\
    m_{01}\ket{0}\numsub{1} + m_{11}\ket{1}\numsub{1}
        &
            i_{[t]} = 1
    \end{cases}
\\[4pt]
    &= \;\;\; m_{i_{[t]}\,i_{[t]}} \ket{i_{[t]}}\numsub{1}
        +
       m_{\LogicalNot i_{[t]}\,i_{[t]}}\ket{\LogicalNot i_{[t]}}\numsub{1}.
       \label{eq:serial_1qb_gate_poopydoopy}
\end{align}
We have used symbol $\LogicalNot$ as \textit{logical not} to flip a single bit.
Substituting Eq.~\ref{eq:serial_1qb_gate_poopydoopy} into Eq.~\ref{eq:serial_1qb_gate_hithere} yields
\begin{align}
    \hat{M}_t & \ket{\psi}
    = 
    \sum\limits_{i}^{2^N} 
    \Big( 
    \\
    &\alpha_i \, m_{i_{[t]}\,i_{[t]}} \, \ket{i_{[N-1]}}\numsub{1} \dots
    \ket{ i_{[t]}}\numsub{1}
    \dots \ket{i_{[0]}}\numsub{1}
    \tag*{}
    \\
    +\;  &\alpha_i \, m_{\LogicalNot i_{[t]}\,i_{[t]}} \, \ket{i_{[N-1]}}\numsub{1} \dots
    \ket{\LogicalNot i_{[t]}}\numsub{1}
    \dots \ket{i_{[0]}}\numsub{1}
    \Big) 
    \tag*{}
    \\
    & \hphantom{\ket{\psi}} = \sum\limits_{i}^{2^N} 
    \alpha_i \, m_{i_{[t]}\,i_{[t]}} \ket{i}\numsub{N} 
    +
    \alpha_i \, m_{\LogicalNot i_{[t]}\,i_{[t]}}
     \ket{i_{\lnot t}}\numsub{N},
\end{align}
where $i_{\neg t}$ denotes the number formed by flipping the $t$-th bit of the integer $i$.
By invoking that $\LogicalNot \LogicalNot b=b$ and $(i_{\neg t})_{\neg t} = i$, we conclude
\begin{align} 
\hat{M}_t\ket{\psi} = \sum\limits_{i}^{2^N} 
    \left( \alpha_i \, m_{i_{[t]}\,i_{[t]}}
     + 
     \alpha_{i_{\neg t}} \, m_{i_{[t]}\, \LogicalNot i_{[t]}}
    \right) 
    \ket{i}\numsub{N}.
    \label{eq:serial_1b_gate_fullform}
\end{align}
This form reveals that the one-target gate $\hat{M}_t$ linearly combines distinct \textit{pairs} of amplitudes, weighted by the elements of $\matr{M}$. The amplitudes of $\ket{\psi}\numsub{N}$ are simultaneously modified to become
\begin{align}
    \alpha_i 
    \;
    \xrightarrow{\hat{M}_t}
    \;
    \alpha_i \; m_{i_{[t]}\,i_{[t]}}
     + 
     \alpha_{i_{\neg t}} \; m_{i_{[t]}\, \LogicalNot i_{[t]}}
     \label{eq:serial_1qb_amp_update_rule}
\end{align}
where each pair of amplitudes $\alpha_i$ and $\alpha_{i_{\neg t}}$ are modified independently from all others. 

\begin{figure}[tb]
    \centering
    \includegraphics[width=.85\columnwidth]{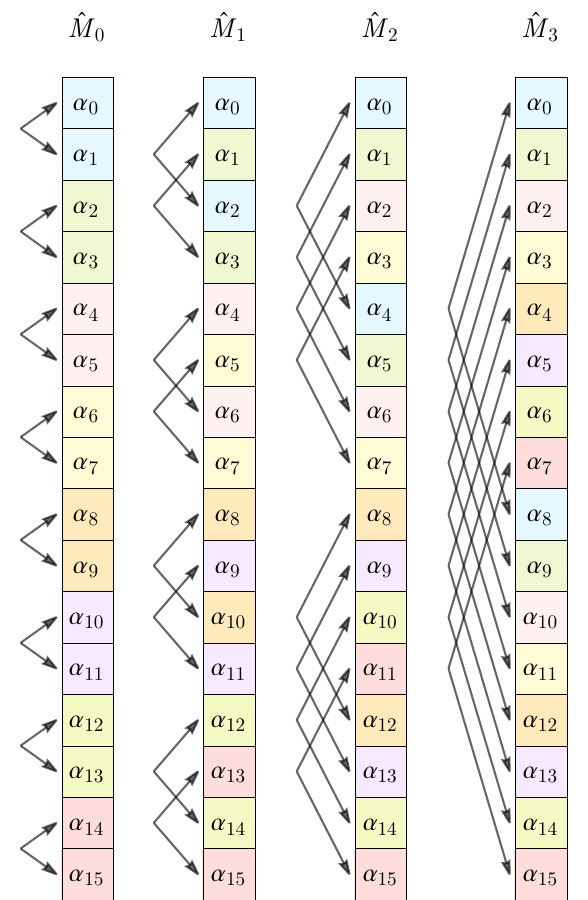}
    \caption{
        The memory access pattern of Alg.~\ref{alg:serial_1qb_gate_alg}'s local simulation of the one-target gate $\hat{M}_t$. 
        Each column of $16$ amplitudes $\alpha_i$ denotes the array representation $\vec{\psi}$ of a $4$-qubit statevector $\ket{\psi}\numsub{4}=\sum_i\alpha_i\ket{i}\numsub{4}$. Amplitudes joined by an arrow (also sharing the same colour for clarity) are linearly combined with one another under the action of $\hat{M}_t$ upon $\vec{\psi}$. This means these elements of $\vec{\psi}$ are accessed within the same iteration of Alg.~\ref{alg:serial_1qb_gate_alg}'s \textbf{\texttt{for}} loop, informing the memory stride and caching behaviour.
        Amplitudes not connected by arrows can be independently processed in parallel. However, the simultaneous modification of amplitudes with intersecting arrows may cause cache conflicts.
        We will use diagrams of this kind throughout this manuscript to demonstrate how varying the target qubits (varied here between columns) of an operator change the memory access pattern.
    }
    \label{fig:serial_1qb_gate_mem}
\end{figure}

Our final task is to clarify precisely the positions of the paired amplitudes in the state's array $\vec{\psi}$ and the resulting memory access pattern.
Because $i_{\neg t}=i\pm2^t$, the paired amplitudes are a distance $2^t$ apart. Bit logic alone already uniquely determines the access pattern, which we visualise in Fig.~\ref{fig:serial_1qb_gate_mem}.
There however remain many ways to implement the amplitude iteration, but few which respect the previous section's HPC considerations. We will now directly derive a specific implementation which favourably avoids branch prediction, enables vectorisation, minimises memory reads and maximises bitwise arithmetic, while hiding all of these nuances.
We simply petition the bits of $i$ into \textit{three} sequences,  letting $\ket{i}\numsub{N} \equiv \ket{k}\numsub{N-t-1}\ket{i_{[t]}}\numsub{1}\ket{j}\numsub{t}$, or equivalently $i \equiv j + i_{[t]}\,2^t + k \,2^{t+1}$.
Without loss of generality, we express
\begin{align}
    \ket{\psi}\numsub{N} \equiv
    \sum\limits_{j}^{2^t}
    \sum\limits_{k}^{2^{N-t-1}}
    \;
    &\beta_{kj} \;
    \ket{k}\numsub{N-t-1} \ket{0}\numsub{1} \ket{j}\numsub{t} \;+
    \tag*{}
    \\
    &\gamma_{kj} \;
    \ket{k}\numsub{N-t-1} \ket{1}\numsub{1} \ket{j}\numsub{t},
\end{align}
where $\beta_{kj},\gamma_{kj}\in\mathbb{C}$ together form the $2^N$ amplitudes of $\ket{\psi}$. Precisely, $\beta_{kj} \equiv \alpha_i \, \implies \, \gamma_{kj} = \alpha_{i_{\neg t}}$.
The middle ket, equivalent to $\ket{i_{[t]}}$, is that targeted by $\hat{M}_t$. Eq.~\ref{eq:serial_1qb_amp_update_rule} prescribes
\begin{align}
    \hat{M}_t \ket{\psi}\numsub{N} \equiv
    \sum\limits_{j}^{2^t}
    \sum\limits_{k}^{2^{N-t-1}}
    &
    \left( \beta_{kj} \, m_{00} + \gamma_{kj} \, m_{01}
    \right)
    \ket{k} \ket{0} \ket{j}
    \tag*{}
    \\
    + \;
    &
    \left( \beta_{kj} \, m_{10} + \gamma_{kj} \, m_{11}
    \right) 
    \ket{k} \ket{1} \ket{j}.
\end{align}
This form suggests a bitwise and trivially parallelisable local strategy to branchlessly locate the paired amplitudes among the $2^N$ of an $N$-qubit statevector's array, and modify them in-place, which we present in Alg.~\ref{alg:serial_1qb_gate_alg}.
It makes use of several bit-twiddling functions defined in Alg.~\ref{alg:bit_twiddles}.

\begin{algorithm}[bt]
\DontPrintSemicolon
\caption[blah blah]{
    \AlgTagLocal \AlgTagStateVector
    \\
        One-target gate $\hat{M}_t$ upon qubit $t$ of $N$-qubit pure state $\vec{\psi}$, where 
            $\matr{M} = \begin{psmallmatrix} m_{00} & m_{01} \\
            m_{10} & m_{11}
            \end{psmallmatrix} \in \mathbb{C}^{2\times 2}$.
    \\
    \protect{\vspace{4pt}\begin{center} $ 
\Qcircuit @C=1em @R=.7em {
& \gate{\hat{M}} & \qw
}$ \end{center} }
    \begin{center}
        \AlgTagBops{$\mathcal{O}(2^N)$}%
        \AlgTagFlops{$\mathcal{O}(2^N)$}\\
        \AlgTagMemOverhead{$\mathcal{O}(1)$}%
        \AlgTagMemWrites{$2^N$}
    \end{center}
}
\label{alg:serial_1qb_gate_alg}

\textbf{local\_oneTargGate}($\vec{\psi}$, $\matr{M}$, $t$): 

    \Indp
    
     $N = \log_2( \, \dim(\vec{\psi})\, )$

    \codegap 
    
    \tcp{loop every $\ket{n}\numsub{N} = \ket{0}\numsub{1} \ket{k}\numsub{N-t-1}\ket{j}\numsub{t}$}
    \AlgThreadComment{multithread}

    \textbf{for} $n$ \textbf{in} \textbf{range}(0, $2^{N}/2$): 
    
    \Indp 

        \codegap 
    
        \tcp{$\ket{i\beta}\numsub{N} = \ket{k}\numsub{N-t-1}\ket{0}\numsub{1}\ket{j}\numsub{t}$}
        
        $i\beta$ = \textbf{insertBit}($n$, $t$, $0$)
        \tcp*{Alg.~\ref{alg:bit_twiddles}}

        \codegap 
        
        \tcp{$\ket{i\gamma}\numsub{N} = \ket{k}\numsub{N-t-1}\ket{1}\numsub{1}\ket{j}\numsub{t}$}
        
        $i\gamma$ = \textbf{flipBit}($i\beta$, $t$)
        \tcp*{Alg.~\ref{alg:bit_twiddles}}
        
        
        $\beta$ = $\vec{\psi}[i\beta]$
        
        $\gamma$ = $\vec{\psi}[i\gamma]$

        \codegap 
        
        \tcp{modify the paired amplitudes}
        
        $\vec{\psi}[i\beta]$ = $m_{00}\, \beta + m_{01}\, \gamma $
        
        $\vec{\psi}[i\gamma]$ = $m_{10} \,\beta + m_{11}\, \gamma$
        
    \Indm 
    \Indm
\end{algorithm}

We finally remark on local parallelisation. Alg.~\ref{alg:serial_1qb_gate_alg} features an exponentially large loop wherein each iteration modifies a unique subset of amplitudes; they can ergo be performed concurrently via multithreading. All variables defined within the loop become thread-private, while the statevector is shared between threads and is simultaneously modifiable only when threads write to separate cache-lines. Otherwise, false-sharing may degrade performance, in the worst case that of serial array access~\cite{bolosky1993false}. We hence endeavour to allocate threads to perform specific iterations which modify disjoint cache-lines. We could simply set
\begin{align}
({\color{gray}
\text{iterations per thread}}) = S \coloneqq \frac{({\color{gray}\text{cache-line size}})}{({\color{gray}\text{amplitude size}})},
\end{align}
where conveniently all sizes are expected to be powers of $2$, and assigned iterations are contiguous. At double precision with a typical $64$~byte cache-line, $S = 4$~\cite{drepper2007every}.
However, we can further avoid some cache-misses incurred by a single thread (due to fetching both $\beta$ and $\gamma$ amplitudes) across its 
 assigned iterations, by setting
\begin{align}
({\color{gray}\text{iterations per thread}}) = 2^{t+1},
\end{align}
when this exceeds $S$. Beware that when $t$ is near its maximum of $N$, such an allocation may non-uniformly divide the work between threads and wastefully leave threads idle. The ideal schedule is ergo the maximum multiple of the cache-line size (in amplitudes) which is less than or equal to the uniform division of total iterations between threads. That is
\begin{align}
({\color{gray}\text{iterations per thread}})
= 
 S \, 
\left\lfloor
\frac{({\color{gray}\text{num iterations}})}{S \; ({\color{gray}\text{num threads}})}\right\rfloor,
\end{align}
where $({\color{gray}\text{num threads}})$ is the maximum concurrently supported by the executing hardware, and is \textit{not} necessarily a power of $2$. Notice however that in Alg.~\ref{alg:serial_1qb_gate_alg}, $({\color{gray}\text{num iterations}}) = 2^N/2$ is a runtime parameter; alas, thread allocation must often be specified at compile-time, such as it is when using OpenMP~\cite{dagum1998openmp}. In such settings, the largest division of contiguous iterations between threads can be allocated, incurring 
$\mathcal{O}(({\color{gray}\text{iterations per thread}}))$
false shares. This is the default static schedule in OpenMP, specified with \texttt{C} precompiler directive
\begin{center}
{
    \AlgThreadCommentColor
\texttt{\# pragma omp for}
}
\end{center}
This is the multithreaded configuration assumed for all algorithms in this manuscript when indicated by directive
\begin{center}
\AlgThreadComment{multithread}
\end{center}

\clearpage

\subsection{Many-control one-target gate} %
\label{sec:intro_manyctrl_gate} %
\noindent 
\begin{equation*}
\Qcircuit @C=1em @R=.7em {
& \ctrl{2} & \qw \\
& \ctrl{1} & \qw \\
& \gate{\hat{M}} & \qw
}
\end{equation*}

The many-control one-target gate introduces one or more control qubits to the previous section's operator. This creates a simple entangling gate, yet even the one-control one-target gate is easily made universal~\cite{lloyd1995universal,deutsch1995universality} and appears as an elementary gate in almost every non-trivial quantum circuit~\cite{barenco1995elem}.
Many-control gates can be challenging to perform experimentally and are traditionally decomposed into a series of one-control gates. But in classical simulation, many-control gates are just as easy to effect directly, and actually 
prescribe \textit{fewer} amplitude modifications and floating-point operations than their non-controlled counterparts.
We here derive a local algorithm to in-place simulate the one-target gate with $s$ control qubits upon an $N$-qubit pure state in $\mathcal{O}(s\, 2^{N-s})$ bops, $\mathcal{O}(2^{N-s})$ flops, $2^{N-s}$ memory writes, and an $\mathcal{O}(1)$ memory overhead. 

We seek to apply $C_{\vec{c}}(\hat{M}_t)$ upon an arbitrary $N$-qubit pure state $\ket{\psi}$, where $\vec{c}=\{c_0,\dots,c_{s-1}\}$ is an arbitrarily ordered list of $s$ unique control qubits, and operator $\hat{M}$ targets qubit $t \notin \vec{c}$ and is described by a $2\times 2$ general complex matrix $\matr{M}$ (as in the previous section).
Let us temporarily assume that the target qubit $t$ is the least significant and rightmost ($t=0$), and the next $s$ contiguous qubits are controlled upon. Such a gate is described by the matrix
\begin{align}
    C_{\{1,\dots,s\}}(\hat{M}_0)
    \tag*{}
\\ & \simeq \;\; \hphantom{2^s}\;
\vcenter{
\begin{tikzpicture}[
    decoration={calligraphic brace},
    every left delimiter/.style={xshift=.4em},
    every right delimiter/.style={xshift=-.4em},
    inner sep=1.5pt, column sep=3.5pt, row sep=2pt
  ]
    \matrix (m) [
        matrix of math nodes,
        left delimiter={(},
        right delimiter={)},
        ampersand replacement=\&
    ] {
        1 \&  \& \&  \& \\
         \& \ddots \&  \&  \& \\
        \hphantom{1} \&   \& 1 \&  \& \\
         \&  \&  \& m_{00} \& m_{01} \\
          \&   \&   \& m_{10}  \& m_{11} \\
    };
    \draw[decorate,transform canvas={xshift=-1.5em},thick] (m-3-1.south west) -- node[left=2pt] {$2^s$} (m-1-1.north west);
\end{tikzpicture}
}
\label{eq:serial_manyqb_ctrl_matrix}
\\
&= \; \; \matr{\mathbb{1}}^{\otimes (s+1)} + \ketbra{1}{1}^{\otimes s} \otimes (\matr{M} - \matr{\mathbb{1}}).
\label{eq:serial_manyqb_ctrl_op_form}
\end{align}
This form prescribes identity (no modification) upon every computational basis state \textit{except} those for which all control qubits are in state $\ket{1}\numsub{1}$; for those states, apply $\hat{M}_0$ as if it were non-controlled, like in the previous section. There are only a fraction $1/2^s$ such states, due to the $2^s$ unique binary assignments of the $s$ control qubits.
This prescription is unchanged when the control and target qubits are reordered or the number of controls is varied, as intuited by swapping rows of Eq.~\ref{eq:serial_manyqb_ctrl_matrix} or inserting additional identities into Eq.~\ref{eq:serial_manyqb_ctrl_op_form}.

\begin{figure}[t]
    \centering
    \includegraphics[width=.85\columnwidth]{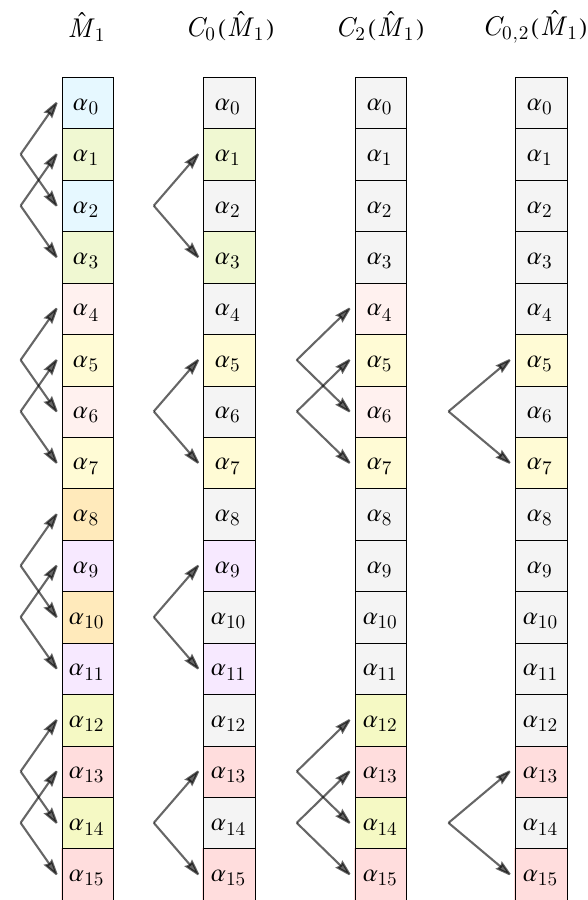}
    \caption{
        The memory access pattern of Alg.~\ref{alg:serial_manyctrl_gate_alg}'s local simulation of the many-control one-target gate $C_{\vec{c}}(\hat{M}_t)$. Amplitudes in grey fail the control condition of Eq.~\ref{eq:serial_manyctrl_ctrl_cond} and are not modified nor accessed.
    }
    \label{fig:serial_manyctrl_gate_mem}
\end{figure}

Therefore, a general $C_{\vec{c}}(\hat{M}_t)$ gate modifies only $2^{N-s}$ amplitudes $\alpha_i$ of the $2^N$ in the statevector array $\vec{\psi}$, which satisfy
\begin{align}
    i_{[c_n]} = 1, \;\; \forall c_n \in \vec{c},
    \label{eq:serial_manyctrl_ctrl_cond}
\end{align}
and does so under the action of the non-controlled $\hat{M}_t$ gate. We illustrate the resulting memory access pattern in Fig.~\ref{fig:serial_manyctrl_gate_mem}.

Ascertaining which amplitudes satisfy (what we dub) the ``\textit{control condition}" of Eq.~\ref{eq:serial_manyctrl_ctrl_cond} must be done efficiently since it may otherwise induce an overhead in every iteration of the exponentially large loop of the non-controlled Alg.~\ref{alg:serial_1qb_gate_alg}. We again leverage HPC techniques to produce a cache-efficient, branchless, vectorisable, bitwise procedure in a derivation which hides all such nuances from the reader.
We seek to iterate only the $2^{N-s}$ amplitudes satisfying the control condition, which have indices $i$ with fixed bits (value $1$) at $\vec{c}$. To do so, we enumerate all $(N-s)$-bit integers $j\in \{0,\,\dots,\,2^{N-s}-1\}$, corresponding to $(N-s)$-qubit basis states
\begin{align}
    \ket{j}\numsub{N-s} \equiv 
    \ket{j_{[N-s-1]}}\numsub{1} \dots 
    \ket{j_{[1]}}\numsub{1}
    \ket{j_{[0]}}\numsub{1}
\end{align}
and predeterminedly interleave $\ket{1}\numsub{1}$ at every index $c_j\in \vec{c}$. It is important these insertions happen at strictly increasing bit indices so as not to displace previously inserted bits. The result is an enumeration of only indices $i$ of the form
\begin{align}
    \ket{i}\numsub{N} 
    =
     \ket{j_{[N-s-1]}}\numsub{1} \dots 
     \ket{1}\numsub{1}
    \ket{j_{[c_0-1]}}\numsub{1}
    \dots
    \ket{j_{[0]}}\numsub{1},
\end{align}
which we \textit{a priori} know satisfy the control condition. Note we should explicitly iterate only \textit{half} of these indices while updating \textit{two} amplitudes per iteration; those paired by the target qubit.
We formalise this in Alg.~\ref{alg:serial_manyctrl_gate_alg}.

    \begin{algorithm}[b]
    \DontPrintSemicolon
    \caption[blah blah]{
        \AlgTagLocal \AlgTagStateVector
        \\
        Many-control one-target gate $C_{\vec{c}}(\hat{M}_t)$ with $s$ unique control qubits $\vec{c}=\{c_0,\dots,c_{s-1}\}$ and target $t \notin \vec{c}$, described by matrix $\matr{M} = \begin{psmallmatrix} m_{00} & m_{01} \\
            m_{10} & m_{11}
            \end{psmallmatrix} \in \mathbb{C}^{2\times 2}
        $,
        applied to an $N$-qubit pure statevector $\vec{\psi}$.
        \\ 
        \protect{\begin{center} 
        $ \Qcircuit @C=1em @R=.7em {
            & \ctrl{2} & \qw \\
            & \ctrl{1} & \qw \\
            & \gate{\hat{M}} & \qw } $ 
        \end{center}}
        \begin{center}
        \AlgTagBops{$\mathcal{O}(s\, 2^{N-s})$}%
        \AlgTagFlops{$\mathcal{O}(2^{N-s})$}\\
        \AlgTagMemOverhead{$\mathcal{O}(1)$}%
        \AlgTagMemWrites{$2^{N-s}$}
    \end{center}
    }
    \label{alg:serial_manyctrl_gate_alg}
    
    \textbf{local\_manyCtrlOneTargGate}($\vec{\psi}, \, \vec{c}, \, \matr{M}, \, t$): 
    
        \Indp
        
        $N = \log_2( \, \dim(\vec{\psi})\, )$
        
        $s = \dim(\vec{c})$
        
        \codegap
        
        \tcp{get all qubits in increasing order}
        
        $\vec{q} = \vec{c} \; \cup \; \{ t \}$
        
        \textbf{sort}($\vec{q}$)
        
        \codegap
        
        \tcp{loop every $\ket{j}\numsub{N-s-1}$}

        \AlgThreadComment{multithread}
        
        \textbf{for} $j$ \textbf{in} \textbf{range}(0, $2^{N}/2^s$): 
        
        \Indp 

            \codegap 
        
            \tcp{produce $\ket{i}\numsub{N}$ where $i_{[c_n]} = i_{[t]}=1$}
        
            $i$ = \textbf{insertBits}($j$, $\vec{q}$, $1$)
            \tcp*{Alg.~\ref{alg:bit_twiddles}}

            \codegap 
            
            \tcp{set ${i\beta}_{[t]}=0$}
            
            $i\beta = \textbf{flipBit}(i, t)$
            \tcp*{Alg.~\ref{alg:bit_twiddles}}
            
            $i\gamma = i$
            
            $\beta = \vec{\psi}[i\beta]$
            
            $\gamma = \vec{\psi}[i\gamma]$

            \codegap 
            
            \tcp{modify the paired amplitudes}
            
            $\vec{\psi}[i\beta] = m_{00}\, \beta + m_{01}\, \gamma $
            
            $\vec{\psi}[i\gamma] = m_{10} \,\beta + m_{11}\, \gamma$
            
        \Indm 
        
    \Indm
    \end{algorithm}

Note that further optimisation is possible.
The one-control gate (the special case of $s=1$ above) permits a bespoke nested-loop routine with an inner-loop accessing explicitly contiguous array elements, enabling explicit vectorisation; we direct the interested reader to Sec.~6.3.2 of Ref.~\cite{jones2022thesis}.


\subsection{Many-target gate} %
\label{sec:serial_manytarg_gate} %
\noindent 
\begin{equation*}
\Qcircuit @C=1em @R=.7em {
& \multigate{1}{\hat{M}} & \qw \\
& \ghost{\hat{M}} & \qw \\
& \qw & \qw \\
}
\end{equation*}

The many-target gate (also known as the ``multi-qubit gate") describes a broad class of digital quantum operators including powerful entangling gates, two-qubit gates like the Mølmer-Sørensen gate natural in trapped ion architectures~\cite{molmer1999multiparticle,haffner2008quantum}, the Barenco gate~\cite{barenco1995universal}, the Berkeley gate~\cite{zhang2004minimum} and the native entangling gate of Google's Sycamore quantum processor~\cite{harrigan2021quantum}. Such gates may defy convenient decomposition into smaller primitives, and are best specified and effected as dense complex matrices. 
Here we derive a local algorithm to in-place simulate an $n$-target gate upon an $N$-qubit pure state in $\mathcal{O}(2^{N+n})$ bops and flops, $\mathcal{O}(2^N)$ memory read/writes, and $\mathcal{O}(2^n)$ memory overhead.
Unitarity of the gate is not required.
Note we distinguish the many-target gate from all-target or full-state operators which act on \textit{all} qubits, as the latter requires special treatment and benefits from a distinct iteration strategy than developed here. We here instead assume $2^n \ll 2^N$.

Let $\hat{M}_{\vec{t}}$ denote an $n$-qubit operator upon unique and arbitrarily ordered target qubits $\vec{t} = \{t_0, \,\dots,\, t_{n-1}\}$, described by matrix $\matr{M} \in \mathbb{C}^{2^n\times 2^n}$ with elements $m_{ij}$. The ordering of indices in $\vec{t}$ corresponds to the ordering of columns in $\matr{M}$. We seek to simulate this operator acting upon an $N$-qubit pure state $\ket{\psi}\numsub{N}$.

Without loss of generality, let
\begin{align}
    \ket{\psi}\numsub{N}
\equiv
    \sum\limits_k^{2^{N-n}}
    \sum\limits_j^{2^n}
    \alpha_{kj} \ket{k,j}\numsub{N},
    \;\;\;\;
    \alpha_{kj} \in \mathbb{C}
    \label{eq:serial_manytrg_psi_form}
\end{align}
where $\ket{k,j}$ is an $N$-qubit computational basis state formed by the $j$-th $n$-qubit substate of qubits in $\vec{t}$, and the $k$-th ($N-n$)-qubit substate of the remaining non-targeted qubits. Scalars $\{\alpha_{kj} : k,j\}$ together form the $2^N$ amplitudes of $\ket{\psi}$. In this manner, we leave the elements and ordering of $\vec{t}$ unspecified. For illustration, if $\vec{t}=\{2,0\}$ (increasing significance), then an example basis state is
\begin{alignat*}{2}
    \ket{
    k=\textcolor{VioletRed}{1},
    j=\textcolor{ProcessBlue}{2}}\numsub{4} 
\;\;
\equiv
\;\;
    \ket{\textcolor{VioletRed}{0}}\numsub{1}
    &\ket{\textcolor{ProcessBlue}{0}}\numsub{1}
    \ket{\textcolor{VioletRed}{1}}\numsub{1}
    &&\ket{\textcolor{ProcessBlue}{1}}\numsub{1}
\;\;
\equiv
\;\;
    \ket{3}\numsub{4}.
    \\
    & \hphantom{x} { \scriptstyle \color{gray} t_0}
    && \hphantom{x} { \scriptstyle \color{gray} t_1}
\end{alignat*}
By definition (through the construction of $\matr{M}$ as a matrix to be multiplied upon a $Z$-basis unit column vector), the many-qubit gate applied to a basis state of its targeted subspace yields
\begin{align}
    \hat{M}\ket{j}\numsub{n}
    =
    \sum\limits_l^{2^n} m_{lj}\ket{l}\numsub{n}.
\end{align}

Applied to the full state, the gate then effects
\begin{align}
    \hat{M}_{\vec{t}} \ket{\psi}\numsub{N} 
=
    \sum\limits_k^{2^{N-n}}
    \sum\limits_j^{2^n}
    \alpha_{kj} 
    \sum\limits_l^{2^n} m_{lj}\ket{k,l}\numsub{N}.
\end{align}
By swapping what have become mere summation labels $j \leftrightarrow l$, we more clearly express the new state as
\begin{align}
    \hat{M}_{\vec{t}} \ket{\psi}\numsub{N} 
=
    \sum\limits_k^{2^{N-n}}
    \sum\limits_j^{2^n}
    \left( 
    \sum\limits_l^{2^n}
    \alpha_{kl} \, 
     m_{jl}
      \right)
      \ket{k,j}\numsub{N}.
\end{align}
Comparing this form to Eq.~\ref{eq:serial_manytrg_psi_form} reveals that the amplitudes have been simultaneously modified to
\begin{align}
    \alpha_{kj} 
    \;\;
    \xrightarrow{\hat{M}_{\vec{t}}}
    \;\;
    \sum\limits_l^{2^n}
    \alpha_{kl} \; 
     m_{jl}\,.
     \label{eq:local_manyqb_amp_update}
\end{align}
We see that under $\hat{M}_{\vec{t}}$, each amplitude becomes a linear combination of $2^n$ amplitudes, weighted by a row of $\matr{M}$, and are independent of all other amplitudes outside the combination.

\begin{figure}[t]
    \centering
    \includegraphics[width=.85\columnwidth]{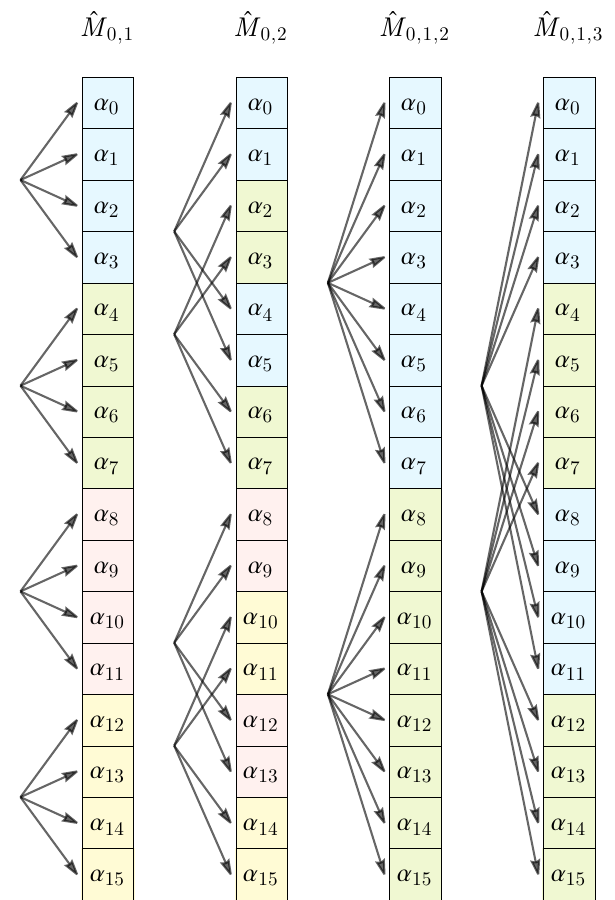}
    \caption{
        The memory access pattern of Alg.~\ref{alg:serial_manyqb_gate_alg}'s local simulation of the many-target gate $\hat{M}_{\vec{t}}$.
    }
    \label{fig:serial_manytarg_gate_mem}
\end{figure}

\begin{algorithm}[t]
\DontPrintSemicolon
\caption[blah]{
    \AlgTagLocal \AlgTagStateVector
    \\
    Many-target gate $\hat{M}_{\vec{t}}$ with $n$ unique target qubits $\vec{t} = \{t_0,\,\dots,\,t_{n-1}\}$, described by matrix $\hat{M}=\{m_{ij}:i,j\} \in \mathbb{C}^{2^n\times 2^n}$, applied to an $N$-qubit pure statevector $\ket{\psi}$.
    \\
    \protect{\vspace{6pt}\begin{center} $ \Qcircuit @C=1em @R=.7em {
& \multigate{1}{\hat{M}} & \qw \\
& \ghost{\hat{M}} & \qw
} $ \end{center}}
    \begin{center}
        \AlgTagBops{$\mathcal{O}(2^{N+n})$}%
        \AlgTagFlops{$\mathcal{O}(2^{N+n})$}\\
        \AlgTagMemOverhead{$\mathcal{O}(2^n)$}
        \AlgTagMemWrites{$\mathcal{O}(2^{N+n})$}
    \end{center}
}
\label{alg:serial_manyqb_gate_alg}

\textbf{local\_manyTargGate}($\vec{\psi}$, $\matr{M}$, $\vec{t}$): 

\Indp

    $N$ = $\log_2( \, \dim(\vec{\psi})\, )$
    
    $n$ = $\dim(\vec{t})\;\;\;$
    \tcp{$ = \log_2(\dim(\matr{M}))$}
    
    \codegap

    $\vec{v}$ = \textbf{new array} of size $2^n$
    
    $\vec{q}$ = \textbf{clone} of $\vec{t}$ (size $n$)
    
    \textbf{sort}($\vec{q}$)
    
    \codegap
    
    \tcp{loop every $\ket{k}\numsub{N-n}$}

    \AlgThreadComment{multithread with private $\vec{v}$ clones}
    
    \textbf{for} $k$ \textbf{in} \textbf{range}($0$, $2^{N-n}$): 
    
    \Indp 

        \codegap
    
        \tcp{form $\ket{z}\numsub{N} \; \equiv \; \ket{k, 0}\numsub{N}$}
    
        $z$ = \textbf{insertBits}($k$, $\vec{q}$, $0$)
        \tcp*{Alg.~\ref{alg:bit_twiddles}}

        \codegap
        
        \tcp{loop every $\ket{j}\numsub{n}$}
        
        \textbf{for} $j$ \textbf{in} \textbf{range($0$, $2^n$)}:
        
        \Indp 
        
            \tcp{form $\ket{i}\numsub{N} \; \equiv \; \ket{k,j}\numsub{N}$}
        
            $i$ = \textbf{setBits($z$, $\vec{t}$, $j$)}
            \tcp*{Alg.~\ref{alg:bit_twiddles}}
            
            \tcp{store amplitudes $\{\alpha_{kl} : k\}$}
            
            $\vec{v}[j]$ = $\vec{\psi}[i]$
            
        \Indm

        \codegap 
        
        \textbf{for} $j$ \textbf{in} \textbf{range($0$, $2^n$)}:
        
        \Indp 
        
            $i$ = \textbf{setBits($z$, $\vec{t}$, $j$)} 
            \tcp*{Alg.~\ref{alg:bit_twiddles}}
            
            \tcp{modify $\alpha_{kj} \rightarrow \sum_l \alpha_{kl} \, m_{kl}$}
            
            $\vec{\psi}[i]$ = $0$
            
            \textbf{for} $l$ \textbf{in} \textbf{range($0$, $2^n$)}:
            
            \Indp 
            
                $\vec{\psi}[i]$ \PlusEq{}  $m_{jl} \;\, \vec{v}[l]$
                
            \Indm 
            
        \Indm
        
    \Indm 
    
\Indm
\end{algorithm}

The precise indices of these amplitudes in the statevector array $\vec{\psi}$ are determined by bit logic. We observe that for a given $k$, the amplitudes $\alpha_{kl}, \; \forall l\in\{0,\dots,2^n-1\}$ are located at indices $\ket{i}\numsub{N}\equiv\ket{k,l}\numsub{N}$ which differ only by the $n$ bits $i_{[q]}$ at indices $q \in \vec{t}$.
\begin{gather} 
\ket{z}\numsub{N} \equiv \ket{k,0}\numsub{N}
\;\; \implies \;\; 
\forall  \ket{i}\numsub{N} \in \{ \ket{k,l}\numsub{N} : l \},
\tag*{}
\\
\exists \; \vec{q} \subseteq \vec{t}
\;\;
    \text{ s.t. } 
\;\;
i = z_{\neg \vec{q}}\,.
\end{gather}
This means the array indices of all amplitudes to be linearly combined together can be produced by flipping bits $q \in \vec{t}$ of the index $z$, where $\ket{z}\numsub{N}=\ket{k,0}\numsub{N}$. We obtain $\ket{z}\numsub{N}$ from a given $\ket{k}\numsub{N-n}$ by inserting $0$ bits into unsigned binary integer $z$ at the indices specified in $\vec{t}$. in strictly increasing order. 
Finally, we simply iterate each of the possible $2^{N-n}$ values of $k$. 
We formalise this protocol in Alg.~\ref{alg:serial_manyqb_gate_alg}, and illustrate its memory access pattern in Fig.~\ref{fig:serial_manytarg_gate_mem}.

Like the previous algorithms of this manuscript (and indeed, most of those to come), its runtime when deployed on modern computers is memory bandwidth bound. That is, the fetching and modification of amplitudes from heap memory dominates the runtime, occluding the time of the bitwise and indexing algebra.
We should expect that targeting lower-index rightmost qubits will see better caching performance than targeting high-index leftmost qubits, because the linearly combining amplitudes in the former scenario lie closer together and potentially within the same cache-lines. 

If we wished to make the memory addresses accessed by the inner $j$ loops of Alg.~\ref{alg:serial_manyqb_gate_alg} be strictly increasing, we would simply initially permute the columns of $\matr{M}$ as per the ordering of $\vec{t}$ to produce matrix $\matr{M}'$, then sort $\vec{t}$. That is, we would leverage that
\begin{align}
    \hat{M}_{\vec{t}} \; \equiv \; {\hat{M}'}_{\textbf{sorted}(\vec{t})}\,.
\end{align}
Doing so also permits array $\textbf{sorted}(\vec{t})$ to be replaced with a bitmask, though we caution optimisation of the indexing will \textit{not} appreciably affect the memory-dominated performance. 

Finally, we caution that multithreaded deployment of this algorithm requires that each simultaneous thread maintains a private $2^n$-length array ($\vec{v}$ in Alg.~\ref{alg:serial_manyqb_gate_alg}), to collect and copy the amplitudes modified by a single iteration. This scales up the temporary memory costs by factor $\mathcal{O}( { \color{gray} \text{num threads} } )$.

\vspace{.08\paperheight}

\section{Distribution}
\label{sec:distribution}

Distributed computing involves dividing the serial task of a single machine into smaller tasks for a multitude of networked machines. Distributed simulation of an $N$-qubit pure state $\ket{\psi}\numsub{N}$ requires partitioning the $2^N$ complex amplitudes of its statevector $\vec{\psi}$ between $W \in \mathbb{N}$ nodes. Here, $W$ is the \textit{world size}~\cite{lusk2009mpi}. A uniform partitioning is only possible between $W=2^w$ nodes for $w\in\mathbb{N}$~\cite{gauss1801disquisitiones} and hence we assume that only a power-of-two number of nodes are ever deployed. This is a typical precondition of distributed quantum simulators~\cite{jones2019quest,guerreschi2020intel}. 
The number of amplitudes stored within each node is then fixed at
\begin{align}
\Lambda \; \coloneqq \; 2^{N-w}.
\end{align}
This naturally upperbounds the world size to simulate an $N$-qubit statevector at $W\le 2^N$ whereby each node stores at least one amplitude. Of course in practical settings, $W \ll 2^N$.

Algorithms~\ref{alg:serial_1qb_gate_alg}, \ref{alg:serial_manyctrl_gate_alg} and \ref{alg:serial_manyqb_gate_alg} presented serial methods for simulating the single-target, many-control and many-target gates respectively. To be performant, they made economical accesses to the state array $\vec{\psi}$.
Distributed simulation of these gates however will require explicit synchronisation and network communication between nodes to even access amplitudes within another node's statevector partition, which we refer to as the ``sub-statevector".
We adopt the message passing interface (MPI)~\cite{lusk2009mpi} ubiquitous in parallel computing. Exchanging amplitudes through messages requires the use of a \textit{communication buffer} to receive and process network-received amplitudes before local modification of a state. For the remainder of this manuscript, we employ fixed labels:
\begin{align*}
    \vec{\psi} &\coloneqq \text{An individual node's sub-statevector.}
    \\
    \vec{\varphi} &\coloneqq \text{An individual node's communication buffer}.
\end{align*}
We identify nodes by their \textit{rank} $r \in \{0,\,\dots,\,W-1\}$. Each node's partition represents an unnormalised substate $\ket{\psi_r}\numsub{N-w}$ of the {full} quantum state represented by the ensemble
\begin{align}
    \ket{\Psi}\numsub{N} 
    \; \equiv \;
    \sum\limits_r^{2^w} \ket{r}\numsub{w}\ket{\psi_r}\numsub{N-w}.
\end{align}
In other words, node $r$ contains global amplitudes $\alpha_i$ of $\ket{\Psi}\numsub{N}$ with indices satisfying
\begin{align}
    i \;\in\; \{ r \Lambda, \; r \Lambda + 1, \; \dots, \; 
        (r+1)\Lambda - 1 \}.
\end{align}
This means that the $j$-th \textit{local} amplitude $\vec{\psi}[j]$ stored in node $r$ corresponds to \textit{global} basis state
\begin{align}
    \ket{i}\numsub{N} \; \equiv \; \ket{r}\numsub{w} \ket{j}\numsub{N-w},
\end{align}
and ergo to global amplitude $\alpha_i$ with index 
\begin{align}
i &= r\,\Lambda + j \\
  &= (r \BitShiftLeft (N-w)) \BitOr j.
\end{align}
We will frequently refer to $\ket{r}\numsub{w}$ and $\ket{j}\numsub{N-w}$ as the ``prefix" and ``suffix" substates of $\ket{i}\numsub{N}$ respectively.
These relationships will enable the convenient and efficient determination of communication and memory access patterns through bitwise algebra.

\subsection{Communication buffer size}

We must decide the size of each node's communication buffer $\vec{\varphi}$, which will share a node's available memory with $\vec{\psi}$.
We will later see that simulating many operators requires an exchange of $\le \Lambda$ amplitudes between pairs of nodes, and so it is prudent to upperbound $\dim(\vec{\varphi}) \le \Lambda$ while choosing $\dim(\vec{\varphi}) = \mathcal{O}(\Lambda)$. A contrary choice of $\dim(\vec{\varphi}) = \mathcal{O}(1)$ necessitates exponentially many communication overheads~\cite{larose2018distributed}.
Distributed simulators like Intel's IQS~\cite{guerreschi2020intel} opt for $\dim(\vec{\varphi})=\Lambda/2$ which requires typical gate simulations perform \textit{two} rounds of amplitude exchange, and may enable the second communication to overlap local state processing for a minor speedup. It also enables the simulation of a single additional qubit on typical node configurations.
In contrast, simulators like Oxford's QuEST~\cite{jones2019quest} fix $\dim(\vec{\varphi})=\Lambda$. 
This restricts simulation to \textit{one} fewer qubits, though induces no appreciable slowdown due to the dominating network speeds.
See References~\cite{jones2022thesis,jones2019quest} for a more detailed comparison of these strategies.

In this work, we fix the buffer to be of the same size as the sub-statevector, i.e.
\begin{align}
    \dim(\vec{\varphi}) \;  \coloneqq \; \Lambda.
\end{align}
This enables the simulation of many advanced operators precluded by a smaller buffer, though means the total memory costs (in bytes) of distributed simulation of an $N$-qubit pure state is $b \, 2^N$ where $b$ is the number of bytes to represent a single complex amplitude ($b=16$ at double precision in the \texttt{C} language). This is double the serial costs shown in Fig.~\ref{fig:serial_total_mem_costs}.

\subsection{Communication costs}

We now discuss the measures of distributed performance.
The local algorithms of Sec.~\ref{sec:local_statevector_algorithms} were optimised to minimise the number of basic operations (bops),
floating-point operations (flops), heap memory writes,
and the memory overhead, while avoid branching, and enabling optimisations like vectorisation.
While these costs remain important in distributed algorithms, they are eclipsed by the overwhelming runtime penalties of network communication~\cite{motlagh1998memory}.


%
%



As such, the primary objectives of high-performance distributed code is to both minimise the number of rounds of serial communication (suppressing synchronisation and latency overheads) and the total data size transferred (suppressing waiting for bandwidth-bound traffic, and mitigating risks of queuing and network saturation). We refer to these quantities respectively as the number of \textit{exchanges} and the number of complex scalars \textit{exchanged}.
We will succinctly summarise a distributed algorithm's costs in its caption with the below tags
\begin{center}
        \AlgTagBops{$a$}%
        \AlgTagFlops{$b$}%
        \AlgTagNumSerialRounds{$c$} \\
        \AlgTagAmpsTransferred{$d$}%
        \AlgTagMemOverhead{$e$}%
        \AlgTagMemWrites{$f$}
\end{center}
Quantities $a,b,f$ are given \textit{per-node} because nodes perform these operations concurrently. Quantities $c,d,e$ are given as totals aggregated between all nodes. 
It will be easy to distinguish these by whether the measures include variable $\Lambda$ (per-node) or $N$ (global).

\subsection{Communication patterns}

This manuscript will present algorithms with drastically differing \textit{communication patterns}, by which we refer to the graph of which nodes exchange messages with one another. 
The metrics of the previous section are useful for comparing two given distributed algorithms, but they cannot alone predict their absolute  runtime performance.
A distributed application's ultimate runtime depends on its emerging communication pattern and the underlying hardware network topology~\cite{gerla1977topological}.

\begin{figure*}[t]
    \centering
    \includegraphics[width=\textwidth]{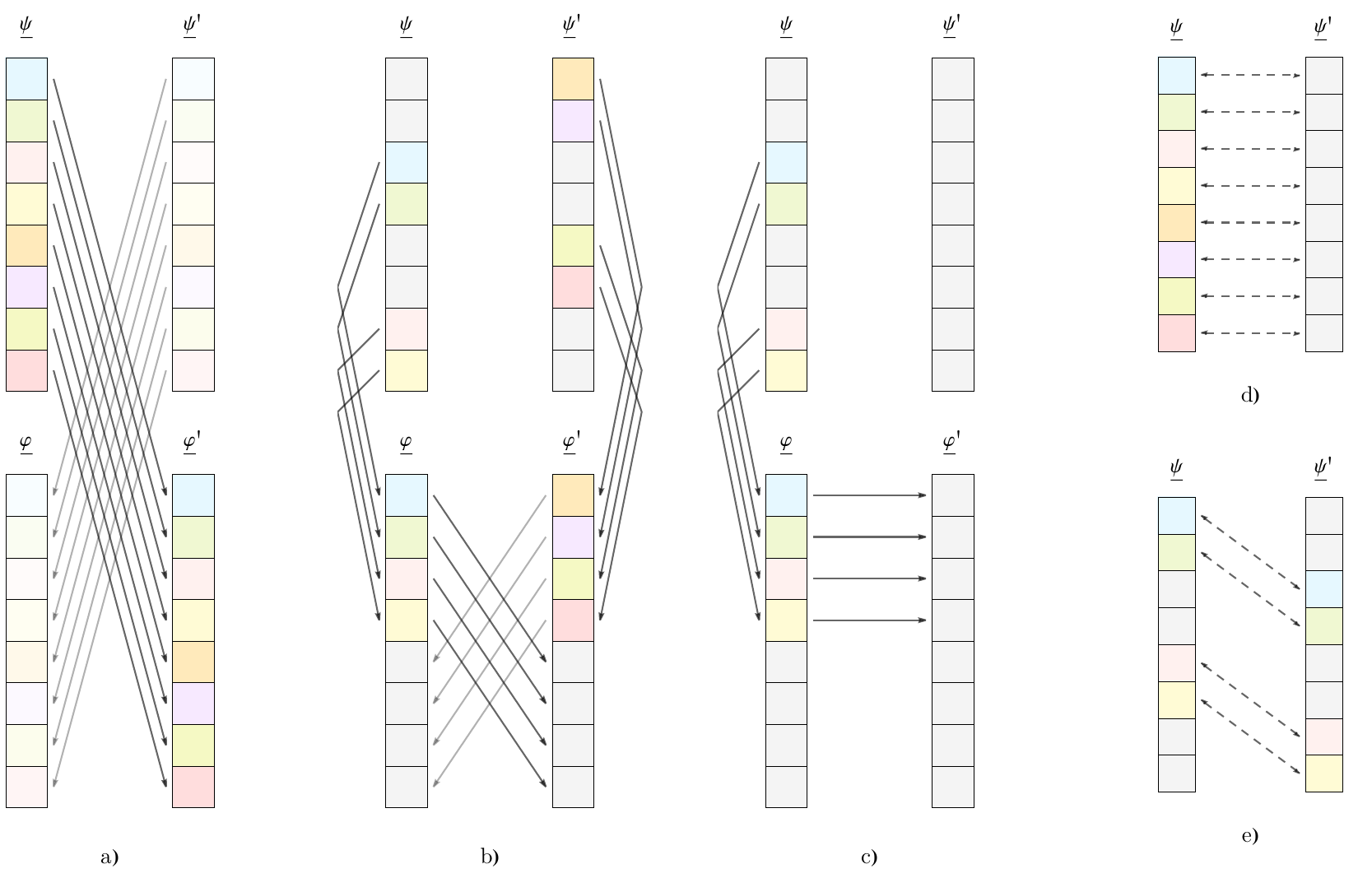}
    \caption[dummy]{
    The five paradigms of pairwise amplitude exchange used in this manuscript. Arrays $\vec{\psi}$ and $\vec{\varphi}$ are a node's sub-statevector and communication buffer respectively, and $\vec{\psi}'$ and $\vec{\varphi}'$ are those of a paired node. During a round of communication, all or some of the nodes will perform one of the below paradigms, with the remaining nodes idle.
    \\
        \textbf{a)} Nodes send their full sub-statevector to directly overwrite their pair node's communication buffer. This permits \\
        \hphantom{\textbf{a)}}
        local modification of received amplitudes in the buffer before integration into the sub-statevector. 
        \\
        \textbf{b)} Nodes pack a subset of their sub-statevector into their buffer before exchanging (a subset of) their buffers. This \\
        \hphantom{\textbf{b)}}
        reduces communication costs from \textbf{a)} when not all pair node amplitudes inform the new local amplitudes.
        Notice
        \\
        \hphantom{\textbf{b)}}
        that since the buffers cannot be directly swapped, amplitudes are sent to the pair node's empty offset buffer.
        \\
        \textbf{c)} One node of the pair packs and sends its buffer, while the receiving pair node sends nothing.
        \\
        \textbf{d)} Nodes intend to directly swap the entirety of their sub-statevector, though must do so via \textbf{a)}.
        \\
        \textbf{e)} Nodes intend to directly swap distinct subsets of their sub-statevectors, though must do so via \textbf{b)}.
    }
    \label{fig:distrib_mem_exchange_patterns}

    \bigskip

    \bigskip

    \begin{algorithm}[H]
\label{alg:mpi_defs}
\caption{
    Some convenience inter-node communication functions used in this manuscript's pseudocodem defined in terms of the MPI \texttt{C} standard~\cite{lusk2009mpi}.
    Note that an actual implementation will require communicating arrays in multiple smaller messages when they would otherwise exceed the MPI maximum message size of approximately $2^{30}$ double-precision amplitudes. It is also prudent to dispatch these messages asynchronously~\cite{adamski2023energy}.
}

    

    \textbf{getRank}():
    
    \Indp 

        \textcolor{gray}{obtain via}  \textbf{MPI\_Comm\_rank}
    
        
        
    
    \Indm 
    
    \codegap 
    
    \textbf{getWorldSize}():
    
    \Indp 

    \textcolor{gray}{obtain via}  \textbf{MPI\_Comm\_size}
    
        
        
    
    \Indm

    \codegap
    
    \tcp{send $n$ elements of $\vec{\text{send}}$ (from index $i$)
        to node $r'$, overwriting $\vec{\text{recv}}$ (from index $j$).
    }
    \tcp{
        Node $r'$ performs the same to this node.}
    
    \textbf{exchangeArrays}($\vec{\text{send}}$, $i$, $\vec{\text{recv}}$, $j$, $n$, $r'$):
    
    \Indp 
    
        \textbf{MPI\_Sendrecv}(
	
    	\Indp
    	
        	(\textcolor{gray}{\textbf{address of}})
        	$\vec{\text{send}}[i]$, $n$, \textbf{MPI\_COMPLEX}, 
        	$r'$, \textbf{MPI\_ANY\_TAG},
        	
        	(\textcolor{gray}{\textbf{address of}}) $\vec{\text{recv}}[j]$, $n$, \textbf{MPI\_COMPLEX},
        	$r'$, \textbf{MPI\_ANY\_TAG}, \textbf{MPI\_COMM\_WORLD})
        	
        \Indm
    
    \Indm 

    \codegap 

    \tcp{send entire $\vec{\psi}$ to node $r$', overwriting $\vec{\varphi'}$, and receive entirety of $\vec{\psi'}$, overwriting $\vec{\varphi}$}
    

    \textbf{exchangeArrays}($\vec{\psi}$, $\vec{\varphi}$, $r'$):
    
    \Indp 
    
        \textbf{exchangeArrays}($\vec{\psi}$, $0$, $\vec{\varphi}$, $0$, $r'$, $\dim(\vec{\psi})$)
    
    \Indm

\end{algorithm}

\end{figure*}

Fortunately, our algorithms share a useful property which simplifies their costs.
We will see that nearly all simulated quantum operators in this manuscript yield \textit{pairwise} communication, whereby a node exchanges amplitudes with a single other unique node. To be precise, if rank $r$ sends amplitudes to $r'$, then it is the only rank to do so, and so too does it receive amplitudes only from $r'$. 
We visualise this below where a circle indicates one of four nodes and an arrow indicates one of four total passed messages:
\begin{center}
    \includegraphics[width=.55\columnwidth]{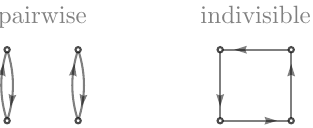}
    \end{center}
%
%
This innocuous property enables significant speedup and simplifications of the communication and synchronisation protocols.
It means all exchanges between paired nodes can happen concurrently without queueing or scheduling, and allows us to abstract almost all communication (MPI) code invoked by this work's algorithms into the few lines of Alg.~\ref{alg:mpi_defs}.

Knowing our communications are pairwise radically reduces the space of possible patterns, makes performance easier to predict, and
may help configure network switches~\cite{papaphilippou2020high}. It also means that communication graphs are trivial to partition into uniformly loaded subgraphs and hence that communication is easy to optimise for restrictive networks like tree topologies~\cite{zhao2020optimized}.
Finally, it simplifies the message passing involved in all algorithms within this manuscript into \textit{five} distinct paradigms, which we visualise in Fig.~\ref{fig:distrib_mem_exchange_patterns}.

\clearpage

\clearpage

\section{Distributed statevector algorithms}
\label{sec:distributed_statevector_algorithms}

This section will derive six novel distributed algorithms to simulate many-control gates, SWAP gates, many-target gates, tensors of Pauli operators, phase gadgets and Pauli gadgets.
As a means of review, we begin however by deriving an existing distributed simulation technique of the one-target gate, generalising the local algorithm of Sec.~\ref{sec:intro_sim_1targ_gate}.

\subsection{One-target gate}
\label{sec:distrib_1qb_gate}
\noindent 
\begin{equation*}
\Qcircuit @C=1em @R=.7em {
& \qw & \qw \\
& \gate{\hat{M}} & \qw \\
& \qw & \qw 
}
\end{equation*}

Let us first revisit the one-target gate, the staple of quantum circuits,
for which we derived a local simulation algorithm in Sec.~\ref{sec:intro_sim_1targ_gate}.
Our distributed simulation of this gate will inform the target performance of all other algorithms in this manuscript.
We here derive a distributed in-place simulation of the one-target gate upon an $N$-qubit pure statevector distributed between $W=2^w$ nodes, with $\Lambda=2^{N-w}$ amplitudes per-node. Our algorithm prescribes $\mathcal{O}(\Lambda)$ bops and flops per-node, exactly $\Lambda$ writes per-node, at most a single round of message passing in which case $2^N$ total amplitudes are exchanged over the network, and an $\mathcal{O}(1)$ memory overhead.

Let $\hat{M}_t$ denote a general one-target gate upon qubit $t\ge 0$, described by a complex matrix 
\begin{align} 
\renewcommand*{\arraystretch}{1}
\matr{M} = \begin{pmatrix} m_{00} & m_{01} \\ m_{10} & m_{11}
\end{pmatrix} \in \mathbb{C}^{2\times 2}.
\end{align}
We will later see bespoke distributed methods for faster simulation of certain families of one-target and separable gates, but we will here assume $\matr{M}$ is completely general and unconstrained. We seek to apply $\hat{M}_t$ upon an arbitrary $N$-qubit pure state $\ket{\Psi}\numsub{N}$ which is distributed between $W=2^w$ nodes, each storing sub-statevector $\vec{\psi}$ of size $\Lambda=2^{N-w}$, and an equal-sized communication buffer $\vec{\varphi}$.

We showed in Eq.~\ref{eq:serial_1qb_amp_update_rule} that $\hat{M}_t$ modifies an amplitude $\alpha_i$ of $\ket{\Psi}\numsub{N}$ to become a linear combination of $\alpha_i$ and $\alpha_{i_{\neg t}}$. Ergo to modify $\alpha_i$, we must determine within which node the paired $\alpha_{i_{\neg t}}$ is stored. Recall that the $j$-th local amplitude stored within node $r\ge 0$ corresponds to global amplitude $\alpha_i$ satisfying
\begin{align}
    \ket{i}\numsub{N}
    & \equiv 
    \ket{i_{[N-1]}}\numsub{1} \dots 
    \ket{i_{[t]}}\numsub{1} \dots 
    \ket{i_{[0]}}\numsub{1}
    \\
    &\equiv \ket{r}\numsub{w}\ket{j}\numsub{N-w}.
\end{align}
The paired amplitude $\alpha_{i_{\neg t}}$ corresponds to basis state
\begin{align}
\ket{i_{\neg t}}\numsub{N} & \equiv \ket{i_{[N-1]}}\numsub{1} \dots 
    \ket{\LogicalNot  i_{[t]}}\numsub{1} \dots 
    \ket{i_{[0]}}\numsub{1}
    \\
    &=
    \ket{r'}\numsub{w} \ket{j'}\numsub{N-w}.
\end{align}
Flipping bit $t$ of integer $i$ must modify either the $w$-bit prefix or the ($N-w$)-bit suffix of $i$'s bit sequence, and ergo modify either $j$ or $r$. Two scenarios emerge:
\begin{enumerate}
    \item When $t < N - w$, then 
    \begin{align}
        r' &= r,
            &
        j' &= j_{\neg t}.
        \label{eq:distrib_1qb_gate_emb_parallel_cond}
    \end{align}
    All paired amplitudes $\alpha_{i \neg t}$ are stored within the same rank $r$ as $\alpha_i$, and so every node already contains all amplitudes which will determine its new values. No communication is necessary, and we say the method is ``embarrassingly parallel". 
    
    Furthermore, since
    \begin{align}
        \hat{M}_{t<N-w} \left( 
        \ket{r}\numsub{w} \ket{\psi}\numsub{N-w}
        \right)
        =
        \ket{r}\numsub{w}
        \left( 
        \hat{M}_{t} 
        \ket{\psi}\numsub{N-w}
        \right),
    \end{align}
    we can modify a node's local sub-statevector $\vec{\psi}$ in an identical manner to the local simulation of $\hat{M}_{t} \ket{\psi}$. We simply invoke Alg.~\ref{alg:serial_1qb_gate_alg} upon $\vec{\psi}$ on every node.

    \item When $t \ge N-w$, then
        \begin{align}
        r' &= r_{\neg(t-(N-w))} 
            &
        j' &= j.
    \end{align}
    Every paired amplitude to those in node $r$ is stored in a single other node $r'$. We call this the ``pair node". Ergo each node must send its full sub-statevector $\vec{\psi}$ to the buffer $\vec{\varphi}'$ of its pair node. This is the paradigm seen in Fig.~\ref{fig:distrib_mem_exchange_patterns}~\textbf{a)}. Thereafter, because $j'=j$ above, each local amplitude $\vec{\psi}[j]$ will linearly combine with the received amplitude $\vec{\varphi}[j]$, weighted as per Eq.~\ref{eq:serial_1qb_amp_update_rule}.
    
\end{enumerate}

\begin{figure}[b]
    \centering
    \includegraphics[width=\columnwidth]{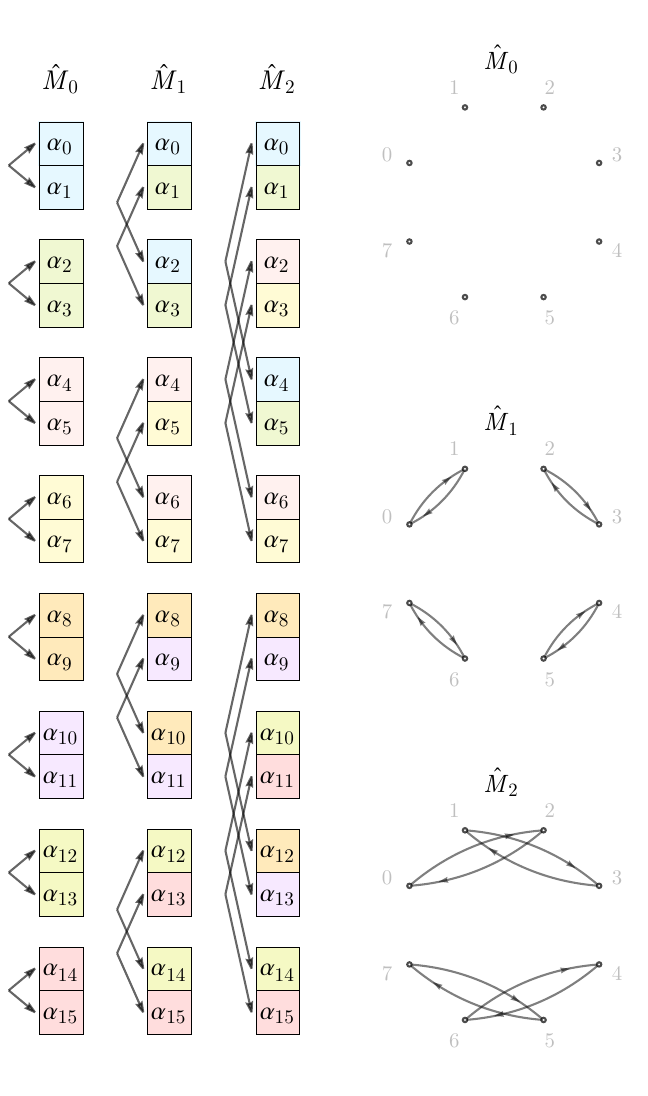}
    \caption{
        The memory access and communication patterns of Alg.~\ref{alg:distrib_1qb_gate}'s distributed simulation of the one-target gate $\hat{M}_{{t}}$. Each (of the left three) column shows the full $4$-qubit statevector $\ket{\Psi}\numsub{4}$ divided between $8$ nodes each containing $\Lambda=2$ amplitudes. Arrows between amplitudes of different nodes indicate the need for communication. The rightmost column shows the resulting communication topology wherein circles represent nodes and directional arrows indicate the sending of amplitudes.
    }
    \label{fig:distrib_1qb_gate_mem_comm}
\end{figure}

The memory access and communication patterns of these scenarios are illustrated in Fig.~\ref{fig:distrib_1qb_gate_mem_comm}.
In general, we cannot know in advance which of scenarios \textcolor{red}{\textbf{1.\ }}or \textcolor{red}{\textbf{2.\ }}will be invoked during distributed simulation, because $N$, $w$ and $t$ are all user-controlled parameters. So our algorithm to simulate $\hat{M}_t$ must incorporate both. We formalise this scheme in Alg.~\ref{alg:distrib_1qb_gate}.

Let us compare the costs of local vs distributed simulation of the one-target gate (i.e. Alg~\ref{alg:serial_1qb_gate_alg} against Alg~\ref{alg:distrib_1qb_gate}). The former prescribed a total of $\mathcal{O}(2^N)$ bops and flops to be serially performed by a single machine. The latter exploits the parallelisation of $W$ distributed nodes and involves only $\mathcal{O}(\Lambda)=\mathcal{O}(2^N/W)$ bops and flops per machine (a uniform load), suggesting an $\mathcal{O}(W)$ speedup. However, when the upper $w$ qubits are targeted, inter-node communication was required and all $\mathcal{O}(2^N)$ amplitudes were simultaneously exchanged over the network in a pairwise fashion. While the relative magnitude of this overhead depends on many physical factors (like network throughput, CPU speeds, cache throughput) and virtual parameters (the number of qubits, the prescribed memory access pattern), distributed simulation of this kind in realistic regimes shows excellent weak scaling and the network overhead is tractable~\cite{jones2019quest}.
This means that if we introduce an additional qubit (doubling the serial costs) while doubling the number of distributed nodes, our total runtime should be approximately unchanged.

As such, our one-target gate distributed simulation sets a salient performance threshold. We endeavour to maintain this weak scaling in the other algorithms of this manuscript.
%

\begin{algorithm}[h]
    \DontPrintSemicolon
    \caption[blah blah]{
        \AlgTagDistributed \AlgTagStateVector
        \\
        One-target gate $\hat{M}_t$, where 
            $\matr{M} = \begin{psmallmatrix} m_{00} & m_{01} \\
            m_{10} & m_{11}
            \end{psmallmatrix}$,
        upon an $N$-qubit pure statevector distributed between $2^w$ nodes as local $\vec{\psi}$ (buffer $\vec{\varphi}$).
        \\
        \protect{\vspace{4pt}\begin{center} $ 
        \Qcircuit @C=1em @R=.7em {
        & \gate{\hat{M}} & \qw
        }$ \end{center} }
%
        \begin{center}
        \AlgTagBops{$\mathcal{O}(\Lambda)$}%
        \AlgTagFlops{$\mathcal{O}(\Lambda)$}%
        \AlgTagNumSerialRounds{$0$ or $1$} \\
        \AlgTagAmpsTransferred{$\mathcal{O}(2^N)$}%
        \AlgTagMemOverhead{$\mathcal{O}(1)$}%
        \AlgTagMemWrites{$\Lambda$}
        \end{center}
    }
    \label{alg:distrib_1qb_gate}

    \textbf{distrib\_oneTargGate}($\vec{\psi}$, $\vec{\varphi}$, $\matr{M}$, $t$): 
    
    \Indp
        
        $r = \textbf{getRank()}$
        \tcp*{Alg.~\ref{alg:mpi_defs}}
        
        $w = \log_2( \textbf{getWorldSize()} )$
        \tcp*{Alg.~\ref{alg:mpi_defs}}
        
        $\Lambda = \dim(\vec{\psi})$
        
        $N =  \log_2(\Lambda) + w$
        
        \codegap

        \tcp{embarrassingly parallel}
        
        \textbf{if} $t < N-w$:
        
        \Indp 
        
            \textbf{local\_oneTargGate}($\vec{\psi}$, $\matr{M}$, $t$)
            \tcp*{Alg.~\ref{alg:serial_1qb_gate_alg}}
        
        \Indm 

        \codegap 

        \tcp{full sub-state exchange is necessary}
        
        \textbf{else}:
        
        \Indp

            \codegap 

            \tcp{exchange with $r'$}
        
            $q = t-(N-w)$
        
            $r' = \textbf{flipBit}$($r$, $q$)
            \tcp*{Alg.~\ref{alg:bit_twiddles}}
            
            \textbf{exchangeArrays}($\vec{\psi}$, $\vec{\varphi}$, $r'$)
            \tcp*{Alg.~\ref{alg:mpi_defs}}

            \codegap 

            \tcp{determine row of $\matr{M}$}
                    
            $b = \textbf{getBit}$($r$, $q$)
            \tcp*{Alg.~\ref{alg:bit_twiddles}}

            \codegap 

            \tcp{modify local amplitudes}

            \AlgThreadComment{multithread}
        
            \textbf{for} $j$ \textbf{in} \textbf{range}($0$, $\Lambda$):
    
            \Indp
                
               $\vec{\psi}[j] \; = \; m_{b,b} \; \vec{\psi}[j] + m_{b, \LogicalNot b} \; \vec{\varphi}[j] $
    
            \Indm 
            
        \Indm 
    
    \Indm     
    
    \end{algorithm}

\clearpage

\subsection{Many-control one-target gate} %
\label{sec:distrib_many_ctrl_one_targ_gate}
\noindent 
\begin{equation*}
\Qcircuit @C=1em @R=.7em {
& \ctrl{2} & \qw \\
& \ctrl{1} & \qw \\
& \gate{\hat{M}} & \qw
}
\end{equation*}

Introducing control qubits to the previous one-target gate empowers it to become entangling and universal~\cite{lloyd1995universal,deutsch1995universality}, which for its relative simplicity, establishes it as the entangling primitive of many quantum algorithms~\cite{barenco1995elem}. Section~\ref{sec:intro_manyctrl_gate} derived a local, serial algorithm to effect the many-control one-target gate which we now adapt for distributed simulation. 
We derive an in-place distributed simulation of the one-target gate with $s$ control qubits upon an $N$-qubit pure state. Our method prescribes as few as $\mathcal{O}(s\Lambda/2^s)$ (at most $\mathcal{O}(\Lambda)$) bops per node, $\mathcal{O}(\Lambda/2s)$ (at most $\mathcal{O}(\Lambda)$) flops and writes per node, at most a \textit{single} round of communication whereby a total of $\mathcal{O}(2^N/2^s)$ amplitudes are exchanged, and a fixed memory overhead.

We consider the operator $C_{\vec{c}}(\hat{M}_t)$ where $\vec{c}=\{c_0,\dots,c_{s-1}\}$ is an arbitrarily ordered list of $s$ unique control qubits, $t \notin \vec{c}$ is the target qubit, and where $\hat{M}$ is described by matrix $\matr{M} \in \mathbb{C}^{2\times 2}$ (as in the previous section).
We seek to apply $C_{\vec{c}}(\hat{M}_t)$
upon an arbitrary $N$-qubit pure state $\ket{\Psi}\numsub{N}$ which is distributed between $W=2^w$ nodes, each with sub-statevector $\vec{\psi}$ of size $\Lambda=2^{N-w}$ and an equal-sized communication buffer $\vec{\varphi}$.

\begin{figure}[t]
    \centering
    \includegraphics[width=.85\columnwidth]{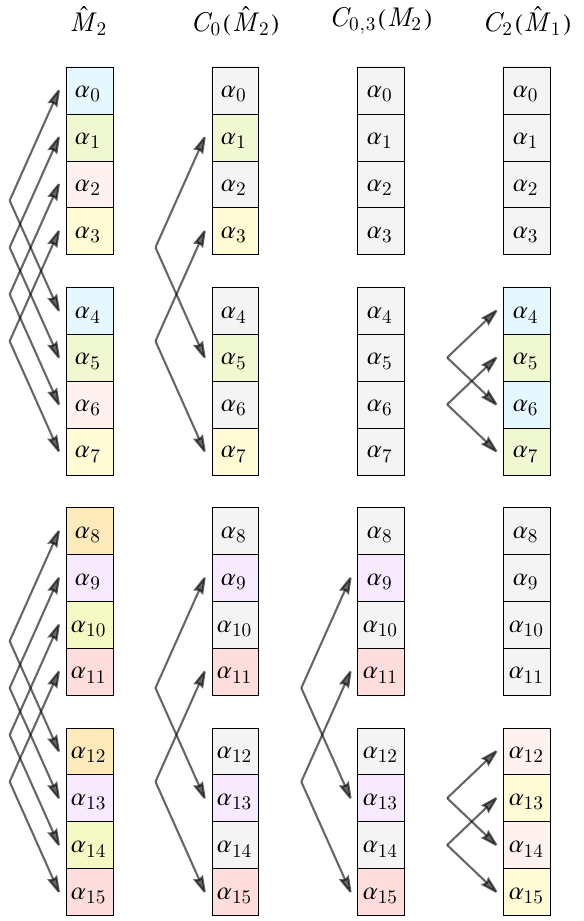}
    \caption{
        The memory access pattern of Alg.~\ref{alg:distrib_manyctrl_gate}'s distributed simulation of the many-control one-target gate $C_{\vec{c}}(\hat{M}_t)$. 
        Each column shows $16$ amplitudes distributed between $4$ nodes.
        If a node contains only grey amplitudes which fail the control condition, it need not perform any communication.
    }
    \label{fig:distrib_manyctrl_gate_mem}
\end{figure}

Section~\ref{sec:intro_manyctrl_gate} established that $C_{\vec{c}}(\hat{M}_t)$ modifies only the $2^{N-s}$ global amplitudes $\alpha_i$ of full-state $\ket{\Psi}\numsub{N}$ which satisfy the control condition
\begin{align}
    i_{[c_n]} = 1, \;\; \forall c_n \in \vec{c},
    \tag{reiteration of \ref{eq:serial_manyctrl_ctrl_cond}}
\end{align}
doing so under the non-controlled action of $\hat{M}_t$. Since all amplitudes failing the control condition are not modified nor are involved in the modification of other amplitudes, they need not be communicated over the network in any circumstance. This already determines the communication pattern, which we illustrate (as a memory access diagram) in Figure~\ref{fig:distrib_manyctrl_gate_mem}.

Explicitly deriving this pattern is non-trivial. We again invoke that each \textit{local} amplitude index $j$ in node $r$ is equivalent to a \textit{global} index $i$ satisfying
\begin{align}
    \ket{i}\numsub{N}
    & \equiv 
    \ket{i_{[N-1]}}\numsub{1} \dots 
    \ket{i_{[t]}}\numsub{1} \dots 
    \ket{i_{[0]}}\numsub{1}
    \\
    & \equiv \ket{r}\numsub{w}\ket{j}\numsub{N-w}.
\end{align}
We call $\ket{r}\numsub{w}$ the prefix substate and $\ket{j}\numsub{N-w}$ the suffix. As per Eq.~\ref{eq:distrib_1qb_gate_emb_parallel_cond}, when $t < N-w$, simulation is embarrassingly parallel regardless of $\vec{c}$. But when $t \ge N-w$, \textit{three} distinct scenarios emerge.

\begin{enumerate}
    \item \textbf{When all controls lie within the prefix}, i.e.
    \begin{align}
        c_n \ge N-w, \;\;\; \forall  c_n \in \vec{c}.
    \end{align}
    The control condition is determined entirely by a node's rank $r$, so that \textit{every} or \textit{none} of the amplitudes therein satisfy it. Ergo, some nodes simulate $\hat{M}_t$ as normal while others do nothing; this is seen in column 4 of Fig.~\ref{fig:distrib_manyctrl_gate_mem}.
    Because only a fraction $1/2^s$ of nodes satisfy $r_{[c_n-(N-w)]}=1 \, \forall c_n \in \vec{c}$, the number of communicating nodes and hence the total data size communicated exponentially shrinks with additional controls. 
    We visualise this communication pattern below where circles indicate nodes, grey symbols indicate their rank, arrows indicate a message, and black symbols indicate the number of amplitudes in each message.
    \begin{center}
        \includegraphics[width=.55\columnwidth]{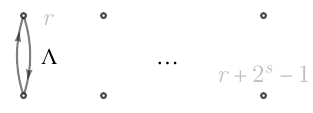}
    \end{center}

    \item \textbf{When all controls lie within the suffix}, i.e.
    \begin{align}
        c_n < N-w, \;\;\; \forall c_n \in \vec{c}.
    \end{align}
    The control condition is independent of rank, and since every node then contains every assignment of bits $\vec{c}$, \textit{all} nodes contain amplitudes which satisfy the condition and need communicating. There are precisely $\Lambda/2^s$ such amplitudes per node. This scenario is seen in column~2 of Fig.~\ref{fig:distrib_manyctrl_gate_mem}, and is illustrated below.
    \begin{center}
        \includegraphics[width=.5\columnwidth]{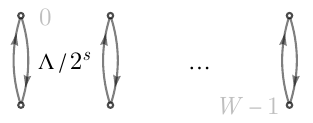}
    \end{center}
    Because not \textit{all} local amplitudes are communicated, it is prudent to pack only those which \textit{are} into the local communication buffer before exchanging the packed buffer subset. This is paradigm \textbf{b)} of Fig.~\ref{fig:distrib_mem_exchange_patterns}, and drastically reduces the total number amplitudes transferred over the network by factor $1/2^s$.
    
\item
    \textbf{When the controls are divded between the prefix and suffix}, i.e.
    \begin{align}
        \exists \;\; c_n \ge N-w \;\;\text{ and }\;\; c_k < N-w.
    \end{align}
    Only a fraction of ranks satisfy the control condition, as do only a fraction of the amplitudes therein; so \textit{some} nodes exchange only \textit{some} of their amplitudes.
    Let us distinguish between the controls acting upon the prefix and suffix substates:
    \begin{gather}
        \vec{c}^{(p)} = \{ c_n \in \vec{c}  : \; c_n \ge N-w  \},
        \\
        \vec{c}^{(s)} = \{ c_n \in \vec{c} \; : \; c_n < N-w   \},
        \\
        s^{(p)} = \dim(\vec{c}^{(p)}),
        \;\;\;\;
         s^{(s)} = \dim(\vec{c}^{(s)}).
    \end{gather}
    A fraction $1/s^{(p)}$ nodes communicate, exchanging fraction $1/s^{(s)}$ of their local amplitudes. This is seen in column~3 of Fig.~\ref{fig:distrib_manyctrl_gate_mem}, and below.
    \begin{center}
        \includegraphics[width=.55\columnwidth]{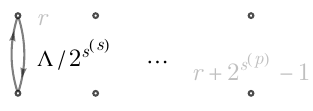}
    \end{center}
    Once again, the communicating nodes will pack the relevant subset of their local amplitudes into buffers before exchanging, as per Fig.~\ref{fig:distrib_mem_exchange_patterns}~\textbf{2)}.
    
\end{enumerate}

    \begin{algorithm}[tb]
        \DontPrintSemicolon
        \caption[bah blah]{
            \AlgTagDistributed \AlgTagStateVector%
    \\
        Many-control one-target gate $C_{\vec{c}}(\hat{M}_t)$ with $s$ unique control qubits $\vec{c}=\{c_0,\dots,c_{s-1}\}$ and target $t \notin \vec{c}$, described by matrix $\matr{M} = \begin{psmallmatrix} m_{00} & m_{01} \\
            m_{10} & m_{11}
            \end{psmallmatrix} \in \mathbb{C}^{2\times 2}
        $,
        applied to an $N$-qubit pure statevector distributed between $2^w$ nodes as local $\vec{\psi}$ (buffer $\vec{\varphi}$).
        \\ 
        \protect{\begin{center} 
        $ \Qcircuit @C=1em @R=.7em {
            & \ctrl{2} & \qw \\
            & \ctrl{1} & \qw \\
            & \gate{\hat{M}} & \qw } $ 
        \end{center}}
        \begin{center}
        \AlgTagBops{best $\mathcal{O}(s \Lambda/2^s)$, worst $\mathcal{O}(\Lambda)$} \\
        \AlgTagFlops{best $\mathcal{O}(\Lambda/2^s)$, worst $\mathcal{O}(\Lambda)$} \\
        \AlgTagNumSerialRounds{$0$ or $1$}%
        \AlgTagAmpsTransferred{$\mathcal{O}(2^N/2^s)$}\\
        \AlgTagMemOverhead{$\mathcal{O}(1)$}%
        \AlgTagMemWrites{best $\Lambda/2^s$, worst $\Lambda$}
        \end{center}
    }
        \label{alg:distrib_manyctrl_gate}
        
        \textbf{distrib\_manyCtrlOneTargGate}($\vec{\psi}$, $\vec{\varphi}$, $\vec{c}$, $\matr{M}$, $t$):
        
        \Indp 
        
            $r = \textbf{getRank()}$
            \tcp*{Alg.~\ref{alg:mpi_defs}}
            
            $\lambda = \log_2( \dim(\vec{\psi}))\;\;\;\;$ \tcp{$= N-w$}
            
            \codegap

            \tcp{separate prefix and suffix controls}
        
            $\vec{c}^{(p)} = \{ q - \lambda  \,: \; q \ge \lambda, \;\;\forall q \in \vec{c} \}$
        
            $\vec{c}^{(s)} = \{ q \; : \; q < \lambda, \;\; \forall q \in \vec{c} \}$
            
            
            
            \codegap 
            
            \tcp{halt if $r$ fails control condition}
            
            \textbf{if} \textbf{not} \textbf{allBitsAreOne}($r$, $\vec{c}^{(p)}$):
            \tcp*{Alg.~\ref{alg:bit_twiddles}}
            
            \Indp 
            
                \textbf{return}
                
            \Indm 
            
            \codegap
            
            \tcp{update as $C_{\vec{c}}(\hat{M}_t)\ket{\Psi}
           = \ket{r} \big( C_{\vec{c}^{(s)}}(\hat{M}_t)\ket{\psi}
           \big)$}
            
            \textbf{if} $t < \lambda$:
            
            \Indp 
                \textbf{local\_manyCtrlOneTargGate}($\vec{\psi},\vec{c}^{(s)},\matr{M}, t$)
                
                \tcp*{Alg.~\ref{alg:serial_manyctrl_gate_alg}}
            
            \Indm
            
            
            \tcp{exchange with $r'$ is necessary}
        
            \textbf{if} $t \ge \lambda$:
        
            \Indp

                \tcp{all local $\alpha_j$ satisfy condition,}
                \tcp{so controls can be disregarded}
        
                \textbf{if} $\dim(\vec{c}^{(s)}) = 0$:
               
                \Indp 
                   \textbf{distrib\_oneTargGate}($\vec{\psi}$, $\matr{M}$, $t$)
                   
                   \tcp*{Alg.~\ref{alg:distrib_1qb_gate}}
                
                \Indm
                
                
                \tcp{only subset of local $\alpha_j$ satisfy,}
                
                \tcp{determined only by suffix controls}
                
                \textbf{else}:
                
                \Indp 

                    $r' = \textbf{flipBit}$($r$, $t-\lambda$)
                    \tcp*{Alg.~\ref{alg:bit_twiddles}}
                
                     \textbf{distrib\_ctrlSub}($\vec{\psi},\,\vec{\varphi},\,r',\,\vec{c}^{(s)},\,\matr{M},\,t$)
                     
                     \tcp*{Alg.~\ref{alg:distrib_manyctrl_gate_subroutine}}

                \Indm
            
            \Indm 
            
        \Indm 

    \end{algorithm}

In all these scenarios, the rank $r'$ of the pair node with which node $r$ communicates (controls permitting) is determined by the target qubit $t$ in the same manner as for the non-controlled gate $\hat{M}_t$.
There are several additional observations to make when decomposing distributed simulation of $C_{\vec{c}}(\hat{M}_t)$ into subtasks which we succinctly summarise in the comments of Alg.~\ref{alg:distrib_manyctrl_gate}.

The performance of this algorithm varies drastically with the configuration of qubits $\vec{c}$ and $t$, but all costs are upperbounded by those to simulate $\hat{M}_t$ via Alg.~\ref{alg:distrib_1qb_gate}. When $t < N-w$, the method is embarrassingly parallel. Otherwise, when none of the left-most $w$ qubits are controlled, every node has an identical task of sending and locally modifying $\Lambda/2^s$ amplitudes. But when those $w$ qubits are controlled, the load per-node varies and in the worst case, a single node may send and modify $\Lambda$ amplitudes while another node does nothing. Note this latter node need \textit{not} wait idly, since synchronisation is not  needed until its next prescribed communication in the program, and it can in the interim proceed with other local tasks (like simulating the next gate in a circuit). Still, we measure the algorithm's performance by its slowest node. In all communicating scenarios, the total number of amplitudes sent over the network is $2^N/2^s$, performed in a single serial round. We summarise these costs in the caption of Alg.~\ref{alg:distrib_manyctrl_gate}.

    \begin{algorithm}[t]
        \DontPrintSemicolon
        \label{alg:distrib_manyctrl_gate_subroutine}
    \caption[bah blah]{A subroutine of Alg.~\ref{alg:distrib_manyctrl_gate}, triggered in aforementioned scenarios \textbf{\textcolor{red}{2.}} and \textbf{\textcolor{red}{3.}}, whereby a subset of the node's sub-statevector is packed and exchanged with the pair node before local modification.
    }
        
        \textbf{distrib\_ctrlSub}($\vec{\psi}$, $\vec{\varphi}$, $r'$, $\vec{c}$, $\matr{M}$, $t$):

        \Indp

            $\Lambda = \dim(\vec{\psi})$
            
            $s = \dim(\vec{c})$

            $l$ = $\Lambda/2^s$

            \textbf{sort}($\vec{c}$)
            
            \codegap
        
            \tcp{pack subset of $\alpha_j$ into buffer}

            \AlgThreadComment{multithread}
            
            \textbf{for} $k$ \textbf{in} \textbf{range}(0, $l$): 
            
            \Indp 
            
                $j$ = \textbf{insertBits}($k$, $\vec{c}$, $1$)
                \tcp*{Alg.~\ref{alg:bit_twiddles}}
                
                $\vec{\varphi}[k] = \vec{\psi}[j]$
                
            \Indm
            
            \codegap

            \tcp{swap buffer subarrays (Fig.~\ref{fig:distrib_mem_exchange_patterns}.2),}
            \tcp{receiving amps at $\vec{\varphi}[l\dots]$}
            
            \textbf{exchangeArrays}($\vec{\varphi},\,0,\,\vec{\varphi},\,l,\,l,\,r'$)
            \tcp*{Alg.~\ref{alg:mpi_defs}}
            
            \codegap 
            
            \tcp{determine row of $\matr{M}$}
            
            $b = \textbf{getBit}$($r$, $t-\log_2(\Lambda)$)
            \tcp*{Alg.~\ref{alg:bit_twiddles}}

            \codegap 

            \tcp{update local amplitudes}

            \AlgThreadComment{multithread}
            
            \textbf{for} $j$ \textbf{in} \textbf{range}(0, $l$):
            
            \Indp 
            
                $k$ = \textbf{insertBits}($j$, $\vec{c}$, $1$)
                \tcp*{Alg.~\ref{alg:bit_twiddles}}
                

                $j'$ = $j+l$
                
               $\vec{\psi}[k]$ = $m_{b,b} \; \vec{\psi}[k] + m_{b, \LogicalNot b} \; \vec{\varphi}[j']$

            \Indm
            
        \Indm
        
    \end{algorithm}

\vspace{5.5cm}

\pagebreak

\subsection{Swap gate}%
\label{sec:swap_gate} %
\noindent 
\begin{equation*}
\Qcircuit @C=1em @R=.7em {
& \qswap & \qw \\
& \qw \qwx & \qw \\
& \qswap \qwx & \qw
}
\end{equation*}

Today's proposed quantum architectures have limited connectivity~\cite{preskill2018quantum}, constraining which qubits can undergo multi-qubit operators like the proceeding section's control gate.
Creating entanglement between arbitrary qubits sometimes requires \textit{swapping} qubits; either physically shuttling them as is typical in ion traps~\cite{murali2020architecting} or effectively exchanging their quantum states through multiple native operations as performed in superconducting platforms~\cite{gokhale2021faster,song2010simultaneous}. 
Through either implementation, the \text{SWAP} operation is of vital importance in the theory of quantum computation, appearing in quantum teleportation~\cite{vaidman1994teleportation}, proofs of universality~\cite{deutsch1995universality}, and in the compilations of generic circuits into those compatible with restricted architectures~\cite{taylor2017study}. We will also later see that distributed simulation of the \text{SWAP} gate itself is a helpful utility in efficiently simulating more advanced gates, motivating that the \text{SWAP} gate itself be made as efficient as possible.
In this section, we develop a distributed simulation of the SWAP gate upon an $N$-qubit statevector which prescribes $\mathcal{O}(\Lambda)$ bops per node, \textit{no} flops at all, a \textit{single} round of communication 
exchanging $2^N/2$ amplitudes total, only $\Lambda/2$ main memory writes per node, and a fixed memory overhead.

We seek to apply gate $\text{SWAP}_{t_1,t_2}$ upon qubits $t_1$ and $t_2>t_1$ of an $N$-qubit pure state $\ket{\Psi}$ distributed between $W=2^w$ nodes, where
\begin{align}
    \renewcommand*{\arraystretch}{1}
    \matr{\text{SWAP}} = \begin{pmatrix}
    1 & 0 & 0 & 0 \\
    0 & 0 & 1 & 0 \\
    0 & 1 & 0 & 0 \\
    0 & 0 & 0 & 1 
    \end{pmatrix}.
\end{align}

While we could naively leverage decomposition
\begin{align}
    \text{SWAP}_{t_1, t_2} \; \equiv \;
    C_{t_1}(\hat{X}_{t_2}) \; 
    C_{t_2}(\hat{X}_{t_1}) \;
    C_{t_1}(\hat{X}_{t_2})
\end{align}
and simulate each control-NOT via the previous section's Alg.~\ref{alg:distrib_manyctrl_gate}, this may (depending on $t_1$ and $t_2$) invoke as many as \textit{three} statevector exchanges and a total of $3/2 \times 2^{N}$ flops, communicated amplitudes, and memory writes. Instead, a greatly superior strategy is possible.

Let us re-express an $N$-qubit basis state of arbitrary state $\ket{\Psi}\numsub{N}=\sum_i\alpha_i\ket{i}\numsub{N}$ as (where $x,y,z\in\mathbb{N}$)
\begin{align}
    \ket{i}\numsub{N} \equiv 
    \ket{z}\numsub{N-t_2-1}\ket{i_{[t_2]}}\numsub{1}\ket{y}\numsub{t_2-t_1-1} \ket{i_{[t_1]}}\numsub{1}\ket{x}\numsub{t_1}.
\end{align}
The SWAP gate swaps the targeted qubits of $\ket{i}$;
\begin{align}
    \text{SWAP}_{t_1,t_2} & \ket{i}\numsub{N} \tag*{}
    \\
    = &   \ket{z}\numsub{N-t_2-1}\ket{i_{[t_1]}}\ket{y}\numsub{t_2-t_1-1} \ket{i_{[t_2]}}\numsub{1}\ket{x}\numsub{t_1}
    \tag*{}
    \\
   = & 
    \begin{cases}
    \ket{i}\numsub{N} & i_{[t_1]} = i_{[t_2]} 
    \\
    \ket{i_{\neg \{t_1,t_2\}}}\numsub{N} & i_{[t_1]} \ne i_{[t_2]} 
   \end{cases} 
\end{align}
This induces a change only when the principal bits differ; swapping them is ergo equivalent to flipping both.
Therefore $\text{SWAP}_{t_1,t_2}$ upon the full state $\ket{\Psi}\numsub{N}=\sum_i\alpha_i\ket{i}\numsub{N}$ swaps a subset of amplitudes;
\begin{align}
    \alpha_i \rightarrow \begin{cases}
        \alpha_i & i_{[t_1]} = i_{[t_2]}
            \\
        \alpha_{i_{\neg \{t_1,t_2\}}} & i_{[t_1]} \ne i_{[t_2]}.
    \end{cases}
\end{align}
We call $\alpha_{i_{\neg \{t_1,t_2\}}} = \alpha_{(i_{\neg t_1})_{\neg t_2}}$ the ``pair amplitude".
Recall that the $j$-th local amplitude stored within node $r$ corresponds to global amplitude $\alpha_i$ satisfying
\begin{align}
    \ket{i}\numsub{N}
    & \equiv \ket{r}\numsub{w}\ket{j}\numsub{\lambda},
    \;\;\;\;
    \text{where}
    \;\;\;\;
    \lambda = N - w.
\end{align}
Three distinct scenarios emerge (assuming $t_2>t_1$).

\begin{figure}[tb]
    \centering
    \includegraphics[width=.7\columnwidth]{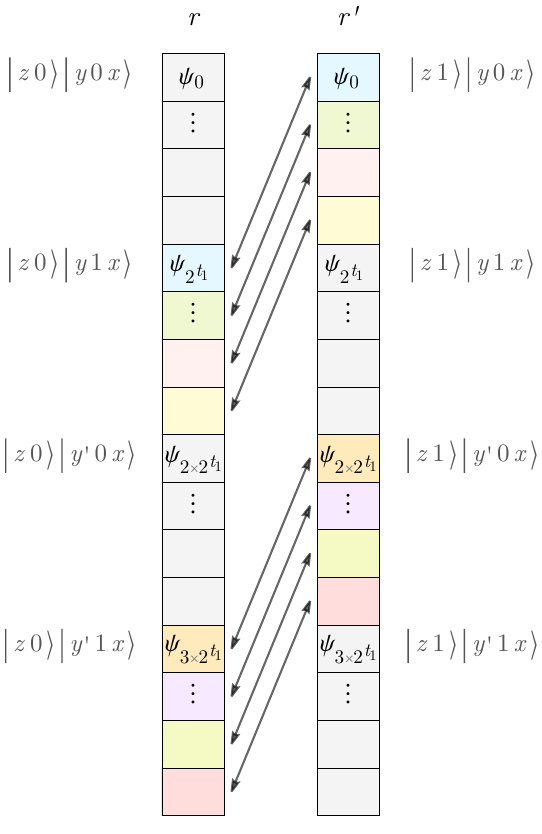}
    \caption{
        The amplitudes which require swapping in scenario \textbf{\textcolor{red}{3.}} of Alg.~\ref{alg:distrib_swap}'s distributed simulation of $\text{SWAP}_{t_1,t_2}$. Arrows (and colour) connect elements of the local sub-statevector $\vec{\psi}$ which must be swapped between nodes $r$ and $r'$. This is the memory exchange pattern \textbf{e)} of Fig.~\ref{fig:distrib_mem_exchange_patterns}.
    }
    \label{fig:distrib_swap_mem_case3}
\end{figure}

\begin{enumerate}
    \item \textbf{When $t_2 < \lambda$}, and consequently
    \begin{align}
        \ket{i_{\neg \{t_1,t_2\}}}\numsub{N} = \ket{r}\numsub{w} \ket{j_{\neg \{t_1,t_2\}}}\numsub{\lambda}.
    \end{align}
    The pair amplitude is contained within the same node and simulation is embarrassingly parallel.
    
    \item \textbf{When $t_1 \ge \lambda$}, such that
    \begin{align}
        \ket{i_{\neg \{t_1,t_2\}}}\numsub{N} = \ket{r_{\neg\{ t_1-\lambda,t_2-\lambda\}}}\numsub{w} \ket{j}\numsub{\lambda}.
    \end{align}
    If $r_{[t_1-\lambda]} \ne r_{[t_2-\lambda]}$ (as satisfied by \textit{half} of all nodes) then all local amplitudes
    in node $r$ must be exchanged with their pair amplitudes within pair node $r' = r_{\neg\{ t_1-\lambda,t_2-\lambda\}}$. No modification is necessary, so the nodes directly swap their sub-statevector $\vec{\psi}$ as per Fig.~\ref{fig:distrib_mem_exchange_patterns}~\textbf{d)}. The remaining nodes contain only amplitudes with global indices $i$ satisfying $i_{[t_1]} = i_{[t_2]}$ and so do nothing. In total $2^N/2$ amplitudes are exchanged in parallel batches of $\Lambda$.
    \begin{center}
        \includegraphics[width=.55\columnwidth]{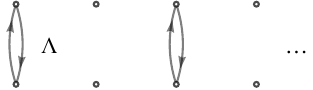}
    \end{center}
    
    \item \textbf{When $t_1 < \lambda$ and $t_2 \ge \lambda$}, in which case
    \begin{align}
        \ket{i_{\neg \{t_1,t_2\}}}\numsub{N} = \ket{r_{\neg (t_2-\lambda)}}\numsub{w} \ket{j_{\neg t_1}}\numsub{\lambda}.
    \end{align}
    Every node $r$ must exchange amplitudes with pair node $r'=r_{\neg (t_2-\lambda)}$, but only those amplitudes of local index $j$ satisfying $j_{[t_1]} \ne r_{[t_2-\lambda]}$; this is \textit{half} of all local amplitudes. In this scenario, a total of $2^N/2$ amplitudes are exchanged in parallel batches of $\Lambda/2$, and the node load is homogeneous. Pairs exchange via Fig.~\ref{fig:distrib_mem_exchange_patterns}~\textbf{e)}.
    \begin{center}
        \includegraphics[width=.5\columnwidth]{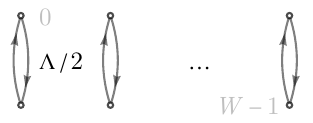}
    \end{center}
    Notice that the destination local address $j'=j_{\neg t_1}$ differs from the source local address; we visualise this in Fig.~\ref{fig:distrib_swap_mem_case3}.
    We expand upon the nuances of this scenario below.

\end{enumerate}

Because scenario \textcolor{red}{\textbf{3.}} requires that paired nodes exchange only \textit{half} their local amplitudes, these amplitudes should first be packed into the communication buffers as per Fig.~\ref{fig:distrib_mem_exchange_patterns}~\textbf{b)}, like was performed for the multi-controlled gate in Sec.~\ref{sec:distrib_many_ctrl_one_targ_gate}.
This means packing every \textit{second} contiguous batch of $2^{t_1}$ amplitudes before swapping the buffers, incurring local memory penalties but \textit{halving} the communicated data. Note too that when $t_1 = \lambda - 1$, packing is unnecessary; the first (or last) contiguous half of a node's sub-statevector can be directly sent to the pair node's buffer, although we exclude this optimisation from our pseudocode.

\begin{algorithm}[b]
\DontPrintSemicolon
\caption[blah blah]{
    \AlgTagDistributed \AlgTagStateVector%
    \\
    $\text{SWAP}_{t_1,t_2}$ gate upon qubits $t_1$ and $t_2>t_1$ of an $N$-qubit pure state distributed between $W=2^w$ nodes as local arrays $\vec{\psi}$ with buffers $\vec{\varphi}$.
    \protect{\begin{center} 
        $ \Qcircuit @C=1em @R=1.4em {
& \qswap & \qw \\
& \qswap \qwx & \qw
} $ 
        \end{center}}
         \begin{center}
        \AlgTagBops{$\mathcal{O}(\Lambda)$}%
        \AlgTagFlops{$0$}%
        \AlgTagNumSerialRounds{$0$ or $1$} \\
        \AlgTagAmpsTransferred{$2^N/2$}%
        \AlgTagMemOverhead{$\mathcal{O}(1)$}%
        \AlgTagMemWrites{$\Lambda/2$}
        \end{center}
}
\label{alg:distrib_swap}

\textbf{distrib\_swapGate}($\vec{\psi}$, $\vec{\varphi}$, $t_1$, $t_2$): 
\tcp*{$t_2>t_1$}

\Indp 

    $\Lambda$ = $\dim(\vec{\psi})$

    $\lambda$ = $\log_2(\Lambda)$
    
    
    \codegap
    
    \tcp{embarrassingly parallel}
    
    \textbf{if} $t_2 < \lambda$:
    
    \Indp 
    
        \tcp{loop $\ket{k}\numsub{\lambda-2} \equiv \ket{z}\numsub{N-t_2-1}\ket{y}\numsub{t_2-t_1-1} \ket{x}\numsub{t_1}$}

        \AlgThreadComment{multithread}
    
        \textbf{for} $k$ \textbf{in} \textbf{range}($0$, $\Lambda/4$):
        
        \Indp 
        
            \tcp{$\ket{j_{ab}}\numsub{\lambda} = \ket{z}\ket{a}\numsub{1}\ket{y}\ket{b}\numsub{1}\ket{x}$}
        
            $j_{11}$ = \textbf{insertBits}($k$, $\{t_1,t_2\}$, $1$)
            
            
            $j_{10}$ = \textbf{flipBit}($j_{11}$, $t_1$)
            
            
            $j_{01}$ = \textbf{flipBit}($j_{11}$, $t_2$)
            
            
            $\vec{\psi}[j_{01}],\; \vec{\psi}[j_{10}]$ = $\vec{\psi}[j_{10}], \; \vec{\psi}[j_{01}]$ 
            \tcp*{swap}
        
        \Indm 
        
    \Indm
    
    \codegap
    
    \tcp{swap entire $\vec{\psi}$ with pair...}
    
    \textbf{else if} $t_1 \ge \lambda$:
    
    \Indp 
    
        $t_1'$ = $t_1-\lambda$
        
        $t_2'$ = $t_2-\lambda$
        
        \tcp{if node contains any $i_{[t_1]} \ne i_{[t_2]}$}
    
        \textbf{if} \textbf{getBit}($r$, $t_1'$) $\ne$ \textbf{getBit}($r$, $t_2'$):
        
        \Indp 
        
            $r'$ = \textbf{flipBits}($r$, $\{t_1',\,t_2'\}$)
            
            \textbf{exchangeArrays}($\vec{\psi}$, $\vec{\varphi}$, $r'$)

            \AlgThreadComment{multithread}

            \textbf{for} $k$ \textbf{in} \textbf{range}($0$, $\Lambda$):

            \Indp 

                $\vec{\psi}[k]$ = $\vec{\varphi}[k]$

            \Indm 
        
        \Indm
    
    \Indm 
    
    \codegap
    
    \tcp{swap only half of $\vec{\psi}$ with pair}
    
    \textbf{else}:
    
    \Indp 

        \tcp{pack half of buffer}

        $b$ = $\LogicalNot$ \textbf{getBit}($r$, $t_2 - \lambda$) 

        \AlgThreadComment{multithread}

        \textbf{for} $k$ \textbf{in} \textbf{range}($0$, $\Lambda/2$):

        \Indp 

            $j$ = \textbf{insertBit}($k$, $t_1$, $b$)

            $\vec{\varphi}[k]$ = $\vec{\psi}[j]$

        \Indm


        \tcp{swap half-buffers}

        $r'$ = \textbf{flipBit}($r$, $t_2 - \lambda$)

        \textbf{exchangeArrays}($\vec{\varphi},\,0,\,\vec{\varphi},\,\Lambda/2,\,\Lambda/2,\,r'$)


        \tcp{overwrite local amps with buffer}

        \AlgThreadComment{multithread}

        \textbf{for} $k$ \textbf{in} \textbf{range}($0$, $\Lambda/2$):

        \Indp 

            $j$ = \textbf{insertBit}($k$, $t_1$, $b$)

            $\vec{\psi}[j]$ = $\vec{\psi}[k + l]$

        \Indm
    
    \Indm 

\Indm

\end{algorithm}

\begin{figure}[b]
    \centering 
    \includegraphics[width=.45\textwidth]{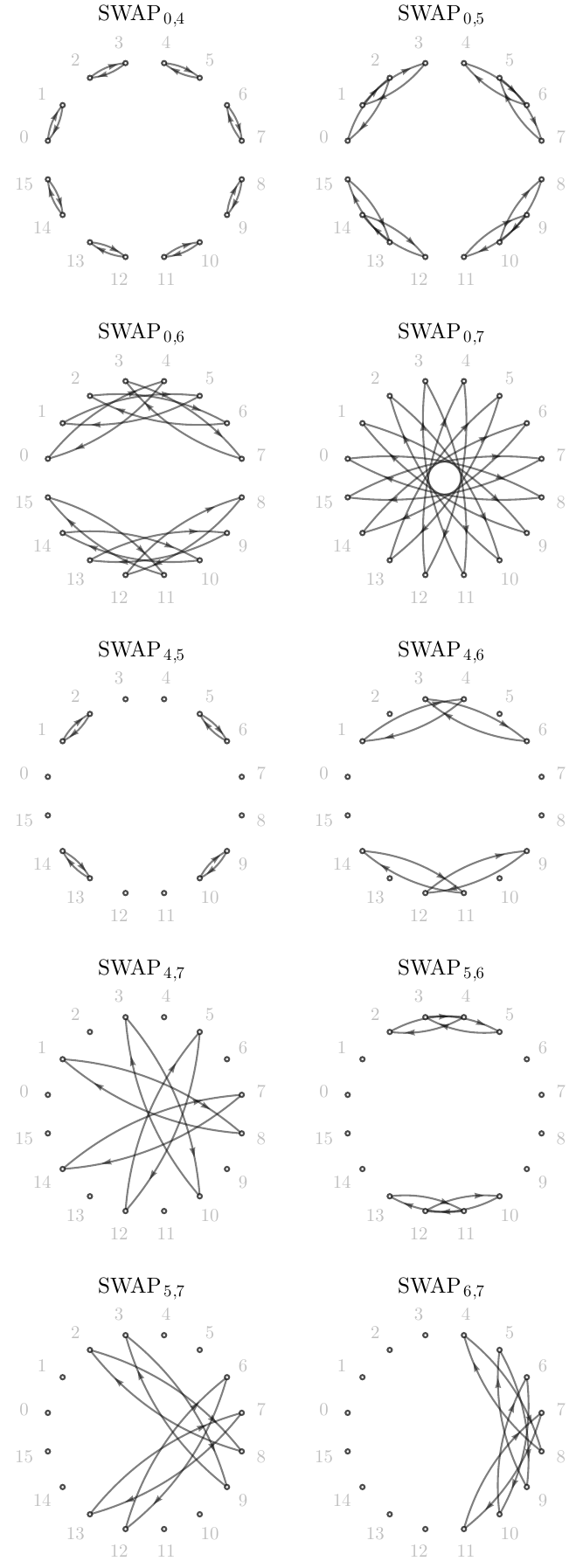}
    \caption{
        Some communication patterns of Alg.~\ref{alg:distrib_swap}'s distributed simulation of the SWAP gate. A circle indicates one of 16 nodes (ranked from $0$) and directional arrows indicate the sending of amplitudes when simulating the SWAP upon an $8$-qubit statevector. 
    }
    \label{fig:distrib_swap_comm}
\end{figure}

The local indices of amplitudes modified by the SWAP gate are determined through the same bit algebra used in the previous algorithms. We present the resulting scheme in Alg.~\ref{alg:distrib_swap}, and its
communication pattern in Fig.~\ref{fig:distrib_swap_comm}.
This algorithm prescribes \textit{no} floating point operations, an exchange of (at most) \textit{half} of all amplitudes, and at most a single round of pairwise communication. Despite being an experimentally fearsome two-qubit gate, we have shown the SWAP gate is substantially \textit{cheaper} to classically simulate than the one-qubit gate of Alg.~\ref{alg:distrib_1qb_gate}.

\clearpage

\subsection{Many-target gate} %
\label{sec:distrib_manytarg_gate}
\noindent 
\begin{equation*}
\Qcircuit @C=1em @R=.7em {
& \multigate{1}{\hat{M}} & \qw \\
& \ghost{\hat{M}} & \qw \\
& \qw & \qw 
}
\end{equation*}

Section~\ref{sec:serial_manytarg_gate} developed a \textit{local} simulation of the many-target (or ``multi-qubit") gate, enabling two-qubit unitaries like the Mølmer-Sørensen~\cite{molmer1999multiparticle,haffner2008quantum}, Barenco~\cite{barenco1995universal} and Berkeley~\cite{zhang2004minimum} gates, as well as any $n$-qubit operator expressible as a dense $2^n\times2^n$ complex matrix.
Alas, directly \textit{distributing} this scheme for $n\ge2$ appears impossible; each node executing (a parallel version of) Alg.~\ref{alg:serial_manyqb_gate_alg} would require more remote amplitudes than can fit in the communication buffer, as we will make concrete below. Fortunately, this limitation can be surpassed with an \textit{indirect} method.
In this section, we derive a distributed in-place simulation of an $n$-qubit general gate upon an $N$-qubit pure state, distributed between $2^w$ nodes (with $\Lambda=2^{N-w}$ amplitudes per-node). Our method prescribes $\mathcal{O}(2^n \Lambda)$ flops, and at most $2n$ rounds of pairwise communication, each exchanging $\Lambda/2$ amplitudes.
We assume that $n$ is sufficiently small so that the gate's $2^{2n}$ element matrix description remains tractable and is safely duplicated in every node. This admits the looser condition that $n \le N-w$ (or that there are fewer matrix columns than statevector amplitudes per-node) which we must enforce as a strict precondition to our method.

Although our derivation here is original, the core mechanic of this section's algorithm (later dubbed ``cache blocking") has appeared frequently in the literature since 2007~\cite{de2007massively}, and has been implemented in the Oxford's QuEST~\cite{jones2019quest}, IBM's Qiskit~\cite{doi2020cache}, Fujitsu's mpiQulacs~\cite{imamura2022mpiqulacs} (which uses the technique as the primary means to distribute Qulacs~\cite{suzuki2021qulacs}) and NVIDIA's cuQuantum~\cite{stanwyck2022cuquantum} simulators.
A recent work~\cite{adamski2023energy} also studied the utility of the technique for reducing energy consumption of quantum simulation in HPC settings.

\begin{figure}[tb]
    \centering
    \includegraphics[width=.45\columnwidth]{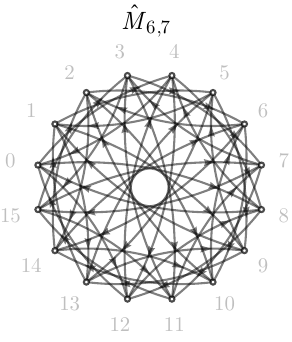}
    \caption{
        The communication pattern of a hypothetical direct simulation of a two-qubit dense gate upon the upper qubits of an $8$-qubit statevector distributed between $W=16$ nodes. Communication is \textit{not} pairwise, and involves sending more amplitudes to a node than can fit in its receiving communication buffer.
    }
    \label{fig:distrib_manytarg_bad_comm_example}
\end{figure}

Let $\hat{M}_{\vec{t}}$ denote an $n$-target operator upon qubits $\vec{t}=\{t_0,\,\dots,\,t_{n-1}\}$ described by matrix $\matr{M}=\{m_{ij} : i,j\} \in \mathbb{C}^{2^n\times 2^n}$. We seek to modify local arrays $\vec{\psi}$ such that their $W=2^w$ distributed ensemble captures the transformation of $N$-qubit state $\ket{\Psi}\numsub{N}$ to
\begin{align}
    \ket{\Psi}\numsub{N} \;\; \rightarrow \;\; \hat{M}_{\vec{t}} \; \ket{\Psi}\numsub{N}.
\end{align}
We first clarify why $\hat{M}_{\vec{t}}$ cannot be \textit{directly} effected under our distribution model before proposing a resolution.
Recall from Eq.~\ref{eq:local_manyqb_amp_update} that $\hat{M}_{\vec{t}}$ modifies each amplitude of $\ket{\Psi}\numsub{N}$ to become a linear combination of $2^n$ amplitudes.
While this was no problem for the local simulation strategy of Alg.~\ref{alg:serial_manyqb_gate_alg}, it is a significant obstacle to a distributed implementation which uses local statevector partitions and buffers of size $\Lambda = 2^{N-w}$. To see why, assume the leftmost $n$ qubits are targeted, and consider the action upon basis state $\ket{i}\numsub{N} \equiv \ket{j}\numsub{n}\ket{0}\numsub{N-n}$.
\begin{gather}
    \hat{M}_{\{N-1, \,N-2, \,\dots, \,N-n\}} \ket{i}\numsub{N}
=
    \matr{M} \ket{j}\numsub{n}
    \otimes 
    \ket{0}\numsub{N-n}
\\
=
    \sum\limits_l^{2^n} m_{lj}\ket{l}\numsub{n} \otimes \ket{0}\numsub{N-n}.
\end{gather}
This basis state, which we will assume has its corresponding amplitude stored in node $r$, has become a ($2^n$)-state superposition.
The very \textit{next} basis state in node $r$, which is $\ket{i+1}\numsub{N}=\ket{j}\numsub{n}\ket{1}\numsub{N-n}$, superposes to a \textit{unique} set of states;
\begin{gather}
    \hat{M}_{\{N-1, \,N-2, \,\dots, \,N-n\}} \ket{i+1}\numsub{N}
=
    \matr{M} \ket{j}\numsub{n}
    \otimes 
    \ket{1}\numsub{N-n}
    \tag*{}
\\
=
    \sum\limits_l^{2^n} m_{lj}\ket{l}\numsub{n} \otimes \ket{1}\numsub{N-n}.
\end{gather}

\begin{figure*}[t]
    \centering
    \includegraphics[width=\textwidth]{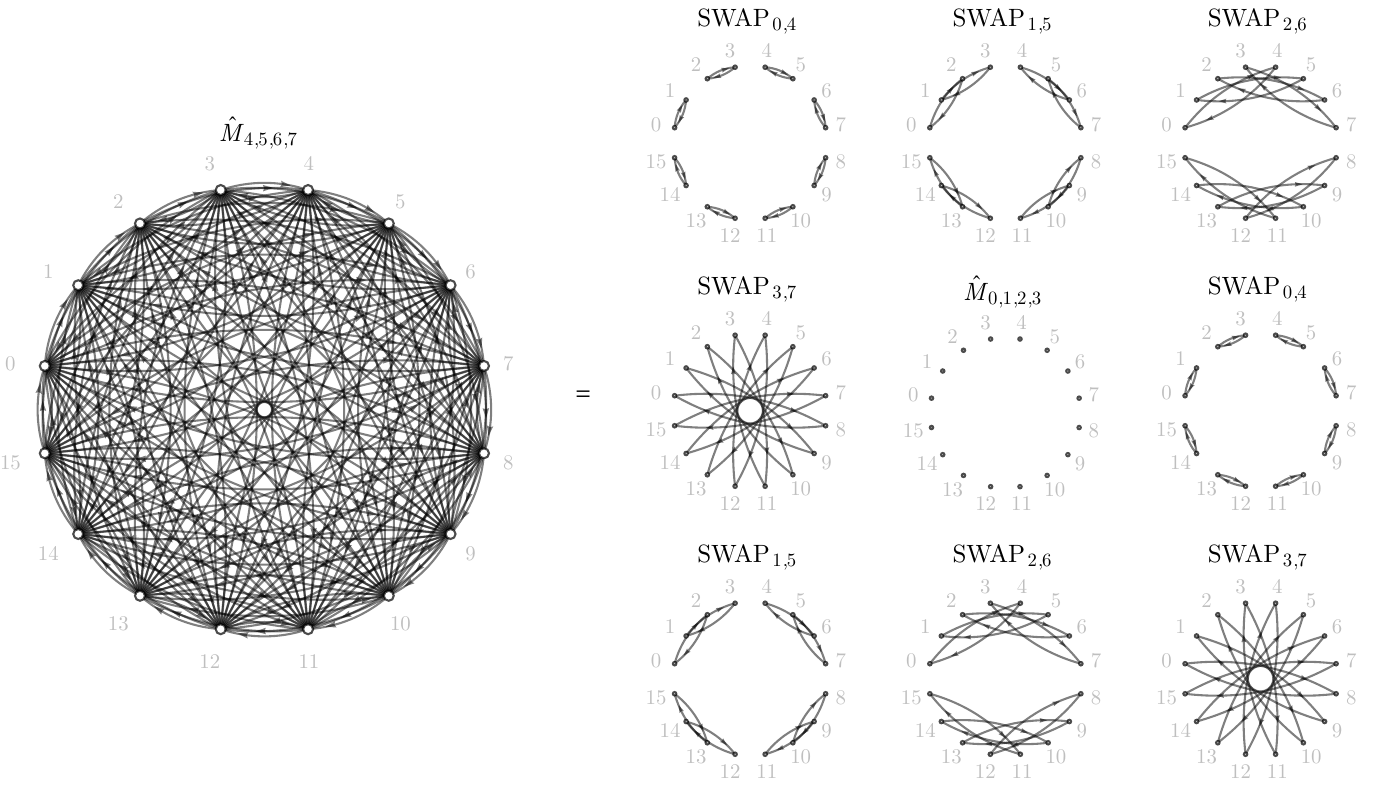}
    \caption{
        The total effective communication pattern (left) and those of each decomposed step (right) of Alg.~\ref{alg:distrib_manytarg_gate}'s distributed simulation of the $4$-target gate $\hat{M}_{\vec{t}}$ upon the upper qubits $\vec{t}=\{4,5,6,7\}$ of an $8$-qubit statevector.
    }
    \label{fig:distrib_manytarg_decomp_example}
\end{figure*}

Continuing, we observe that every state of global index $i \in \{0,2^{N-\max(n,w)}\}$ within node $r$ becomes a unique ($2^n$)-state superposition. Hence, the total number of unique global amplitudes which determine the updated local amplitudes in $r$ can be as many as $2^{\min(n,w)}\Lambda$.
This exceeds the buffer size of $\Lambda$; the remote amplitudes cannot be obtained within a single round of communication, nor can they be simultaneously stored within a node. 
This problem arises even for two-qubit ($n=2$) gates! In precise terms, if two or more qubits with indices $t \ge N-w$ are targeted, the many-qubit gate requires buffer-exceeding communication; an example is shown in Fig.~\ref{fig:distrib_manytarg_bad_comm_example}.
It \textit{is} principally possible to decompose the problematic communication into several serial steps, each using tractable buffer-size messages. Devising such a decomposition algebraically is non-trivial and beyond the talents of the author; but we can fortunately devise it through other means.

Consider now the contrary scenario of targeting the \textit{rightmost} $n$ qubits (where $n<N-w$) of a basis state $\ket{i}\numsub{N}=\ket{r}\numsub{w}\ket{j}\numsub{N-w}$, which has its amplitude stored within rank $r$.
\begin{align}
    \hat{M}_{\{0,\dots,n-1\}} \ket{i}\numsub{N} 
&=
    \ket{r}\numsub{w} \otimes \sum\limits_l^{2^{n}} m_{lj} \ket{l}\numsub{N-w}.
\end{align}
All $2^n$ amplitudes which inform each new amplitude within the node are already contained within the node. Simulating this gate is embarrassingly parallel and will resemble the local strategy of Alg.~\ref{alg:serial_manyqb_gate_alg}.

If we were given $n$ arbitrary target qubits $\vec{t}$ (where $n<N-w$) and we could somehow transform $\hat{M}_{\vec{t}}$ into an alternate operation $\hat{M}_{\vec{t}'}$ which targets the lower qubits (where $q<N-w,\;\forall q\in\vec{t'}$), then we could simulate $\hat{M}_{\vec{t}}$ via embarrassingly parallel simulation of $\hat{M}_{\vec{t}'}$. If such a transformation were possible, we should also wish it is \textit{cheap}. Enter our hero, the SWAP gate, whose especially efficient simulation we derived in the previous section!

\begin{figure}[t]
    \centering
    \includegraphics[width=\columnwidth]{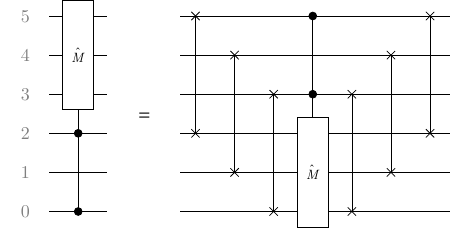}
    \caption{
        Decomposition of a many-control many-target gate targeting the upper qubits into a sequence of SWAPs and a gate targeting the lower qubits.
    }
    \label{fig:circ_multi_ctrl_multi_target_via_swaps}
\end{figure}

We utilise that
\begin{align}
    \hat{M}_{q_1} \;\; \equiv \;\; \text{SWAP}_{q_1,q_2} \;\; \hat{M}_{q_2} \;\; \text{SWAP}_{q_1,q_2},
\end{align}
to re-express an upper-targeting gate in terms of a lower-targeting one;
\begin{gather}
    \hat{M}_{\vec{t}} \bigg\rvert_{t_q \ge N-w, \; \forall q}
    \equiv \\
    \left( \bigotimes_q^n \text{SWAP}_{q,t_q} \right)
    \,
    \hat{M}_{\{0, \; \dots, \;n-1\}}
    \,
    \left( \bigotimes_q^n \text{SWAP}_{q,t_q} \right).
    \tag*{}
\end{gather}
Given arbitrary targets $\vec{t}$, we swap any of index $t_q > N-w$ with a non-targeted lower qubit. We then simulate the embarrassingly parallel lower-targeting many-qubit gate, and finally undo our swaps.
This scheme incurs the costs of simulating $2 \eta$ SWAP gates, where $\eta = \dim(\{ q : q\in\vec{t}, \; q\ge N-w\})$. 
We clarify the effective circuit prescribed by our decomposition (shown with control qubits) in Fig.~\ref{fig:circ_multi_ctrl_multi_target_via_swaps}. 
The communication pattern of an example decomposition is shown in Fig.~\ref{fig:distrib_manytarg_decomp_example}.

Which low-index qubits should we swap high-index targets to? Definitively, the \textit{smallest}/\textit{rightmost} untargeted qubits (with indices $<N-w$), starting at $0$.
This minimises the stride between subsequently superposed amplitudes during the local invocation of the many-target gate (Alg.~\ref{alg:serial_manyqb_gate_alg}), achieving its best caching performance.
The choice of the low-index qubit has no effect on the communication pattern of the SWAP gate.

How costly are these SWAP gates?
Each swap encounters scenario \textbf{\textcolor{red}{3.}} of Sec.~\ref{sec:swap_gate}, whereby every node exchanges $\Lambda/2$ amplitudes with its pair node, and does so with \textit{no} floating-point operations. We ergo communicate $\eta\Lambda$ total amplitudes in $2\eta$ serial rounds, and induce the same flops as the \textit{local} many-target gate.
We formalise our algorithm and its resource costs in Alg.~\ref{alg:distrib_manytarg_gate}.

We remark that a many-\textit{controlled} many-target gate can be simulated in an almost identical fashion (recalling from Sec.~\ref{sec:distrib_many_ctrl_one_targ_gate} that control qubits do not induce any communication) though we must exercise care when swapping targets with control qubits.

We now discuss several potential optimisations to our scheme.
\begin{itemize}
    
    \item Our algorithm swaps \textit{any} target qubit of index $q \ge N-w$, but in fact having only a \textit{single} high-index target will yield pairwise buffer-compatible direct simulation. So in principle, one fewer SWAP than suggested above is necessary.
    However, the bespoke implementation of this scenario (a pairwise-communicating many-qubit gate) will yield the \textit{same} communication complexity as our method, and can serve only to shrink caching and memory-movement costs. We anticipate these savings to be modest, and not worth the added algorithmic complexity for the treatment of this scenario as an explicit edge-case.

\item Our algorithm performs each SWAP gate of the decomposition one by one. This means a single amplitude might be moved by multiple SWAPs, and exchanged between nodes multiple times before arriving at its final node.
    In principle, all SWAPs (on the same side of the embarrassingly parallel simulation of $\hat{M}_{0,1,\dots}$) can be combined into a single operation, reducing the total network traffic by ensuring amplitudes are sent directly to their final node. 
    This communication would \textit{not} be pairwise, and would require more sophisticated logic to efficiently pack and exchange amplitudes than seen in Alg.~\ref{alg:distrib_swap}.
    Such a method, making use of so-called ``fused-swaps", is discussed in Ref.~\cite{de2007massively}, implemented in Fujitsu's mpiQulacs~\cite{imamura2022mpiqulacs}, and made possible in NVIDIA's cuQuantum simulator~\cite{stanwyck2022cuquantum} through function \texttt{custatevecDistIndexBitSwapScheduler}.
    
    \item 
    Consider introducing control qubits $\vec{c}$. The decomposition of $C_{\vec{c}}(\hat{M}_{\vec{t}})$ suggested by Fig.~\ref{fig:circ_multi_ctrl_multi_target_via_swaps} applies \textit{non}-controlled SWAPs and ergo communicate amplitudes back and forth which are ultimately not modified by $C_{\vec{c}}(\hat{M}_{\vec{t}})$. 
    An optimised method would develop and use a bespoke controlled-SWAP gate, where each swap inherits controls $\vec{c}$, exponentially reducing communication costs in the same manner described in Sec.~\ref{sec:distrib_many_ctrl_one_targ_gate}.
    
    \item Control qubits also introduce many new opportunities for direct pairwise simulation of $C_{\vec{c}}(\hat{M}_{\vec{t}})$ without decomposition because they may decrease the number of external amplitudes needed by a node to become tractable and fit within the buffer. Note that such an optimisation presents an overwhelming number of edge-cases.

\end{itemize}

Even without these optimisations, Alg.~\ref{alg:distrib_manytarg_gate} to simulate an $n$-qubit gate prescribes only a factor $\mathcal{O}(2^n)$ more flops than the one-qubit gate of Alg.~\ref{alg:distrib_1qb_gate}; this is a fundamental minimum due to the $\mathcal{O}(2^{2n})$ elements of input matrix $\matr{M}$. 

Because Alg.~\ref{alg:distrib_manytarg_gate} did not impose unitarity or other constraints upon $\matr{M}$, it can in principle be invoked to simulate any ($N-w$)-qubit digital quantum operation upon a statevector. It can ergo effect all subsequent statevector operators derived in this manuscript.
However, we will instead develop a variety of bespoke simulations for advanced operators which use different (and more convenient) input parameterisations and which yield exponentially faster simulation.

\begin{algorithm}[t]
\DontPrintSemicolon
\caption[blah blah]{
    \AlgTagDistributed \AlgTagStateVector%
    \\
    Many-target gate $\hat{M}_{\vec{t}}$ upon $n$ unique target qubits $\vec{t}=\{t_0,\,\dots,\,t_{n-1}\}$ described by matrix $\matr{M} \in \mathbb{C}^{2^n\times 2^n}$, applied to an $N$-qubit pure statevector distributed between $2^w$ nodes as local $\Lambda$-length array $\vec{\psi}$ (with buffer $\vec{\varphi}$).
    \protect{\begin{center}
        $
            \Qcircuit @C=1em @R=.7em {
            & \multigate{1}{\hat{M}} & \qw \\
            & \ghost{\hat{M}} & \qw
            }
        $
    \end{center}}
    \begin{center}
        \AlgTagBops{$\mathcal{O}(2^n\Lambda)$}%
        \AlgTagFlops{$\mathcal{O}(2^n\Lambda)$} \\
        \AlgTagNumSerialRounds{best: $0$, worst: $2\min(n,w)$} \\
        \AlgTagAmpsTransferred{worst: $2^N\,\min(n,w)$} \\
        \AlgTagMemOverhead{$\mathcal{O}(2^n)$}%
        \AlgTagMemWrites{$\mathcal{O}(2^n \Lambda)$}
    \end{center}
}
\label{alg:distrib_manytarg_gate}

\textbf{distrib\_manyTargGate}($\vec{\psi}$, $\vec{\varphi}$, $\matr{M}$, $\vec{t}$):

\Indp 

    $\lambda$ = $\log_2(\dim(\vec{\psi}))$
    \tcp*{$=N-w$}

    \codegap

    \tcp{need fewer targets than suffix qubits}
    
    $n$ = $\dim(\vec{t})$

    \textbf{if} $n > \lambda$:
    
    \Indp 
    
        \textbf{fail}
    
    \Indm 

    \codegap
    
    \tcp{locate lowest non-targeted qubit}
    
    $b$ = \textbf{getBitMask}($\vec{t}$)
    \tcp*{Alg.~\ref{alg:bit_twiddles}}
    
    $q$ = $0$
    
    \textbf{while} \textbf{getBit}($b$, $q$) \EqEq{} $1$:
    \tcp*{Alg.~\ref{alg:bit_twiddles}}
    
    \Indp 
    
        $q$ \PlusEq{} $1$
    
    \Indm
    
    \codegap 
    
    \tcp{record which targets require swapping}
    
    $\vec{t}' = \{ \}$
    
    \textbf{for} $i$ \textbf{in} \textbf{range}($0$, $n$):
    
    \Indp 
    
        \textbf{if} $\vec{t}[i] < \lambda$:
        
        \Indp 
        
            $\vec{t}'[i]$ = $\vec{t}[i]$
        
        \Indm 
        
        \textbf{else}:
        
        \Indp
        
            $\vec{t}'[i]$ = $q$

            $q$ \PlusEq{} $1$
            
            \textbf{while} \textbf{getBit}($b$, $q$) \EqEq{} $1$:
            \tcp{Alg.~\ref{alg:bit_twiddles}}
    
            \Indp 
                
                $q$ \PlusEq{} $1$
                
            \Indm
            
        \Indm 
    
    \Indm 
    
    \codegap
    
    \tcp{perform necessary swaps}
    
    \textbf{for} $i$ \textbf{in} \textbf{range}($0$, $n$):
    
    \Indp
    
        \textbf{if} $\vec{t}'[i] \ne \vec{t}[i]$:
        
        \Indp 
        
            \textbf{distrib\_swapGate}($\vec{\psi}$, $\vec{\varphi}$, $\vec{t}'[i]$, $\vec{t}[i]$)
            
            \tcp*{Alg.~\ref{alg:distrib_swap}}
        
        \Indm 
    
    \Indm 
    
    \tcp{embarrassingly parallel $\hat{M}_{\vec{t}'}$}
    
    \textbf{local\_manyTargGate}($\vec{\psi}$, $\matr{M}$, $\vec{t}'$)
    \tcp*{Alg.~\ref{alg:serial_manyqb_gate_alg}}
    
    \codegap 
    
    \tcp{undo swaps}
    
    \textbf{for} $i$ \textbf{in} \textbf{range}($0$, $n$):
    
    \Indp
    
        \textbf{if} $\vec{t}'[i] \ne \vec{t}[i]$:
        
        \Indp 
        
            \textbf{distrib\_swapGate}($\vec{\psi}$, $\vec{\varphi}$, $\vec{t}'[i]$, $\vec{t}[i]$)
            
            \tcp*{Alg.~\ref{alg:distrib_swap}}
        
        \Indm 
    
    \Indm 

\Indm 

\end{algorithm}

\pagebreak

\hphantom{.}

\pagebreak

\subsection{Pauli tensor}%
\label{sec:pauli_tensor}
\noindent %
\begin{equation*}
\Qcircuit @C=1em @R=.2em {
& \gate{\hat{X}} & \qw \\
& \gate{\hat{Y}} & \qw \\
& \gate{\hat{Z}} & \qw 
}
\end{equation*}

The one-target Pauli operators $\hat{X}$, $\hat{Y}$ and $\hat{Z}$ are core primitives of quantum information theory, and their controlled and many-target extensions appear extensively in experimental literature.
For instance, Toffoli gates~\cite{toffoli1980reversible}, fan-out gates~\cite{khazali2020fast}, many-control many-target $\hat Z$ gates~\cite{molmer2011efficient} and others appear as natural primitive operators for Rydberg atom computers~\cite{saffman2016quantum}.
Many-control many-target $\hat X$ gates appear in quantum arithmetic circuits~\cite{vedral1996quantum}. But most significantly, Pauli tensors form a natural basis for Hermitian operators like Hamiltonians, and their efficient simulation would enable fast calculation of quantities like expectation values.
%
Rapid, direct simulation of a Pauli tensor will also enable efficient simulation of more exotic operators like Pauli gadgets, as we will explore in Sec.~\ref{sec:distrib_pauli_gadget}.

As always, let $\Lambda$ be the size of each node's sub-statevector.
In this section, we derive a distributed in-place simulation of the $n$-qubit Pauli tensor which prescribes $\mathcal{O}(n\,\Lambda)$ bops, $\Lambda$ flops, $\Lambda$ memory writes, no memory overhead and at most a single round of communication. 
This makes it more efficient than the one-target gate of Alg.~\ref{alg:distrib_1qb_gate} despite prescribing a factor $\mathcal{O}(n)$ more bops.
Further, all amplitude modifications happen to be multiplication with unit scalars $\pm 1$ or $\pm \iu$, which may enable optimisation on some systems and with some amplitude types.


We consider the separable $n$-qubit unitary
\begin{align}
    \hat{U}_{\vec{t}} = \bigotimes\limits_{q}^n
    \hat{\sigma}^{(q)}_{t_q}
\end{align}
composed of Pauli operators $\hat{\sigma}^{(q)} \in \{\hat{X},\hat{Y},\hat{Z}\}$
acting upon target qubits $\vec{t}=\{t_0,\,\dots,\,t_{n-1}\}$ of an $N$-qubit pure state $\ket{\Psi}\numsub{N}$.
A naive and inefficient method to simulate $\hat{U}_{\vec{t}}\ket{\Psi}\numsub{N}$ is to construct matrix
\begin{gather}
    \matr{M} = 
    \bigotimes\limits_{q}^n
    \matr{\sigma}^{(q)}
    \;\;
    \in \mathbb{C}^{2^n\times 2^n},\;\;
    \text{ where}
    \\
    \renewcommand*{\arraystretch}{1}
    \matr{X} = \begin{pmatrix}
    0 & 1 \\ 1 & 0
    \end{pmatrix},
    \;\;\;
    \matr{Y} = \begin{pmatrix}
    0 & -\iu \\ \iu & 0
    \end{pmatrix},
    \;\;\;
    \matr{Z} = \begin{pmatrix}
    1 & 0 \\ 0 & -1
    \end{pmatrix},
    \tag*{}
\end{gather}
and effect it as a dense $n$-qubit gate $\hat{M}_{\vec{t}}$ via Alg.~\ref{alg:distrib_manytarg_gate}. This would involve $2^n\,\Lambda$ flops and writes, and up to $2n$ rounds of communication.

Alternatively, we could simulate each one-target operator in turn, utilising
\begin{align}
    \hat{U}_{\vec{t}}\ket{\Psi}\numsub{N} 
    &=
    \hat{\sigma}^{(n-1)}_{t_{n-1}}
    \left( 
    \dots \,
    \hat{\sigma}^{(1)}_{t_{1}}
    \left( 
    \hat{\sigma}^{(0)}_{t_{0}}
    \ket{\Psi}\numsub{N}
    \right) 
    \dots
    \right),
\end{align}
and perform a total of $n$ invocations of Alg.~\ref{alg:distrib_1qb_gate}. This would cost $n\,\Lambda$ flops and writes, and at most $n$ rounds of communication. Still, a superior method can accelerate things by a factor $n$.

Neither of these naive methods leverage two useful properties of the Pauli matrices; that they have unit complex amplitudes (as do their Kronecker products) and that they are diagonal or anti-diagonal. These properties respectfully enable simulation of $\hat{U}_{\vec{t}}$ with no arbitrary floating-point operations (all instead become sign flips and complex component swaps) and in (at most) a single round of communication. 
We now derive such a scheme.

A single Pauli operator $\hat{\sigma}_q$ maps an $N$-qubit basis state $\ket{i}\numsub{N}$ to
\begin{align}
    \hat{X}_q \ket{i}\numsub{N} &= \ket{i_{\neg q}}\numsub{N},
    \\
    \hat{Y}_q \ket{i}\numsub{N} &=
    \iu  \,
    (-1)^{i_{[q]}}
    \,
    \ket{i_{\neg q}}\numsub{N},
    \\
    \hat{Z}_q \ket{i}\numsub{N} &=
    (-1)^{i_{[q]}}
    \,
    \ket{i}\numsub{N}.
    \label{eq:pauli_tensor_Z_op_on_basis}
\end{align}

Let $\vec{t}^{x} \subseteq \vec{t}$ contain the indices of qubits targeted by an $\hat{X}$ operator in the full tensor $\hat{U}_{\vec{t}}$. 
Formally,
\begin{align} 
\vec{t}^{x} = \{t_j : \hat{\sigma}^{(j)}=\hat{X} \}. 
\end{align}
Let $\vec{t}^{x,y} \supseteq \vec{t}^{x}$ contain all qubits targeted by either $\hat X$ \textit{or} $\hat Y$. We can then express the action of the full tensor upon a basis state as
\begin{align}
    \hat{U}_{\vec{t}}
    \ket{i}\numsub{N}
    &=
        \prod\limits_{q\in\vec{t}^{z}}
    (-1)^{i_{[q]}}
    \;
    %
        \prod\limits_{q\in\vec{t}^{y}}
        \iu  \,
    (-1)^{i_{[q]}}
    \;
    %
    \ket{i_{\neg \vec{t}^{x,y}}}\numsub{N}
    \\
    &=
    (-1)^{f(i)} \, \eta \,  
    \ket{i_{\neg \vec{t}^{x,y}}}\numsub{N}
    \label{eq:pauli_tensor_action_on_basis_state}
\end{align}
where we have defined constant $\eta = \iu^{\dim(\vec{t}^y)} \in \{\pm1,\pm \iu\}$ and basis state dependent function
$f(i) = \dim(\{ q\in\vec{t}^{y,z} : q_{[i]}=1\}) \in \mathbb{N}$. This function returns the number of qubits in $\ket{i}\numsub{N}$ which are targeted by either $\hat{Y}$ or $\hat{Z}$ and which are in their $\ket{1}\numsub{1}$ state. The expression $(-1)^{f(i)}$ simplifies to $\pm 1$ as determined by the \textit{parity} of that number, which we will eventually compute with efficient bit logic.

\begin{figure*}[t]
    \centering
    \includegraphics[width=\textwidth,trim={0 4.5cm 0 0},clip]{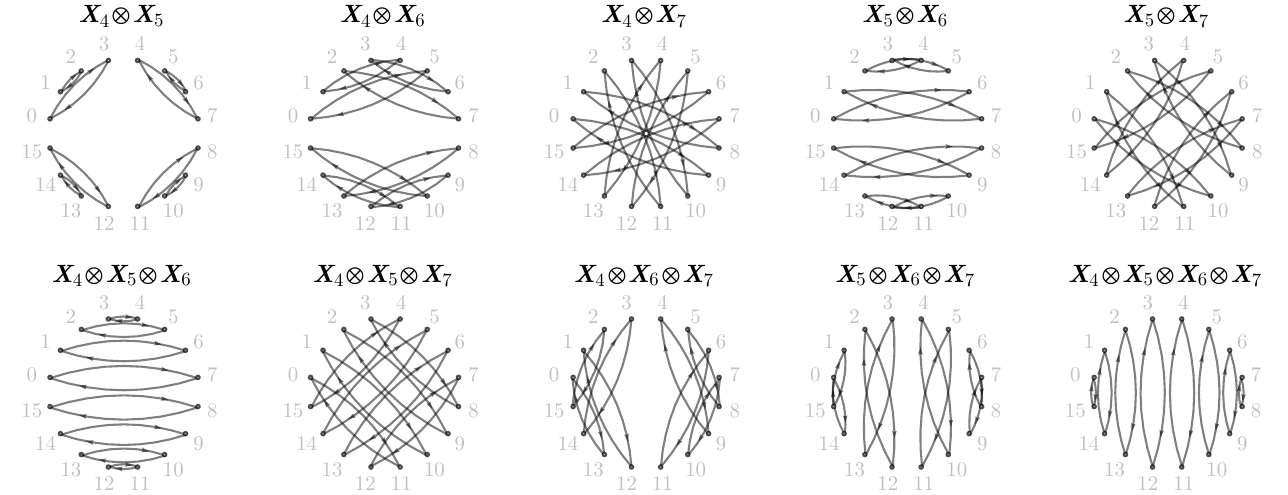}
    \caption{
        Some communication patterns of Alg.~\ref{alg:distrib_pauli_tensor}'s simulation of the Pauli tensor upon an $8$-qubit statevector distributed between $16$ nodes.
        Only $\hat{X}$ and $\hat{Y}$ targeting qubits $t \ge 4$ trigger communication, and do so identically, so only $\hat{X}$ is demonstrated. This is incidentally the same pattern admitted by the Pauli gadget of Alg.~\ref{alg:distrib_pauli_gadget} (Sec.~\ref{sec:distrib_pauli_gadget}).
    }
    \label{fig:distrib_pauli_tensor_comm}
\end{figure*}

The full Pauli tensor upon an arbitrary $N$-qubit pure state $\ket{\Psi}\numsub{N} = \sum_i \alpha_i \ket{i}\numsub{N}$ is therefore expressable as
\begin{align}
    \hat{U}_{\vec{t}} \ket{\Psi}\numsub{N} 
    &=
    \sum\limits_i^{2^N}
    \;
    \beta_i 
    \;
    \alpha_i
    \; 
    \ket{i_{\neg \vec{t}^{x,y}}}\numsub{N}
    \\
    &=
    \sum\limits_i^{2^N}
    \beta_{i_{\neg \vec{t}^{x,y}}}
    \;
    \alpha_{i_{\neg \vec{t}^{x,y}}}
    \ket{i}\numsub{N},
    \\
    \text{where}\;\;\; \beta_i &= (-1)^{f(i)} \; \eta 
    \;\;
    \in \{\pm 1\, \pm \iu \}.
\end{align}
The explicit action of the Pauli tensor is to modify the amplitudes under
\begin{align}
    \alpha_i 
    \;\;\;
    \xrightarrow{\hat{U}_{\vec{t}}}
    \;\;\;
    \beta_{i_{\neg \vec{t}^{x,y}}}
    \;
    \alpha_{i_{\neg \vec{t}^{x,y}}}
    \label{eq:pauli_tensor_amp_update}
\end{align}
where complex unit $\beta_i$ is trivially bitwise evaluable independent of any amplitude. Notice too that unlike the previously presented gates in this manuscript, Eq.~\ref{eq:pauli_tensor_amp_update} does not prescribe any superposition or linear combination of the amplitudes with one another; only the \textit{swapping} of amplitudes and their multiplication with a complex unit. 

\begin{figure}[tb]
    \centering
    \includegraphics[width=.45\columnwidth]{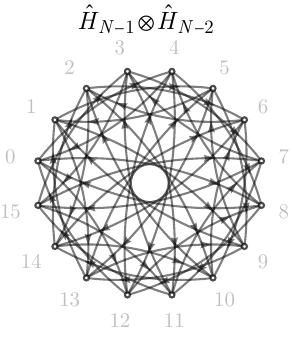}
    \caption{
        The communication necessary to simulate a pair of Hadamard gates on the uppermost qubits. While each Hadamard is pairwise simulable by Alg.~\ref{alg:distrib_1qb_gate}, their direct tensor is not. This is unlike Pauli tensors which are always pairwise simulable.
    }
    \label{fig:distrib_pauli_tensor_hadamard_comm}
\end{figure}

We now derive the communication strategy to effect Eq.~\ref{eq:pauli_tensor_amp_update}. Recall that the $j$-th local amplitude in node 
$r < 2^w$
corresponds to global amplitude $\alpha_i$ of basis state
\begin{align}
    \ket{i}\numsub{N} 
    \equiv \ket{r}\numsub{w}\ket{j}\numsub{\lambda},
    \;\;\;\;\;
    \lambda = N-w.
\end{align}
Under $\hat{U}_{\vec{t}}$, this amplitude swaps with that of index
\begin{gather}
    \ket{
        i_{\neg \vec{t}^{x,y}}}\numsub{N} 
    \equiv \ket{
        r_{
            \neg \vec{t}''
        }
        }\numsub{w}\ket{
        j_{
            \neg \vec{t}'
        }
    }\numsub{\lambda},
    \label{eq:pauli_tensor_global_index_update}
    \\
    \shortintertext{where we have defined}
    \vec{t}' = 
        \{ q \hphantom{-\lambda} \;\;\,\, : \;
                q < \lambda, 
                \;\;\;
                q \in \vec{t}^{x,y}
                \},
    \\
    \vec{t}'' =
    \{ q-\lambda \; : \;
                q \ge \lambda, 
                \;\;\;
                q \in \vec{t}^{x,y}
                \}.
\end{gather}
Every amplitude within node $r$ is exchanged with a fixed pair node $r' = r_{\neg \vec{t}''}$ as determined by only the qubits targeted by $\hat{X}$ or $\hat{Y}$ and with indices $\ge N-w$. By idempotency of bit flips, communication is pairwise and tractable; we visualise the communication pattern in Fig.~\ref{fig:distrib_pauli_tensor_comm}.
This is indeed a result of the (anti-)diagonal form of the Pauli operators, and \textit{not} due to the separability of the tensor into one-qubit gates, as we make evident in Fig~\ref{fig:distrib_pauli_tensor_hadamard_comm}.
Notice the algorithm is embarrassingly parallel whenever all operators are $\hat{Z}$, \textit{or} target only the lowest $N-w$ qubits.

Eq.~\ref{eq:pauli_tensor_global_index_update} also reveals that the local position $j'$ of a swapped amplitude (of previous local index $j$) is $j' = j_{\neg \vec{t}'}$, and can be computed in $\mathcal{O}(1)$ using a bitmask.
Our final consideration is how to locally update amplitudes during the embarrassingly parallel scenario of $r' = r$, i.e. when $t'' = \{ \}$. We will require swapping amplitudes (and multiplying by $\beta_i$) which differ in array indices only at the targeted bits. Our iteration of these amplitudes should avoid branching, and should not visit an amplitude more than once. Our method will leverage the the bitwise complement, and its property:
\begin{align}
    \ket{\BitComp i}\numsub{N} = 
    \bigotimes_q^{N}
    \ket{\LogicalNot i_{[q]}}\numsub{1}
    = \ket{2^N-i-1}\numsub{N}.
\end{align}
To clarify this property, consider the ordered three-bit sequences below which share a colour if they are bitwise complements of one another.
\begin{align*}
    \ket{0}\numsub{3} &=
    \color[rgb]{0.266122, 0.486664, 0.802529} 000 \\
    \ket{1}\numsub{3}
    &=
    \color[rgb]{0.513417, 0.72992, 0.440682}
    001 \\
    \ket{2}\numsub{3}
    &=
    \color[rgb]{0.863512, 0.670771, 0.236564}
    010 \\
    \ket{3}\numsub{3}
    &=
    \color[rgb]{0.857359, 0.131106, 0.132128}
    011 \\
    \ket{\BitComp 3}\numsub{3}
    &=
    \color[rgb]{0.857359, 0.131106, 0.132128}
    100 \\
    \ket{\BitComp 2}\numsub{3}
    &=
    \color[rgb]{0.863512, 0.670771, 0.236564}
    101 \\
    \ket{\BitComp 1}\numsub{3}
    &=
    \color[rgb]{0.513417, 0.72992, 0.440682}
    110 \\
    \ket{\BitComp 0}\numsub{3}
    &=
    \color[rgb]{0.266122, 0.486664, 0.802529}
    111
\end{align*}
When a subset $\vec{t}$ of all qubits are targeted, it holds that
\begin{gather}
    \ket{i}\numsub{N} \equiv \ket{k,j}\numsub{N}
    \implies
    \ket{i_{\neg \vec{t}}}\numsub{N}
    =
    \ket{k,2^{\dim(\vec{t})}-j-1},
\end{gather}
where binary integer $j$ spans the classical states of $\vec{t}$.
This means we can iterate the first $2^{\dim(\vec{t})}/2$ of the $\dim(\vec{t})$-length bit-sequences (each corresponding to amplitude $\alpha_j$) and flip all bits in $\vec{t}'$ to access the paired amplitude $\alpha_{j_{\neg \vec{t}'}}$. The result is that all pairs of amplitudes (to be swapped) are iterated precisely once.
We formalise this scheme in Alg.~\ref{alg:distrib_pauli_tensor}.

\begin{algorithm}[b]
\DontPrintSemicolon
\caption[blah blah]{
    \AlgTagDistributed \AlgTagStateVector  \\
    Pauli tensor $\hat{U}_{\vec{t}}=\bigotimes_q^n \hat{\sigma}_{t_q}^{(q)}$ consisting of $n$ Pauli operators $\hat{\sigma}^{(q)}$ (encoded into array $\vec{\sigma}$) upon unique targets $\vec{t}=\{t_0,\,\dots,\,t_{n-1}\}$, applied to an $N$-qubit statevector distributed between $2^w$ $\Lambda$-length arrays $\vec{\psi}$ (with buffer $\vec{\varphi}$).
    %
    \protect{\begin{center} 
    $\Qcircuit @C=1em @R=.2em {
& \gate{\hat{X}} & \qw \\
& \gate{\hat{Y}} & \qw \\
& \gate{\hat{Z}} & \qw 
}   $
    \end{center}}
    \begin{center}
        \AlgTagBops{$\mathcal{O}(n\,\Lambda)$}%
        \AlgTagFlops{$\Lambda$}%
        \AlgTagNumSerialRounds{$0$ or $1$}\\
        \AlgTagAmpsTransferred{$2^N$}%
        \AlgTagMemOverhead{$\mathcal{O}(1)$}%
        \AlgTagMemWrites{$\Lambda$}
    \end{center}
}
\label{alg:distrib_pauli_tensor}

\codegap 

\textbf{distrib\_pauliTensor}($\vec{\psi}$, $\vec{\varphi}$, $\vec{\sigma}$, $\vec{t}$):

\Indp 
    
    $\lambda$ =  $\log_2(\dim(\vec{\psi}))$
    
    $n$ = $\dim(\vec{t})\;\;\;$ 
    \tcp{$=\dim(\vec{\sigma})$}
    
    $r$ = \textbf{getRank()}
    \tcp*{Alg.~\ref{alg:mpi_defs}}
    
    \codegap 
    
    \tcp{prepare $\eta = \iu^{\dim(\vec{t}^{y})}$}
    
    $\eta$ = $1$
    
    \textbf{for} $q$ \textbf{in} \textbf{range}($0$, $n$):
    
    \Indp 
    
        \textbf{if} $\vec{\sigma}[q]$ \textbf{is} $\hat{Y}$:
        
        \Indp 
        
            $\eta$ \TimesEq{} $\iu$
        
        \Indm
    
    \Indm
    
    \codegap
    
    \tcp{determine pair rank}
    
    $r'$ = $r$
    
    \textbf{for} $q$ \textbf{in} \textbf{range}($0$, $n$):
    
    \Indp 
    
        \textbf{if} $\vec{t}[q] \ge \lambda$ \textbf{and} ($\vec{\sigma}[q]$ \textbf{is} $\hat{X}$ or $\hat{Y}$):
        
        \Indp 
        
            $r'$ = \textbf{flipBit}($r'$, $\vec{t}[q] - \lambda$)
            \tcp*{Alg.~\ref{alg:bit_twiddles}}
        
        \Indm 
    
    \Indm
    
    \codegap 
    
    \tcp{prepare 
    index bit mask (for $j\rightarrow j'$)}
    
       $\vec{t}'$ = $\{ \}$
    
    \textbf{for} $q$ \textbf{in} \textbf{range}($0$, $n$):
    
    \Indp 
    
        \textbf{if}
        $t[q] < \lambda$ \textbf{and} ($\vec{\sigma}[q]$ \textbf{is} $\hat{X}$ or $\hat{Y}$):
        
        \Indp 
        
            \textcolor{gray}{append} $\vec{t}[q]$ \textcolor{gray}{to} $\vec{t}'$
        
        \Indm 
        
    \Indm 

    \codegap
    
    $b_{xy}$ = \textbf{getBitMask}($\vec{t}'$)
    \tcp*{Alg.~\ref{alg:bit_twiddles}}
    
    \codegap
    
    \tcp{prepare coefficient bit mask (for $\beta_i$)}
    
    $b_{yz}$ = $0$

    \textbf{for} $q$ \textbf{in} \textbf{range}($0$, $n$):
    
    \Indp 
    
        \textbf{if} ($\vec{\sigma}[q]$ \textbf{is} $\hat{Y}$ or $\hat{Z}$):
        
        \Indp 
        
            $b_{yz}$ = \textbf{flipBit}($b_{yz}$, $\vec{t}[q]$)
            \tcp*{Alg.~\ref{alg:bit_twiddles}}
        
        \Indm 
    
    \Indm
    
    \codegap

    \tcp{invoke subroutine (Alg.~\ref{alg:distrib_pauli_tensor_subroutine})}
    
    \textbf{if} $r' = r$:
    
    \Indp 
    
        \textbf{local\_tensorSub}($\vec{\psi},\,\vec{t}',\,r,\,\eta,\,b_{xy},\,b_{yz},\,\lambda$)
    
    \Indm

    \textbf{else}:

    \Indp
    
        \textbf{distrib\_tensorSub}($\vec{\psi},\,\vec{\varphi},\,r',\,\eta,\,b_{xy},\,b_{yz},\,\lambda$)
        
    \Indm

\Indm

\end{algorithm}

\begin{algorithm}[b]
\DontPrintSemicolon
\caption[blah blah]{
    Subroutines of Alg.~\ref{alg:distrib_pauli_tensor}\\
}
\label{alg:distrib_pauli_tensor_subroutine}

\codegap 

\textbf{local\_tensorSub}($\vec{\psi},\,\vec{t}',\,\,r,\,\,\eta,\,\,b_{xy},\,\,b_{yz},\,\lambda$):
    
    \Indp

        $\Lambda$ = $\dim(\vec{\psi})$
        
        $n'$ = $\dim(\vec{t}')$

        \textbf{sort}($\vec{t}'$)

        \codegap
        
        \tcp{loop $\ket{k}\numsub{\lambda-n'}$}

        \AlgThreadComment{multithread}
    
        \textbf{for} $k$ \textbf{in} \textbf{range}($0$, $\Lambda/2^{n'}$):
        
        \Indp 

            $h$ = \textbf{insertBits}($k$, $\vec{t}'$, $0$)
            \tcp*{Alg.~\ref{alg:bit_twiddles}}

            \codegap 
            
            \tcp{loop $\ket{j}\numsub{\lambda} = \ket{k,l}\numsub{\lambda}$ (first half)}

            \AlgThreadComment{multithread (flatten with above)}
            
            \textbf{for} $l$ \textbf{in} \textbf{range}($0$, $2^{n'}/2$):
            
            \Indp
            
                $j$ = \textbf{setBits}($h$, $\vec{t}'$, $l$)
                \tcp*{Alg.~\ref{alg:bit_twiddles}}

                \codegap
                
                \tcp{find pair amp and coeffs}
                
                $i$ = $(r \BitShiftLeft \lambda) \BitOr j$
                
                $p$ = \textbf{getMaskParity}($i \BitAnd b_{yz}$)
                \tcp*{Alg.~\ref{alg:bit_twiddles}}
                
                $\beta$ = $(1-2 \, p) \, \eta$
                
                $j'$ = $j \BitXor b_{xy}$
                
                $i'$ = $(r \BitShiftLeft \lambda) \BitOr j'$
                
                $p'$ = \textbf{getMaskParity}($i' \BitAnd b_{yz}$)
                \tcp{Alg.~\ref{alg:bit_twiddles}}
            
                $\beta'$ = $(1-2 \, p') \, \eta$

                \codegap 
                
                \tcp{swap (scaled) amps}
                
                $\gamma$ = $\vec{\psi}[j]$
                
                $\vec{\psi}[j]$ = $\beta' \vec{\psi}[j']$
                \label{algline:pauli_tensor_A}
                \tcp*{times unit-complex}

                $\vec{\psi}[j']$ = $\beta \;  \gamma$
                \label{algline:pauli_tensor_B}
                \tcp*{times unit-complex}

            \Indm
        
        \Indm 
        
    \Indm

\codegap 

\codegap 

\codegap

\textbf{distrib\_tensorSub}($\vec{\psi},\,\,\vec{\varphi},\,\,r',\,\,\eta,\,\,b_{xy},\,\,b_{yz},\,\lambda$):

\Indp

        $\Lambda$ = $\dim(\vec{\psi})$

        \codegap 

        \tcp{exchange full sub-statevectors}
    
        \textbf{exchangeArrays}($\vec{\psi}$, $\vec{\varphi}$, $r'$)
        \tcp*{Alg.~\ref{alg:mpi_defs}}

        \codegap

        \AlgThreadComment{multithread}
        
        \textbf{for} $j$ \textbf{in} \textbf{range}($0$, $\Lambda$):
        
        \Indp 

            \codegap 
            
            \tcp{update local $j$ as $\ket{r}\ket{j} \rightarrow \ket{r'}\ket{j'}$}
        
            $j'$ = $j \BitXor b_{xy}$
            
            $i'$ = $(r' \BitShiftLeft \lambda) \BitOr j'$
            
            $p$ = \textbf{getMaskParity}($i' \BitAnd b_{yz}$)
            \tcp*{Alg.~\ref{alg:bit_twiddles}}
            
            $\beta$ = $(1-2 \, p) \, \eta$

            \codegap 
            
            $\vec{\psi}[j]$ = $\beta \; \vec{\varphi}[j']$
            \label{algline:pauli_tensor_C}
            \tcp*{times unit-complex}
        
        \Indm 
        
    \Indm 

\end{algorithm}

Our simulation of the $n$-target Pauli tensor is as efficient as the one-qubit gate of Alg.~\ref{alg:distrib_1qb_gate}, and prescribes a factor $n$ fewer communications than if each Pauli was simulated in turn, or as a dense matrix via Alg.~\ref{alg:distrib_manytarg_gate}.
Yet, we have neglected several optimisations possible for specific input parameters. For instance, when \textit{all} of the Paulis are $\hat{Z}$, the entire operator is diagonal and can be simulated in an embarrassingly parallel manner as a phase gadget $ \hat{Z}^{\otimes} \simeq e^{\iu \frac{\pi}{2} \hat{Z}^{\otimes}}$ as described in the next section.
Further, when \textit{none} of the lowest (rightmost) $N-w$ qubits are targeted by $\hat{X}$ or $\hat{Y}$ gates, then after amplitudes are exchanged, the local destination address $j'$ is equal to the original local index $j$. The sub-statevectors $\vec{\psi}$ are effectively directly swapped, as per the paradigm in Fig.~\ref{fig:distrib_mem_exchange_patterns}~\textbf{d)}, before local scaling (multiplying by factor $\beta_i$). Bespoke iteration of this situation, avoiding the otherwise necessary bitwise arithmetic, can improve caching performance and auto-vectorisation.

\clearpage

\subsection{Phase gadget}%
\label{sec:distrib_phase_gadget} %
\noindent 
\begin{equation*}
\Qcircuit @C=1em @R=.7em {
& \multigate{2}{ e^{\iu  \, \theta \, \hat{Z}_1 \, \hat{Z}_2 \, \dots } } & \qw \\
& \ghost{  e^{\iu  \, \theta \, \hat{Z}_1 \, \hat{Z}_2 \, \dots }  } & \qw \\
& \ghost{ e^{\iu  \, \theta \, \hat{Z}_1 \, \hat{Z}_2 \, \dots }  } & \qw 
}
\end{equation*}

The $\hat{Z}$-phase gadget, also known as $ZZ(\theta)$ and as the many-qubit $Z$ rotation, appears frequently in quantum machine learning~\cite{gao2017efficient}, recompilation~\cite{kisskiss_noncliff,cowtan2019phase} and variational eigensolving literature~\cite{gu2021efficient,cowtan2020generic} as a simple parameterised entangling gate. 
It also forms the basic evolution operator in Trotterised Ising Hamiltonians~\cite{trotter1959product,biamonte2008realisable,cervera2018exact}.
Though the phase gadget is a high-fidelity native gate in neutral-atom~\cite{xia2015randomized} and ion-trap experiments~\cite{harty2014high,schindler2013quantum}, it is regularly compiled into simpler operations on other platforms~\cite{mottonen2004quantum,martinez2016compiling}, such as a sequence of controlled-NOT gates~\cite{debroy2020extended}.
In classical simulation however, its $\hat{Z}$-basis matrix is diagonal and trivial to effect. 
In this section, we derive an embarrassingly parallel in-place simulation of the many-qubit phase gadget upon a distributed $N$-qubit pure state (with $\Lambda$ amplitudes per-node), prescribing $\mathcal{O}(\Lambda)$ bops, flops and writes per-node.

We define the phase gadget as an $n$-qubit unitary operator with real parameter $\theta \in \mathbb{R}$,
\begin{align}
    \hat{U}_{\vec{t}}(\theta) = \exp\left( \iu  \, \theta \bigotimes\limits_j^n \hat{Z}_{t_j} \right),
\end{align}
acting upon target qubits $\vec{t} = \{t_0, \, \dots, \, t_{n-1}\}$. We seek to apply $\hat{U}_{\vec{t}}(\theta)$ upon an $N$-qubit statevector $\ket{\Psi}\numsub{N}$ distributed between $W=2^w$ nodes.

A treatment of $\hat{U}_{\vec{t}}(\theta)$ as a dense $2^n\times 2^n$ complex matrix (simulated via the many-target gate of Alg.~\ref{alg:distrib_manytarg_gate}) is needlessly wasteful and incurs a $\mathcal{O}(2^n)\times$ slowdown over the optimum. 
One might otherwise be tempted to leverage the Euler equality
\begin{align}
    \hat{U}_{\vec{t}}(\theta)\ket{\Psi}\numsub{N} 
    &=
    \cos(\theta)\ket{\Psi}\numsub{N} + \iu \, \sin(\theta) \big( \bigotimes_j^n \hat{Z}_{t_j}  \big) \ket{\Psi}\numsub{N}
\end{align}
and effect $\bigotimes_j^n \hat{Z}_{t_j}$ through the previous section's embarrassingly parallel (when all $\hat{Z}$) scheme to simulate Pauli tensors (Alg.~\ref{alg:distrib_pauli_tensor}). One would then sum the $\iu \sin(\theta)$ scaled result (also an embarrassingly parallel operation) with a $\cos(\theta)$ scaled clone of $\ket{\Psi}\numsub{N}$. This \textit{not-in-place} scheme wastefully incurs an $\mathcal{O}(2^N)$ memory overhead which \textit{could} make use of the communication buffers but which still prescribes needlessly many flops and memory writes. A superior strategy is possible.

We simply leverage that our computational basis states $\ket{i}\numsub{N}$ are composed of eigenstates of $\hat{Z}$.
\begin{gather}
    \ket{\Psi}\numsub{N} \equiv \sum\limits_{i}^{2^N} \alpha_i \ket{i}\numsub{N},
    \hspace{.3cm}
    \ket{i}\numsub{N} = \bigotimes\limits_q^N \ket{i_{[q]}}\numsub{1}, 
    \\
    \hat{Z}\ket{0} = \ket{0}, 
    \hspace{.5cm}
    \hat{Z} \ket{1} = - \ket{1}
    \\
    \implies 
    \hat{Z}_q\ket{i}\numsub{N} = (-1)^{i_{[q]}}\ket{i}\numsub{N}.
    \tag{reiteration of Eq.~\ref{eq:pauli_tensor_Z_op_on_basis}}
\end{gather}
By the spectral theorem~\cite{neumann1955mathematical},
\begin{gather}
     \exp(\iu \, \theta \hat{Z}_q) \ket{i}\numsub{N} = \exp(\iu \, \theta \, (-1)^{i_{[q]}} )
    \ket{i}\numsub{N},
    \\
    \therefore \;
    \hat{U}_{\vec{t}}(\theta)\ket{i}\numsub{N}
    =
    \exp(\iu \, \theta \, (-1)^{\sum_q^n i_{[t_q]}} )
    \ket{i}\numsub{N}.
\end{gather}
The phase gadget upon the state therefore merely multiplies a complex scalar upon each amplitude,
\begin{align}
    \alpha_i 
    \;\;\;
    \xrightarrow{\hat{U}_{\vec{t}}(\theta)}
    \;\;\;
    \exp( \iu \, \theta \, s_i ) \;
    \alpha_i,
    \;\;\;\; s_i = \pm 1
    \label{eq:phase_gadget_amp_change}
\end{align}
where sign $s_i$ is determined by the parity of the number of targeted $\ket{1}\numsub{1}$ substates in the basis state $\ket{i}\numsub{N}$. 
Our pseudocode assumes this quantity is $\mathcal{O}(1)$ evaluable by function \texttt{\textbf{getMaskParity()}}~\cite{anderson2005bit}; such a function is provided by many \texttt{C} compilers.

We formalise a embarrassingly parallel scheme to evaluate Eq.~\ref{eq:phase_gadget_amp_change} in Alg.~\ref{alg:distrib_phase_gadget}. It is as cheap as the best case of the one-qubit gate simulation of Alg.~\ref{alg:distrib_1qb_gate}, though is expected to have superior caching performance.

\begin{algorithm}[H]
\DontPrintSemicolon
\caption[blah blah]{
    \AlgTagDistributed \AlgTagStateVector  \\
    Phase gadget
$
    \hat{U}_{\vec{t}}(\theta) = \exp( \iu  \, \theta \bigotimes_q^n \hat{Z}_{t_q} )
$
upon $n$ unique target qubits $\vec{t}$ of an $N$-qubit statevector distributed between partitions $\vec{\psi}$.
    \protect{ 
    \begin{center}
        $
        \Qcircuit @C=1em @R=.7em {
        & \multigate{1}{ e^{\iu  \, \theta \, \hat{Z}^{\otimes n} } } & \qw \\
        & \ghost{ e^{\iu  \, \theta \, \hat{Z}^{\otimes n} } } & \qw
        }
        $
    \end{center}
    }
    %
    %
    \begin{center}
        \AlgTagBops{$\mathcal{O}(\Lambda)$}%
        \AlgTagFlops{$\mathcal{O}(\Lambda)$}%
        \AlgTagNumSerialRounds{$0$}\\
        \AlgTagAmpsTransferred{$0$}%
        \AlgTagMemOverhead{$\mathcal{O}(1)$}%
        \AlgTagMemWrites{$\Lambda$}
    \end{center}
}
\label{alg:distrib_phase_gadget}

\textbf{distrib\_phaseGadget}($\vec{\psi}$, $\vec{t}$, $\theta$):

\Indp 

    $\Lambda$ = $\dim(\vec{\psi})$

    
    $r'$ = \textbf{getRank}() $\BitShiftLeft \; \log_2(\Lambda)$
    \tcp*{Alg.~\ref{alg:mpi_defs}}

    $b$ = \textbf{getBitMask}($\vec{t}$)
    \tcp*{Alg.~\ref{alg:bit_twiddles}}
    
    
    $u$ = $\exp(\iu \, \theta)$
    
    $v$ = $\exp(-\iu \, \theta)$

    \codegap
    
    \tcp{loop all $\ket{j}\numsub{\lambda}$}

    \AlgThreadComment{multithread}
    
    \textbf{for} $j$ \textbf{in} \textbf{range}($0$, $\Lambda$):
    
    \Indp 

    
    
        $i$ = $r' \BitOr j$
        \tcp*{$\ket{i}\numsub{N} = \ket{r}\numsub{w}\ket{j}\numsub{\lambda}$}

        \codegap 
        
        \tcp{find $p= \sum_q^n i_{[t_q]} \mod 2$ in $\mathcal{O}(1)$}
        
        $p$ = \textbf{getMaskParity}($i \, \BitAnd \, b$)

        
        
        $\vec{\psi}[j]$ \TimesEq{} $v \, p + u (\LogicalNot p)$
        
        
    \Indm

\Indm

\end{algorithm}

Note a potential optimisation is possible. The main loop, which evaluates the sign $s_i=\pm 1$ per-amplitude, can be divided into \textit{two} loops which each iterate only the amplitudes \textit{a priori} known to yield even (and odd) parity (respectively), pre-determining $s_i$. This may incur more caching costs from the twice iteration of the state array, but enables a compiler to trivially vectorise the loop's array modification.

\vspace{.08\paperheight}

\subsection{Pauli gadget}%
\label{sec:distrib_pauli_gadget} %
\noindent 
\begin{equation*}
\Qcircuit @C=1em @R=.7em {
& \multigate{2}{ e^{\iu  \, \theta \, \hat{X}_1 \, \hat{Y}_2 \, \hat{Z}_3\, \dots } } & \qw \\
& \ghost{  e^{\iu  \, \theta \, \hat{X}_1 \, \hat{Y}_2 \, \hat{Z}_3\, \dots }  } & \qw \\
& \ghost{ e^{\iu  \, \theta \, \hat{X}_1 \, \hat{Y}_2 \, \hat{Z}_3\, \dots }  } & \qw 
}
\end{equation*}

The Pauli gadget, also known as the Pauli exponential~\cite{cowtan2020generic} and the multi-qubit multi-axis rotation, is a powerful paramaterised many-qubit entangling unitary gate fundamental in Trotterisation and real-time simulation~\cite{trotter1959product,biamonte2008realisable}, variational algorithms~\cite{cowtan2019phase,gu2021efficient} and error correction~\cite{gottesman2010introduction}.
It is a generalisation of the previous section's phase gadget to rotations around an arbitrary axis, and describes the M{\o}lmer–-S{\o}rensen gate (MS or $XX(\theta)$)~\cite{molmer1999multiparticle} and its global variant (GMS)~\cite{maslov2018use} as well as other exponentiated Pauli gates ~\cite{van2021constructing} natural to ion-trap computers~\cite{harty2014high,schindler2013quantum}.
In this section, we derive an in-place distributed simulation of an $n$-target Pauli gadget with the same costs as the one-qubit gate of Alg.~\ref{alg:distrib_1qb_gate}, albeit a factor $n$ more bops, though which is expected insignificant in the ultimate runtime.

We define the Pauli gadget as an $n$-qubit unitary operator with parameter $\theta \in \mathbb{R}$,
\begin{align}
    \hat{U}_{\vec{t}}(\theta) = \exp\left( \iu  \, \theta \bigotimes\limits_j^n \hat{\sigma}^{(j)}_{t_j} \right),
\end{align}
comprised of Pauli operators $\hat{\sigma}^{(j)}\in\{\hat{X},\hat{Y},\hat{Z}\}$ and acting upon target qubits $\vec{t} = \{t_0, \, \dots, \, t_{n-1}\}$. We seek to apply $\hat{U}_{\vec{t}}(\theta)$ upon an $N$-qubit statevector $\ket{\Psi}\numsub{N}$ distributed between $2^w$ nodes.

Once again, it is prudent to avoid constructing an exponentially large matrix description of $\hat{U}_{\vec{t}}(\theta)$ and simulating it through the multi-target gate of Alg.~\ref{alg:distrib_manytarg_gate}. A clever but ultimately unsatisfactory solution, inspired by our scheme to effect the phase gadget of the previous section, is to rotate the target qubits of the gadget into the eigenstates of the corresponding Pauli operator. For example, 
\begin{align}
     \exp(\iu \, \theta \, \hat X_0 \hat Y_1 \hat Z_2)
    \equiv \;\; &
     {\hat{R}_{Y_0}} \left(\frac{\pi}{2}\right) \hat{R}_{X_1}\left(-\frac{\pi}{2}\right)
     \\ 
     \times & 
    \exp(\iu \theta \hat{Z}_0 \hat{Z}_1 \hat{Z}_2)
    \tag*{}
    \\ 
     \times & 
    \hat{R}_{Y_0}\left(-\frac{\pi}{2}\right) \hat{R}_{X_1}\left(\frac{\pi}{2}\right),
    \tag*{}
\end{align}
where the $\hat{Z}$-phase gate is simulated by Alg.~\ref{alg:distrib_phase_gadget} and each one-target rotation gate by Alg.~\ref{alg:distrib_1qb_gate} using matrix representations
\begin{align}
    R_{\hat{X}}(\pm \frac{\pi}{2} ) &= \frac{1}{\sqrt{2}} 
    \renewcommand*{\arraystretch}{1}
    \begin{pmatrix}
    1 & \mp 1 \\ \pm 1 & 1
    \end{pmatrix}, 
    \\
    R_{\hat{Y}}(\pm \frac{\pi}{2} ) &=
    \frac{1}{\sqrt{2}} 
    \renewcommand*{\arraystretch}{1}
    \begin{pmatrix}
    1 & \mp \iu \\ \mp \iu & 1
    \end{pmatrix}.
\end{align}
Such a scheme involves $\mathcal{O}(n\,2^N)$ total flops and could invoke as many as $2\, n$ rounds of statevector exchange. A superior strategy admitting at most a single exchange is possible, which we now derive.

We first re-express the Pauli gadget into an Euler form made possible by the idempotency of the Pauli operators.
\begin{align}
    \hat{U}_{\vec{t}}(\theta) 
    \; \equiv \;
    \cos(\theta) \hat{\mathbb{1}}^{\otimes n} + \iu \, \sin(\theta)\bigotimes\limits_j^n \hat{\sigma}^{(j)}_{t_j}.
\end{align}
We decided in Sec.~\ref{sec:distrib_phase_gadget} that this was not a useful form to optimally simulate the $\hat{Z}$-phase gadget, though it will here prove worthwhile; it allows us to invoke the results of Sec.~\ref{sec:pauli_tensor} which showed that a Pauli tensor upon basis state $\ket{i}\numsub{N}$ produces
\begin{gather}
\bigotimes\limits_j^n \hat{\sigma}^{(j)}_{t_j}
\ket{i}\numsub{N}
=
\beta_i \,  
    \ket{i_{\neg \vec{t}^{x,y}}}\numsub{N},
    \tag{Eq~.\ref{eq:pauli_tensor_action_on_basis_state}}
    \\
\text{where } \;\;\;
\beta_i =
(-1)^{f(i)} \, \eta 
\;\;\;\;
\in \{\pm 1\, \pm \iu \}
    \tag*{}
\end{gather}
and where $\vec{t}^{x,y} \subseteq \vec{t}$ are the target qubits with corresponding $\hat{X}$ or $\hat{Y}$ operators, $\eta = \iu^{\dim(\vec{t}^y)} \in \{\pm1,\pm \iu\}$ and function $f(i) = \dim(\{ q\in\vec{t}^{y,z} : q_{[i]}=1\}) \in \mathbb{N}$ counts the $\hat{Y}$ or $\hat{Z}$ targeted qubits in the $\ket{1}\numsub{1}$ state within $\ket{i}\numsub{N}$.
The Pauli gadget ergo modifies the $i$-th global amplitude of general state $\ket{\Psi}=\sum_i \alpha_i\ket{i}\numsub{N}$ under
\begin{align}
    \alpha_i 
    \;\,
    \xrightarrow{\hat{U}_{\vec{t}}(\theta)}
    \;\,
    \cos(\theta) \, \alpha_i + 
    \iu \sin(\theta) \, \beta_{i_{\neg \vec{t}^{x,y}}} \, \alpha_{i_{\neg \vec{t}^{x,y}}}.
\end{align} 
This is similar to the modification prescribed by the Pauli tensor as per Eq.~\ref{eq:pauli_tensor_amp_update} and features the same pair amplitude $\alpha_{i_{\neg \vec{t}^{x,y}}}$ with factor $\beta_{i_{\neg \vec{t}^{x,y}}}$, although now the new amplitude also depends on its old value of $\alpha_i$. 
The admitted communication pattern of the Pauli gadget is therefore identical to that of the Pauli tensor. However, floating-point multiplication with generally non-integer quantities $\cos(\theta)$ and $\sin(\theta)$ have been introduced.

We formalise this scheme in Alg.~\ref{alg:distrib_pauli_gadget} which merely describes a small revision to the Pauli tensor simulation of Alg.~\ref{alg:distrib_pauli_tensor}. The Pauli gadget admits $\mathcal{O}(\Lambda)$ flops but otherwise identical scaling costs to the Pauli tensor, and is expected the same ultimate runtime performance of the one-qubit gate.


\begin{algorithm}[b]
\DontPrintSemicolon
\caption[blah blah]{
    \AlgTagDistributed \AlgTagStateVector  \\
    Pauli gadget $\hat{U}_{\vec{t}}(\theta) = \exp\left( \iu  \, \theta \bigotimes_q^n \hat{\sigma}^{(q)}_{t_q} \right)$ consisting of $n$ Pauli operators $\hat{\sigma}^{(q)}$ (encoded into array $\vec{\sigma}$) upon unique targets $\vec{t}=\{t_0,\,\dots,\,t_{n-1}\}$, applied to an $N$-qubit statevector distributed as $\vec{\psi}$ local arrays, with buffers $\vec{\varphi}$.
    \protect{\begin{center} 
        $
        \Qcircuit @C=1em @R=.7em {
        & \multigate{1}{ e^{\iu  \, \theta \, \hat{X}_1 \hat{Y}_2 \dots } } & \qw \\
        & \ghost{ e^{\iu  \, \theta \, \hat{X}_1 \hat{Y}_2 \dots } } & \qw
        }
        $
    \end{center}}
    \begin{center}
        \AlgTagBops{$\mathcal{O}(n \, \Lambda)$}%
        \AlgTagFlops{$\mathcal{O}(\Lambda)$}%
        \AlgTagNumSerialRounds{$0$ or $1$}\\
        \AlgTagAmpsTransferred{$2^N$}%
        \AlgTagMemOverhead{$\mathcal{O}(1)$}%
        \AlgTagMemWrites{$\Lambda$}
    \end{center}
}
\label{alg:distrib_pauli_gadget}

\textbf{distrib\_pauliGadget}($\vec{\psi}$, $\vec{\varphi}$, $\vec{\sigma}$, $\vec{t}$, $\theta$):

\Indp 

    $a$ = $\cos(\theta)$

    $b$ = $\sin(\theta)$

    \codegap 

    \tcp{Identical to Alg.~\ref{alg:distrib_pauli_tensor} except lines~\ref{algline:pauli_tensor_A}-\ref{algline:pauli_tensor_B} of subroutine~\ref{alg:distrib_pauli_tensor_subroutine} become:
    }

    \Indp 

    $\vec{\psi}[j]$ = $a \; \vec{\psi}[j] \;\; + \;\; \iu \; b \; \beta' \; \vec{\psi}[j'] 
    $

    $\vec{\psi}[j']$ = $a \; \vec{\psi}[j'] \;\; + \;\; \iu \; b \; \beta \; \gamma 
    $

    \Indm 

    \codegap 

    \tcp{and line~\ref{algline:pauli_tensor_C} becomes:
    }

    \Indp 

    $\vec{\psi}[j] = a\; \vec{\psi}[j] \;\; + \;\; \iu \; b \; \beta \; \vec{\varphi}[j']$

    \Indm 

    \codegap

    \tcp{Note the subroutine of Alg.~\ref{alg:distrib_pauli_tensor_subroutine} must be modified in a similar, trivial way.}

\Indm 
 \end{algorithm}

\pagebreak


\section{Distributed density matrix algorithms}
\label{sec:distributed_dens_matr_algorithms}

The algorithms presented so far in this manuscript simulated idealised purity-preserving operators upon pure states numerically instantiated as statevectors.
We now seek a treatment of realistic operations and noise, as their simulation is essential in the development of quantum computers~\cite{obenland1998simulating}.
We consider processes modelled as \textit{channels}; linear, completely-positive, trace-preserving maps between density matrices. Such maps describe a very general family of physical processes~\cite{nielsen2002quantum}. 

In order to precisely simulate decoherence and the evolution of a mixed state, we will numerically instantiate and evolve a density matrix. We must replace our $2^N$-length array $\vec{\Psi}$, which encoded $N$-qubit statevector $\ket{\Psi}\numsub{N}$, with an encoding of the \textit{squared larger} density matrix $\bm{\rho}\numsub{N}$.
This appears an intimidating new task; one may expect to create a new distributed matrix data-structure, then re-derive all previous algorithms for simulating unitaries upon it, in addition to deriving new simulations of decoherence channels. Incredibly, this gargantuan effort is not necessary. By leveraging the Choi--Jamio\l{}kowski isomorphism~\cite{choi1975completely,jamiolkowski1972linear} (also referred to as the channel-state duality~\cite{min_jiang_channel_state}), all distributed statevector algorithms presented in this manuscript can be repurposed for numerically simulating unitaries upon a density matrix. This was first very briefly described by the authors in Ref.~\cite{jones2019quest}.

In this section, we will explicitly demonstrate this correspondence, then derive \textit{thirteen} novel distributed algorithms for simulating density matrices. These algorithms will effect unitaries and noise channels, and calculate expectation values and partial traces.
In all these schemes, we assume each of $W=2^w$ nodes contains a state partition labelled $\vec{\rho}$ (described below) and an equally sized communication buffer $\vec{\varphi}$. For a state of $N$-qubits, these arrays have size $\Lambda = 2^{2N-w}$. We also introduce the constraint that $w \le N$, also elaborated upon below.


\subsection{State representation}
\label{sec:dens_state}

Sec.~\ref{sec:local_statevector_algorithms} modelled an $N$-qubit pure state $\ket{\Psi}\numsub{N}$ as a dense statevector in the $\hat{Z}$-basis, numerically instantiated as a complex array $\vec{\Psi}$.
\begin{align}
    \ket{\Psi}\numsub{N} &= \sum\limits_{i}^{2^N} \alpha_i \ket{i}\numsub{N}
    \tag*{}
    \;\;\;\;
    \leftrightarrow
    \\
    \vec{\Psi} &= \{\alpha_0, \; \dots, \; \alpha_{2^N-1} \}.
    \tag{repetition of \ref{eq:local_statevec_model}}
\end{align}
Sec.~\ref{sec:distribution} then distributed $\vec{\Psi}$ between $W=2^w$ nodes, each with a local sub-statevector $\vec{\psi}$ of length $2^{N-w}$. We must now perform a similar procedure for representing a density matrix.

Consider a general $N$-qubit density matrix $\bm\rho\numsub{N}$ with elements $\alpha_{kl} \in \mathbb{C}$ corresponding to structure
\begin{align}
    \bm\rho\numsub{N} &= \sum\limits_{k}^{2^N} \sum\limits_{l}^{2^N} \, \alpha_{kl} \, \ket{k}\bra{l}\numsub{N}.
\end{align}

The basis projector $\ket{k}\bra{l}\numsub{N}$ \textit{can} be numerically instantiated as a 
$\mathbb{R}^{2^N\times 2^N}$ matrix.
Instead, we vectorise it under the  Choi--Jamio\l{}kowski isomorphism~\cite{choi1975completely,jamiolkowski1972linear}, admitting the same form as a $2N$-qubit basis ket of index $i = k + l\,2^N$,
\begin{align}
\choivec{i}\numsub{2N} \,\simeq\, \ket{l}\numsub{N}\ket{k}\numsub{N},
\end{align}
which we numerically instantiate as an array. We have used notation $\choivec{i}\numsub{2N}$ to indicate a state which is described with a $2N$-qubit statevector but which is \textit{not} necessarily a pure state of $2N$-qubits; it will instead generally be unnormalised and describe a mixed state of $N$ qubits. We henceforth refer to this as a ``Choi-vector".

Our state $\bm{\rho}\numsub{N}$ is then expressible as
\begin{align}
    \choivec{\bm\rho}\numsub{2N}
    &=
     \sum\limits_{k}^{2^N} \sum\limits_{l}^{2^N} \, \alpha_{kl} \, \ket{l}\numsub{N}\ket{k}\numsub{N}
    \\
    &\equiv
    \sum\limits_{i}^{2^{2N}} \;
    \beta_i 
    \choivec{i}\numsub{2N},
    \;\;\;\;
    \beta_i = \alpha_{i \BitMod 2^N, \floor{i/2^N}},
\end{align}
and in local simulation, is numerically instantiated as a $(2^{2N})$-length complex array. This can be imagined as concatenating the columns of matrix $\matr{\rho}$ into a column vector. This means we can leverage an existing statevector data structure to store $\bm{\rho}$, though the next section will reveal a greater benefit of our representation.

We distribute the array encoding a density matrix between compute nodes in an identical fashion to the statevector array in Sec.~\ref{sec:distribution}, as if it were indeed a $2N$-qubit statevector.
Precisely, we uniformly distribute the $2^{2N}$ elements $\beta_i$ between $W=2^w$ nodes, each storing a sub-array $\vec{\rho}$ of length $\Lambda=2^{2N-w}$. 
The $j$-th local element $\vec{\rho}[j]$ in node $r$ stores the element $\beta_{j+r\,\Lambda}$, which is the coefficient of global basis projector
\begin{align}
\choivec{i}\numsub{2N}  \equiv \choivec{r}\numsub{w} \choivec{j}\numsub{\lambda}.
\end{align}

In theory, we can instantiate density matrices with as few as $N=\lceil w/2\rceil$ qubits, prescribing $1$ or $2$ amplitudes per-node. Of course this is a ridiculous and impractical regime ill-suited for distribution. 
Even employing more nodes than there are \textit{columns} of the encoded density matrix is a considerable waste of parallel resources.
We are ergo safe to impose an important precondition for this section's algorithms:
 \begin{align}
N \ge w.
\label{eq:densmatr_amps_per_node_min_constraint}
 \end{align}
That is, we assume the number of density matrix elements stored in each node, $\Lambda$, is at least $2^N$, or \textit{one column's worth}. The \textit{smallest} density matrix that $W$ nodes can cooperatively simulate is then $N \ge \log_2(W)$.
This is in no way restrictive;
employing $32$ nodes, for example, would require we simulate density matrices of at least $5$ qubits, and $1024$ nodes demand at least $10$ qubits; both are trivial tasks for a \textit{single} node.
Furthermore, using $4096$ nodes of ARCHER2~\cite{archer2_hardware}, our precondition requires we employ at least $\times 10^{-6}\,\%$ of local memory (to simulate a measly $12$ noisy qubits); it is prudent to use over $50\%$!
It is fortunate that precondition~\ref{eq:densmatr_amps_per_node_min_constraint} is safely assumed, since it proves critical to eliminate many tedious edge-cases from our subsequent algorithms .

Initialising our distributed Choi-vector  into canonical pure states is as trivial as it is for statevectors;
%
\begin{align}
&\ket{0}\bra{0}\numsub{N}  && \simeq  \ket{0}\numsub{2N} 
\\
&\ket{+}\bra{+}\numsub{N} && \simeq  \frac{1}{2^{N/2}} \ket{+}\numsub{2N} 
\\
 &\ket{i}\bra{i}\numsub{N} && \simeq\; 
  \ket{i(2^N+1)}\numsub{2N} 
  \\
 &\ket{\psi}\bra{\psi}\numsub{N} && \simeq  \sum\limits_{j,k}^{2^N} \psi_j \psi_k^* \ket{j+k\,2^N}\numsub{2N}
\end{align}

For concision, our density matrix algorithms will often invoke the below subroutine (Alg.~\ref{alg:distrib_dens_convenience_funcs}).

\begin{algorithm}[htb]
\DontPrintSemicolon
\caption{
    A convenience subroutine of our distributed density-matrix algorithms which merely infers the number of qubits described by a local Choi-vector array $\vec{\rho}$.
}
\label{alg:distrib_dens_convenience_funcs}

\textbf{getNumQubits}($\vec{\rho}$):

\Indp

    $W$ = \textbf{getWorldSize()}
    \tcp*{Alg.~\ref{alg:mpi_defs}}

    $N$ = $\log_2(\dim(\vec{\rho}) \, W  )/2$

    \textbf{return} $N$

\Indm 

\end{algorithm}

\pagebreak

\subsection{Unitary gates} %
\label{sec:dens_unitaries} %
\noindent %
\begin{equation}
\Qcircuit @C=1em @R=.7em {
& \multigate{1}{\hat{U}} & \qw \\
& \ghost{\hat{U}} & \qw \\
& \qw & \qw 
}
\end{equation}

The ability to simulate unitary, purity-preserving operations upon a density matrix is as important as simulating decoherence.

In quantum computing literature, wherein a precise single-channel description of an experimental process is often unobtainable (and otherwise computationally intractable),
the execution of a quantum circuit on a noisy device is traditionally modelled as an intertwined sequence of unitary and mixing operations~\cite{aharonov2000quantum}. T
It is ergo prudent to develop distributed simulations of this manuscript's previous unitary operations upon density matrices. Fortunately, as first described in Ref.~\cite{jones2019quest}, the Choi--Jamio\l{}kowski isomorphism~\cite{choi1975completely,jamiolkowski1972linear} permits us to repurpose our previous statevector algorithms upon our vectorised density matrix. In this section, we derive a distributed simulation of any unitary $\hat{U}$ upon an $N$-qubit density matrix $\bm{\rho}\numsub{N}$ represented as a Choi-vector $\choivec{\bm{\rho}}\numsub{2N}$, in approximately \textit{double} the time to simulate the same unitary upon a $2N$-qubit pure state via the schemes of Section~\ref{sec:distributed_statevector_algorithms}. 
In its generic $n$-qubit unitary form, we require $n \le N-\lceil \log_2(W) \rceil$, where $W$ is the number of nodes.
We also explicitly develop bespoke methods to simulate the SWAP gate, Pauli tensor, phase gadget and Pauli gadget.
We note that unitarity of the matrix representation of $\hat{U}$ is not actually required by our algorithms.

Algorithms~\ref{alg:distrib_1qb_gate}~to~\ref{alg:distrib_pauli_gadget} modelled the evolution of an $N$-qubit pure state $\ket{\Psi}\numsub{N}$
under an operator $\hat{U} \in SU(2^N)$, whereby
\begin{align}
\ket{\Psi}\numsub{N} \rightarrow \hat{U} \ket{\Psi}\numsub{N}.
\end{align}
These algorithms also did not require unitarity of $\hat{U}$ when specified as a dense matrix.
Under the same operation, a density matrix $\bm{\rho}$ becomes
\begin{align}
    \bm\rho\numsub{N} \rightarrow \hat{U} \bm\rho\numsub{N} \, \hat{U}^\dagger,
\end{align}
equivalently described by Choi-vector
\begin{align}
\choivec{\hat{U} \bm\rho \, \hat{U}^\dagger}\numsub{2N} &= \left( \hat{U}^* \otimes \hat{U} \right) \choivec{\bm\rho}\numsub{2N}
\\
&=
\left( \hat{U}^* \otimes \vec{\mathbb{1}}^{\otimes N} \right) 
\left( \vec{\mathbb{1}}^{\otimes N} \otimes \hat{U} \right) 
\choivec{\bm\rho}\numsub{2N}.
\end{align}
Assume that $\hat{U}$ targets specific qubits with indices $\vec{t}$, effecting identity upon the remaining. Then
\begin{align}
\choivec{\hat{U}_{\vec{t}}\, \bm\rho \, \hat{U}_{\vec{t}}^\dagger}\numsub{2N} 
&=
\hat{U}_{\vec{t}+N}^* \, \hat{U}_{\vec{t}} \, \choivec{\bm\rho}\numsub{2N},
\label{eq:unitary_on_vectorised_dens}
\end{align}
where $\vec{t}+N$ notates array $\vec{t}$ with $N$ added to each element.
We next appreciate that when $\choivec{\bm\rho}$ is instantiated as an unnormalised statevector, effecting $\choivec{\bm\rho}\numsub{2N} \rightarrow \hat{U}_{\vec{t}}\,\choivec{\bm\rho}\numsub{2N}$ can be done through an identical process to the updating of a pure state $\ket{\Psi}\numsub{2N} \rightarrow \hat{U}_{\vec{t}}\ket{\Psi}\numsub{2N}$.
Eq.~\ref{eq:unitary_on_vectorised_dens} then bespeaks a simple strategy to simulate a unitary upon a density matrix by sequentially simulating \textit{two} gates ($\hat{U}_{\vec{t}}$ then $\hat{U}^*_{\vec{t}+N}$) upon its Choi-vector, treated as an unnormalised statevector of $2N$ qubits, using the statevector algorithms of Sec.~\ref{sec:distributed_statevector_algorithms}. 

We formalise the general case, when $\hat{U}$ is a many-target unitary specified element-wise, in Alg.~\ref{alg:distrib_dens_matr_unitary}. This is a density matrix generalisation of the equivalent statevector scheme in Alg.~\ref{alg:distrib_manytarg_gate}, which is incidentally invoked twice within, and ergo prescribes only twice as many flops and writes than the latter algorithm upon a pure state of twice as many qubits. Further, since Alg.~\ref{alg:distrib_dens_matr_unitary}'s first invocation of Alg.~\ref{alg:distrib_manytarg_gate} modifies at most the lowest $N$ qubits of $\choivec{\bm\rho}\numsub{2N}$ (and because $N \ge w$ by precondition), it achieves the best-case scenario and is embarrassingly parallel. 
The second invocation of Alg.~\ref{alg:distrib_manytarg_gate} invokes at most $\min(w, \dim(\vec{t}))$ rounds of communication.

\begin{algorithm}[b]
\DontPrintSemicolon
\caption[blah blah]{
    \AlgTagDistributed \AlgTagDensityMatrix  \\
    Unitary $\hat{U}_{\vec{t}}$ with $n$ target qubits $\vec{t}$, specified element-wise as a dense matrix $\matr{U}\in \mathbb{C}^{2^{n}\times 2^{n}}$, acting upon an $N$-qubit density matrix with Choi-vector distributed between $W=2^w$ nodes as
    local arrays $\vec{\rho}$ of size $\Lambda=2^{2N-w}$. Each node also stores a $\Lambda$-sized buffer $\vec{\varphi}$.
    Control qubits are trivial to introduce, and are simply shifted like $\vec{t}$.
    \protect{\begin{center} 
        $
\Qcircuit @C=1em @R=.7em {
& \multigate{1}{\hat{U}} & \qw \\
& \ghost{\hat{U}} & \qw 
}
        $
    \end{center}}
    \begin{center}
        \AlgTagBops{$\mathcal{O}(2^n\Lambda)$}%
        \AlgTagFlops{$\mathcal{O}(2^n\Lambda)$} \\
        \AlgTagNumSerialRounds{best: 0, worst: $\min(n,w)$} \\
        \AlgTagAmpsTransferred{worst: $\min(n,w)\,2^{2N}/2$} \\
        \AlgTagMemOverhead{$\mathcal{O}(2^n)$}%
        \AlgTagMemWrites{$\mathcal{O}(2^n\Lambda )$}
    \end{center}
}
\label{alg:distrib_dens_matr_unitary}

\textbf{distrib\_density\_manyTargGate}($\vec{\rho}$, $\vec{\varphi}$, $\matr{U}$, $\vec{t}$):

\Indp 

    \codegap 

    \tcp{$\choivec{\bm\rho} \rightarrow \hat{U}_{\vec{t}} \choivec{\bm\rho}$}

    \textbf{distrib\_manyTargGate}($\vec{\rho},\,\vec{\varphi},\,\matr{U},\,\vec{t}$)
    \tcp{Alg.~\ref{alg:distrib_manytarg_gate}}

    \codegap

    \tcp{$\hat{U}_{\vec{t}} \rightarrow \hat{U}^*_{\vec{t}+N}$}

    $N$ = \textbf{getNumQubits}($\vec{\rho}$)
    \tcp*{Alg.~\ref{alg:distrib_dens_convenience_funcs}}

    \textbf{for} $q$ \textbf{in} \textbf{range}($0$, $\dim(\vec{t})$):

    \Indp 

        $\vec{t}[q]$ \PlusEq{} $N$

    \Indm 

    $\matr{U}$ = $\matr{U}^{\,*}$

    \codegap

    \tcp{$\choivec{\bm\rho} \rightarrow \hat{U}^*_{\vec{t}+N} \choivec{\bm\rho}$}

    \textbf{distrib\_manyTargGate}($\vec{\rho},\,\vec{\varphi},\,\matr{U},\,\vec{t}$)
    \tcp{Alg.~\ref{alg:distrib_manytarg_gate}}

\Indm 
 \end{algorithm}

\begin{algorithm}[b]
\DontPrintSemicolon
\caption[blah blah]{
    \AlgTagDistributed \AlgTagDensityMatrix  \\
    The SWAP gate, Pauli tensor, phase gadget and Pauli gadget upon an $N$-qubit density matrix with Choi-vector distributed between arrays $\vec{\rho}$ and same-sized per-node buffer $\vec{\varphi}$. These are special cases of Alg.~\ref{alg:distrib_dens_matr_unitary} which avoid construction of a unitary matrix.
    \protect{\begin{center} 
        $
\Qcircuit @C=1em @R=.7em {
& \qswap     & \gate{\hat{X}}  & \multigate{1}{e^{\iu\,\theta\,\hat{Z}^{\otimes}}}  & \multigate{1}{e^{\iu\,\theta\,\hat{\sigma}^{\otimes}}} & \qw \\
& \qswap \qwx &  \gate{\hat{Y}}  & \ghost{{e^{\iu\,\theta\,\hat{Z}^{\otimes}}}}  & \ghost{{e^{\iu\,\theta\,\hat{\sigma}^{\otimes}}}} & \qw 
}
        $
    \end{center}}
}
\label{alg:distrib_dens_matr_specific_unitaries}

\tcp{for clarity, assume $N$ is global}

$N$ = \textbf{getNumQubits}($\vec{\rho}$)
\tcp*{Alg.~\ref{alg:distrib_dens_convenience_funcs}}

\codegap

\textbf{distrib\_density\_swapGate}($\vec{\rho}$, $\vec{\varphi}$, $t_1$, $t_2$):

\Indp 

    \textbf{distrib\_swapGate}($\vec{\rho}$, $\vec{\varphi}$, $t_1$, $t_2$)
    \tcp*{Alg.~\ref{alg:distrib_swap}}

    \textbf{distrib\_swapGate}($\vec{\rho}$, $\vec{\varphi}$, $t_1+N$, $t_2+N$)

\Indm 

\codegap 

\codegap

\textbf{distrib\_density\_pauliTensor}($\vec{\rho}$, $\vec{\varphi}$, $\vec{\sigma}$, $\vec{t}$):

\Indp 

    \textbf{distrib\_pauliTensor}($\vec{\rho},\,\vec{\varphi},\,\vec{\sigma},\,\vec{t}$)
    \tcp*{Alg.~\ref{alg:distrib_pauli_tensor}}

    \codegap 

    $m$ = $0$

    \textbf{for} $q$ \textbf{in} \textbf{range}($0$, $\dim(\vec{t})$):

    \Indp 

        $\vec{t}[q]$ \PlusEq{} $N$

        \textbf{if} $\vec{\sigma}[q]$ \textbf{is} $\hat{Y}$:

        \Indp 

            $m$ = $\LogicalNot m$

        \Indm 

    \Indm 

    \codegap 

    \textbf{distrib\_pauliTensor}($\vec{\rho},\,\vec{\varphi},\,\vec{\sigma},\,\vec{t}$)
    \tcp*{Alg.~\ref{alg:distrib_pauli_tensor}}

    \codegap

    \textbf{if} $m$ \EqEq{} $1$:

    \Indp 

        \AlgThreadComment{multithread}

        \textbf{for} $j$ \textbf{in} \textbf{range}($0$, $\dim(\vec{\rho})$):

        \Indp 

            $\vec{\rho}[j]$ \TimesEq{} $-1$

        \Indm 

    \Indm
    
\Indm 

\codegap 

\codegap 

\textbf{distrib\_density\_phaseGadget}($\vec{\rho}$, $\vec{t}$, $\theta$):

\Indp 

    \textbf{distrib\_phaseGadget}($\vec{\rho}$, $\vec{t}$, $\theta$)
    \tcp*{Alg.~\ref{alg:distrib_phase_gadget}}

    \textbf{for} $q$ \textbf{in} \textbf{range}($0$, $\dim(\vec{t})$):

    \Indp 

        $\vec{t}[q]$ \PlusEq{} $N$

    \Indm 

    \textbf{distrib\_phaseGadget}($\vec{\rho}$, $\vec{t}$, $-\theta$)
    \tcp*{Alg.~\ref{alg:distrib_phase_gadget}}

\Indm 

\codegap 

\codegap 

\textbf{distrib\_density\_pauliGadget}($\vec{\rho}$, $\vec{\varphi}$, $\vec{\sigma}$, $\vec{t}$, $\theta$):

\Indp 

    \textbf{distrib\_pauliGadget}($\vec{\rho},\,\vec{\varphi},\,\vec{\sigma},\,\vec{t},\,\theta$)
    \tcp{Alg~\ref{alg:distrib_pauli_gadget}}

    \codegap 
    

    $\theta$ \TimesEq{} $-1$

    \textbf{for} $q$ \textbf{in} \textbf{range}($0$, $\dim(\vec{t})$):

    \Indp 

        $\vec{t}[q]$ \PlusEq{} $N$

        \textbf{if} $\vec{\sigma}[q]$ \textbf{is} $\hat{Y}$:

        \Indp 


            $\theta$ \TimesEq{} $-1$

        \Indm 

    \Indm 

    \codegap 

    \textbf{distrib\_pauliGadget}($\vec{\rho},\,\vec{\varphi},\,\vec{\sigma},\,\vec{t},\,\theta$)
    \tcp{Alg~\ref{alg:distrib_pauli_gadget}}

\Indm 

 \end{algorithm}

We can improve on this for specific unitaries.
The previous section's bespoke statevector algorithms for simulating SWAP gates (Alg.~\ref{alg:distrib_swap}), Pauli tensors (Alg.~\ref{alg:distrib_pauli_tensor}), phase gadgets (Alg.~\ref{alg:distrib_phase_gadget}) and Pauli gadgets (Alg.~\ref{alg:distrib_pauli_gadget}) were markedly more efficient than their simulation as general many-target gates via Alg.~\ref{alg:distrib_manytarg_gate}. We can repurpose every one of these algorithms for simulating the same operations upon density matrices represented as Choi-vectors. Observe:
\begin{align}
&\hat{U} = \text{SWAP} && \implies && \hat{U}^* = \hat{U}
\tag*{}
\\
&\hat{U} = \hat{X}^\otimes \hat{Y}^{\otimes m}\hat{Z}^\otimes 
&&\implies &&
\hat{U}^* = (-1)^m \hat{U}
\label{eq:distrib_pauli_tensor_conj}
\tag*{}
\\
&\hat{U}(\theta) = \text{exp}( \iu \,  \theta \hat{Z}^{\otimes})
 && \implies  &&
\hat{U}(\theta)^* = \hat{U}(-\theta)
\tag*{}
\\
&\hat{U}(\theta) = \text{exp}(\iu \, \theta \hat{X}^{\otimes} \hat{Y}^{\otimes m} \hat{Z}^{\otimes})
&& \implies &&
\hat{U}(\theta)^* = \hat{U}((-1)^{m+1}\theta).
\tag*{}
\end{align}
The above $\hat{U}^*$ can be directly effected upon a Choi-vector through $\hat{U}$'s statevector algorithm, without needing a matrix construction. We make this strategy explicit in Alg.~\ref{alg:distrib_dens_matr_specific_unitaries}.


\vspace{2cm}

\subsection{Kraus map}%
\label{sec:kraus_map}%
\noindent %
\begin{equation}
\Qcircuit @C=1em @R=.7em {
& \multigate{1}{\mathcal{E}_{\{\hat{K}^{(m)}\}}} & \qw \\
& \ghost{\mathcal{E}_{\{\hat{K}^{(m)}\}}} & \qw \\
& \qw & \qw 
}
\end{equation}

The power of a density matrix state description is its ability to capture classical uncertainty as a result of decohering processes.
A Kraus map, also known as an operator-sum representation of a channel~\cite{nielsen2002quantum}, allows an operational description of an open system's dynamics which abstracts properties of the environment. A Kraus map can describe any completely-positive trace-preserving channel, and ergo capture almost all quantum noise processes of practical interest.
In this section, we derive a distributed simulation of an $n$-qubit Kraus map of $M$ operators, each specified as general matrices, acting upon an $N$-qubit density matrix represented as a Choi-vector.
We assume $M \ll 2^{2N}$ such that $M 2^{4n} \ll 2^{2N-w}$; i.e. that the descriptions of the Kraus channels are much smaller than the distributed density matrix, so are safely duplicated on each node. We will strictly require that $n \le N- \lceil w/2 \rceil$ to satisfy memory preconditions imposed by an invoked subroutine (incidentally, Alg.~\ref{alg:distrib_manytarg_gate}'s simulation of the many-target gate upon a statevector).

An $n$-qubit Kraus map consisting of $M$ Kraus operators $\{\hat{K}_{\vec t}^{(m)}:m<M\}$, each described by a matrix $\matr{K}^{(m)} \in \mathbb{C}^{2^n \times 2^n}$,
operating upon qubits $\vec{t}$ (where $n=\dim(\vec{t})$) modifies an $N$-qubit density matrix $\bm\rho$ via
\begin{gather}
\bm\rho \; \rightarrow \;
    \mathcal{E}(\bm\rho) 
    =
    \sum\limits_m^M \hat{K}_{\vec t}^{(m)} \, \bm\rho \; \hat{K}_{\vec t}^{(m)\dagger}.
\end{gather}
A valid Kraus map satisfies
\begin{gather} 
    \sum\limits_m^M \hat{K}_{\vec t}^{(m)\dagger} \hat{K}_{\vec t}^{(m)} = \hat{\mathbb{1}}^{\otimes N},
\end{gather}
though our algorithm does not require this.
A naive strategy may seek to clone $\bm{\rho}$, modify $\bm{\rho}\rightarrow \hat{K}_{\vec{t}}^{(m)} \bm\rho \, \hat{K}_{\vec{t}}^{(m)\dagger}$ for a particular $m$ via Alg.~\ref{alg:distrib_dens_matr_unitary} (which, recall, did not require unitarity), and matrix-sum the results across $m$. 
Such a strategy requires $\mathcal{O}(2^{2N})$ additional memory and yields a $\mathcal{O}(M \, 2^{2n} \, 2^{2N})$ runtime, inducing as many as $\mathcal{O}(M \, n)$ rounds of communication. A superior strategy is possible.

Equ.~\ref{eq:unitary_on_vectorised_dens} informs us that the action of a single Kraus operator upon the Choi-vector of $\bm\rho\numsub{N}$ is
\begin{align}
    \choivec{ \hat{K}_{\vec t}^{(m)}\, \bm\rho \; \hat{K}_{\vec t}^{(m) \dagger}}\numsub{2N}
    &= \;
    \hat{K}_{\vec t + N}^{(m)*}  \; \hat{K}_{\vec{t}}^{(m)} \choivec{\bm\rho}\numsub{2N}.
\end{align}
By linearity, the full Kraus map produces
\begin{align}
    \choivec{ \mathcal{E}(\bm\rho)}\numsub{2N} &=
    \left( 
    \sum_m^M \hat{K}_{\vec t + N}^{(m)*}  \; \hat{K}_{\vec{t}}^{(m)}
    \right) 
    \choivec{\bm\rho}\numsub{2N}
    \\
    &= \hat{S}_{\vec{t} \,\cup\, \vec{t}+N} \, \choivec{\bm\rho}\numsub{2N},
\end{align}
described as the action of a single 
$2n$-qubit superoperator
\begin{align}
\hat{S}_{\vec{t} \,\cup\, \vec{t}+N} &=
\sum\limits_m^M \left( \hat{K}^{(m)*} \otimes \hat{K} \right)_{\vec{t} \,\cup\, \vec{t}+N}.
\label{eq:superop_from_kraus_map}
\end{align}
A dense matrix representation of $\hat{S}$ can ergo be obtained by numerically evaluating $M$ Kronecker products of $\mathbb{C}^{2^n\times2^n}$ matrices, and $M$ sums of $\mathbb{C}^{2^{2n}\times2^{2n}}$ matrices. This is a trivial overhead when $M$ and $n$ are small, as assumed.
Then,
$\hat{S}$ could be simulated upon $\choivec{\bm\rho}\numsub{2N}$ via Alg.~\ref{alg:distrib_manytarg_gate} as if it were an (unnormalised) $2n$-qubit unitary gate acting upon an (unnormalised) $2N$-qubit statevector.

We formalise this scheme in Alg.~\ref{alg:distrib_dens_matr_kraus_map}. It admits the same asymptotic costs as distributed simulation of a unitary targeting \textit{twice} as many qubits upon a statevector containing \textit{twice} as many qubits.

Note that for clarity, we assumed that the construction of the non-distributed $2n$-qubit superoperator matrix $\matr{S}$ is serially tractable. However, as it involves $\mathcal{O}(M 16^n)$ floating-point operations, it \textit{may} quickly grow to be a non-negligible overhead even when $\matr{S}$ remains tractable in memory. In that scenario, the five nested loops of Alg.~\ref{alg:distrib_dens_matr_kraus_map} can be trivially flattened into a single embarrassingly parallel iteration of the elements of $\matr{S}$, and locally parallelised with multithreading.

\begin{algorithm}[tb]
\DontPrintSemicolon
\caption[blah blah]{
    \AlgTagDistributed \AlgTagDensityMatrix  \\
    Kraus map of $M$ operators $\{\hat{K}_{\vec{t}}^{(m)}:m\}$, each described by a matrix $\matr{K}^{(m)}\in\mathbb{C}^{2^n\times 2^n}$, targeting $n$ qubits $\vec{t}$ of an $N$-qubit density matrix with Choi-vector distributed between $W=2^w$ nodes as
    local arrays $\vec{\rho}$ of size $\Lambda=2^{2N-w}$. Each node also stores a $\Lambda$-sized buffer $\vec{\varphi}$.
    \protect{\begin{center} 
        $
\Qcircuit @C=1em @R=.7em {
& \multigate{1}{\mathcal{E}_{\{\hat{K}^{(m)}\}}} & \qw \\
& \ghost{\mathcal{E}_{\{\hat{K}^{(m)}\}}} & \qw \\
& \qw & \qw 
}
        $
    \end{center}}
    \begin{center}
        \AlgTagBops{$\mathcal{O}(2^{2n}\Lambda)$}%
        \AlgTagFlops{$\mathcal{O}(2^{2n}\Lambda)$} \\
        \AlgTagNumSerialRounds{best: 0, worst: $\min(2n,w)$} \\
        \AlgTagAmpsTransferred{worst: $\min(2n,w)\,2^{2N}/2$} \\
        \AlgTagMemOverhead{$\mathcal{O}(2^{4n})$}%
        \AlgTagMemWrites{$ \mathcal{O}(2^{2n}\Lambda)$}
    \end{center}
}
\label{alg:distrib_dens_matr_kraus_map}

\textbf{distrib\_krausMap}($\vec{\rho}$, $\vec{\varphi}$, $\{\matr{K}^{(m)}\}$, $\vec{t}$):

\Indp 

    $n$ = $\dim(\vec{t})$

    \codegap 

    \tcp{create $\matr{S}=\sum_m \;\matr{K}^{(m)*}\otimes\matr{K}^{(m)}$}

    $\matr{S} = \matr{0}^{2^{2n} \times 2^{2n}}$
    
    \textbf{for} $m$ \textbf{in} \textbf{range}($0$, $M$):
    
        \Indp 
        
        \textbf{for} $i$ \textbf{in} \textbf{range}($0$, $2^n$):
        
            \Indp
            \textbf{for} $j$ \textbf{in} \textbf{range}($0$, $2^n$):
            
                \Indp
                \textbf{for} $k$ \textbf{in} \textbf{range}($0$, $2^n$):
                
                    \Indp
                    \textbf{for} $l$ \textbf{in} \textbf{range}($0$, $2^n$):
                    
                        \Indp
                        
                        $r = i \, 2^n+k$
                        
                        $c = j \,2^n+l$
                        
                        $v = \matr{K}^{(m)}[i,j]^* \;  \matr{K}^{(m)}[k,l]$
                        
                        $\matr{S}[r, \, c]$ \PlusEq{} $v$
                    
                    \Indm
                \Indm 
            \Indm 
        \Indm
    \Indm

    \codegap 

    \tcp{create $\vec{t}' = \vec{t}\,\cup\,\vec{t}+N$}
    
    $N$ = \textbf{getNumQubits}($\vec{\rho}$)
    \tcp*{Alg.~\ref{alg:distrib_dens_convenience_funcs}}

    $\vec{t}'$ = $\vec{0}^{2n}$

    \textbf{for} $q$ \textbf{in} \textbf{range}($0$, $n$):

    \Indp 

        $\vec{t}'[q]$ = $\vec{t}[q]$

        $\vec{t}'[q+n]$ = $\vec{t}[q] + N$

    \Indm 

    \codegap 

    \tcp{$\choivec{\bm\rho} \rightarrow \hat{S}_{\vec{t}\,\cup\,\vec{t}+N} \choivec{\bm\rho}$}

    \textbf{distrib\_manyTargGate}($\vec{\rho},\,\vec{\varphi},\,\matr{S},\,\vec{t}'$)
    \tcp{Alg.~\ref{alg:distrib_manytarg_gate}}

\Indm 

 \end{algorithm}

\clearpage

\subsection{Dephasing channel} %
\label{sec:dephasing}
\noindent %
\begin{equation}
\Qcircuit @C=1em @R=.7em {
& \gate{\mathcal{E}_\phi} & \multigate{1}{\mathcal{E}_\phi} & \qw \\
& \qw & \ghost{\mathcal{E}_\phi} & \qw \\
& \qw & \qw  & \qw
}
\end{equation}

 The dephasing channel, also known as the phase damping channel~\cite{nielsen2002quantum}, is arguably the simplest and most commonly deployed noise model appearing in quantum computing literature.
It describes the loss of phase coherence~\cite{huibers1998dephasing}, or the loss of quantum information without energy loss~\cite{nielsen2002quantum}
and can be conceptualised discretely in time as a probabilistic, erroneous $\hat{Z}$ operation(s). 
We here consider the one and two qubit dephasing channels. While both can in-principle be described by Kraus maps and simulated through Alg.~\ref{alg:distrib_dens_matr_kraus_map}, their Kraus operators are sparse and suggest a superior, bespoke treatment.
In this section, we derive embarrassingly parallel distributed algorithms to simulate both the one and two qubit depashing channels upon an $N$-qubit density matrix, distributed between $2^w$ Choi-vectors of local size $\Lambda=2^{2N-w}$ in $\mathcal{O}(\Lambda)$ bops, flops and memory writes.

\subsubsection{One qubit}

A dephasing channel describing a single erroneous $\hat{Z}$ occurring on qubit $t$ of an $N$--qubit density matrix $\bm\rho$ with probability $p$ produces state
\begin{align}
    \mathcal{E}(\bm\rho) &= (1 - p) \bm\rho + p \, \hat{Z}_t \, \bm\rho \, \hat{Z}_t.
\end{align}
We \textit{could} express this as a Kraus map with operators $\hat{K}_t^{(1)} = \sqrt{p}\;\hat{Z}$ and $\hat{K}_t^{(2)} = \sqrt{1-p}\;\hat{\mathbb{1}}$, and simulate it via Alg.~\ref{alg:distrib_dens_matr_kraus_map} in as many as $\sim 4 \Lambda$ memory writes and $2$ rounds of communication. Or, we could unitarily apply $\hat{Z}_t$ to a clone of the state via Alg.~\ref{alg:distrib_dens_matr_specific_unitaries} then linearly combine the states, doubling our memory costs and caching penalties.  
A superior strategy is possible.

The action of $\hat{Z}_t$ on a basis projector $\ket{k}\bra{l}$ is
\begin{align}
\hat{Z}_t \ket{k}\bra{l} \hat{Z}_t
&= (-1)^{k_{[t]}} (-1)^{l_{[t]}} \ket{k}\bra{l},
\label{eq:z_on_dens_basis}
\end{align}
inducing sign $s_{kl} = \pm 1$ determined by bits $k_{[t]}$, $l_{[t]}$. The channel ergo maps a general state $\bm\rho = \sum\limits_{kl}\alpha_{kl}\ket{k}\bra{l}$ to
\begin{align}
    \mathcal{E}(\bm\rho) &= \sum\limits_{k}^{2^N}\sum\limits_{l}^{2^N} \alpha_{kl} (1-p+p\,s_{kl}) \ket{k}\bra{l},
\end{align}
thereby modifying the amplitudes as
\begin{align}
\alpha_{kl} \rightarrow \begin{cases}
\alpha_{kl}, & k_{[t]} = l_{[t]}, \\
(1 - 2p) \alpha_{kl}, & k_{[t]} \ne l_{[t]}.
\end{cases}
\end{align} 
Our density matrix $\bm\rho\numsub{N}$ encoded as a Choi-vector $\choivec{\bm\rho}\numsub{2N} = \sum_{kl}\alpha_{kl}\ket{l}\numsub{N}\ket{k}\numsub{N} \equiv \sum_i \beta_i \choivec{i}\numsub{2N}$ sees
\begin{align}
\beta_i \rightarrow 
\begin{cases}
\beta_i, & i_{[t]} = i_{[t+N]}, \\
(1 - 2p) \beta_i, & i_{[t]} \ne i_{[t+N]}.
\end{cases}
\end{align}
This reveals that dephasing upon a Choi-vector resembles a diagonal operator upon a statevector, and that distributed simulation will be embarrassingly parallel. Our next task is to efficiently identify which local amplitudes satisfy $i_{[t]} \ne i_{[t+N]}$ once distributed. We refer to $i_{[t]}$ and $i_{[t+N]}$ as the ``principal bits".

Recall that the $2^{2N}$ amplitudes are uniformly distributed between arrays $\vec{\rho}$ among $2^w$ nodes (where $N \ge w$)
such that the $j$-th local amplitude $\vec{\rho}[j]\equiv \beta_i$ of node $r$ corresponds to global basis state
\begin{align}
\choivec{i}\numsub{2N}
&\equiv
\choivec{r}\numsub{w}\choivec{j}\numsub{2N-w}
\\
&\equiv \;
{
    \color{lightgray}
\underbrace{
    \mystrut{7pt}
    \color{black}
\hfsetbordercolor{subketborder2}
\tikzmarkin[subketstyle2]{d}
\choivec{r}\numsub{w}
\tikzmarkend{d} \;\;\;
\tikzmarkin[subketstyle1]{b}
    \choivec{  \dots 
    }\numsub{N-w}
\tikzmarkend{b}
}_{\color{black}
t+N
}
\;\;\,
\underbrace{
    \mystrut{7pt}
    \color{black}
\tikzmarkin[subketstyle2]{c}
    \choivec{\dots}\numsub{w}
\tikzmarkend{c} \;\;\;
\tikzmarkin[subketstyle1]{a}
\choivec{
    \dots 
}\numsub{N-w}
\tikzmarkend{a}}_{\color{black} 
t
 }
}
\tag*{}
\end{align}
We have highlighted the distinct subdomains of $t$ (and corresponding $t+N$) as pink and blue.
This form suggests there are \textit{two} distinct scenarios in the determination of which local amplitudes satisfy $i_{[t]} \ne i_{[t+N]}$, informed by qubit $t$:

\begin{enumerate}
    \item \textbf{When $t < N-w$} (\textcolor{subketborder1}{pink} above), the principal bits of $i$ are determined entirely by local index $j$,
    \begin{align}
        \text{i.e.}\;\;\;
        i_{[t]} &= j_{[t]} \;\;\;\text{and}\;\;\;
        i_{[t+N]} = j_{[t+N]}.
    \end{align}
    We can directly enumerate indices $j$ satisfying $j_{[t]} \ne j_{[t+N]}$, modifying $\vec{\rho}[j]$.

    \item \textbf{When $t \ge N-w$} (\textcolor{subketborder2}{blue} above), the rank $r$ determines the bit $i_{[t+N]}$, which is fixed for across all local indices $j$ in the node. We enumerate only $j$ satisfying
    \begin{align}
        j_{[t]} \ne r_{[t-(N-w)]}.
    \end{align}

\end{enumerate}

This informs a simple, embarrassingly parallel distributed algorithm to simulate the one-qubit dephasing gate, which we formalise in Alg.~\ref{alg:distrib_dens_matr_one_qb_dephasing}. It uses bit algebra and avoids branching in a similar logic to the SWAP gate of Sec.~\ref{sec:swap_gate}.

\begin{algorithm}[tb]
\DontPrintSemicolon
\caption[blah blah]{
    \AlgTagDistributed \AlgTagDensityMatrix  \\
    One-qubit dephasing of qubit $t$ with probability $p$ of an $N$-qubit density matrix with Choi-vector distributed between arrays $\vec{\rho}$ of length $\Lambda$ among $2^w$ nodes.
    \protect{\begin{center} 
        $
\Qcircuit @C=1em @R=.7em {
& \gate{\mathcal{E}_\phi} & \qw
}
        $
    \end{center}}
    \begin{center}
        \AlgTagBops{$\mathcal{O}(\Lambda)$}%
        \AlgTagFlops{$\Lambda/2$}%
        \AlgTagNumSerialRounds{$0$}\\
        \AlgTagAmpsTransferred{$0$}%
        \AlgTagMemOverhead{$\mathcal{O}(1)$}%
        \AlgTagMemWrites{$\Lambda/2$}
    \end{center}
}
\label{alg:distrib_dens_matr_one_qb_dephasing}

\textbf{distrib\_oneQubitDephasing}($\vec{\rho}$, $t$, $p$):

\Indp 

    $\Lambda$ = $\dim(\vec{\rho})$

    $N$ = \textbf{getNumQubits}($\vec{\rho}$)
    \tcp*{Alg.~\ref{alg:distrib_dens_convenience_funcs}}

    $w$ = $\log_2(\textbf{getWorldSize()})$
    \tcp*{Alg.~\ref{alg:mpi_defs}}

    $c$ = $1 - 2\,p$


    \textbf{if} $t \ge N-w$:

    \Indp 

        $r$ = \textbf{getRank}()
        \tcp*{Alg.~\ref{alg:mpi_defs}}

        $b$ = \textbf{getBit}($r$, $t-(N-w)$) 
        \tcp*{Alg.~\ref{alg:bit_twiddles}}


        \AlgThreadComment{multithread}

        \textbf{for} $k$ \textbf{in} \textbf{range}($0$, $\Lambda/2$):

        \Indp  

            $j$ = \textbf{insertBit}($k$, $t$, $\LogicalNot\,b$)
            \tcp*{Alg.~\ref{alg:bit_twiddles}}

            $\vec{\rho}[j]$ \TimesEq{} $c$

        \Indm

    \Indm 


    \textbf{else}:

    \Indp 

        \AlgThreadComment{multithread}

        \textbf{for} $k$ \textbf{in} \textbf{range}($0$, $\Lambda/4$):

        \Indp 

            $j$ = \textbf{insertBit}($k$, $t$, $1$)
            \tcp*{Alg.~\ref{alg:bit_twiddles}}

            $j$ = \textbf{insertBit}($j$, $t+N$, $0$)

            $\vec{\rho}[j]$ \TimesEq{} $c$


            $j$ = \textbf{insertBit}($k$, $t$, $0$)
            \tcp*{Alg.~\ref{alg:bit_twiddles}}

            $j$ = \textbf{insertBit}($j$, $t+N$, $1$)

            $\vec{\rho}[j]$ \TimesEq{} $c$

        \Indm 

    \Indm 

\Indm 

 \end{algorithm}

\subsubsection{Two-qubit}
\label{sec:two_qubit_dephasing}

We can derive a similar method for a \textit{two}-qubit dephasing channel, inducing $\hat{Z}$ on either or both of qubits $t_1$ and $t_2$ with probability $p$. Assume $t_2>t_1$. The channel upon a general state $\bm\rho = \sum\limits_{kl}\alpha_{kl}\ket{k}\bra{l}$ produces
\begin{align}
    \varepsilon(\bm\rho) 
    &= (1 - p)\bm\rho + \frac{p}{3} \left( 
    \begin{aligned}
    &\hat{Z}_{t_1} \bm\rho \hat{Z}_{t_1} + 
     \hat{Z}_{t_2} \bm\rho \hat{Z}_{t_2}  \\
    & + \hat{Z}_{t_1} \hat{Z}_{t_2} \bm\rho \hat{Z}_{t_1}  \hat{Z}_{t_2}
      \end{aligned}
    \right)
    \\
    &=
    \sum\limits_{kl} \alpha_{kl} \ket{k}\bra{l}
    \left( 
    1 - p + \frac{p}{3} \left( 
    \begin{aligned}
    & s_{kl}^{(t_1)} + s_{kl}^{(t_2)} \\
    & +  s_{kl}^{(t_1)}\, s_{kl}^{(t_2)} 
    \end{aligned}
    \right)
    \right)
\end{align}
where $s_{kl}^{(t)} = (-1)^{k_{[t]}+l_{[t]}} = \pm 1$. This suggests
\begin{align}
\alpha_{kl}  \rightarrow 
\begin{cases}
    \alpha_{kl}, & k_{[t_1]}=l_{[t_1]} \wedge k_{[t_2]}=l_{[t_2]},
    \\
    \left(1 - \frac{4\,p}{3}\right) \alpha_{kl},
    & 
    \text{otherwise},
    \label{eq:dephasing_2qb_update_rule}
\end{cases}
\end{align}
and that an amplitude $\beta_i$ of the Choi-vector $\choivec{\bm\rho}\numsub{2N} = \sum_i \beta_i \choivec{i}\numsub{N}$ is multiplied by $(1-4\,p/3)$ if either $i_{[t_1]} \ne i_{[t_1+N]}$ or $i_{[t_2]} \ne i_{[t_2+N]}$ (or both).

Like the one-qubit dephasing channel, we've shown two-qubit dephasing is also diagonal and embarrassingly parallel. We distribute the $2^{2N}$ amplitudes uniformly between $2^w$ nodes, such that
the $j$-th local amplitude $\vec{\rho}[j]\equiv \beta_i$ of node $r$ again corresponds to global basis state $\choivec{i}=\choivec{r}\choivec{j}$, where
\begin{align}
\choivec{i}\numsub{2N}
&\equiv \;
{
    \color{lightgray}
\underbrace{
    \mystrut{7pt}
    \color{black}
\hfsetbordercolor{subketborder2}
\tikzmarkin[subketstyle2]{hello}
\choivec{r}\numsub{w}
\tikzmarkend{hello} \;\;\;
\tikzmarkin[subketstyle1]{goodbye}
    \choivec{  \dots 
    }\numsub{N-w}
\tikzmarkend{goodbye}
}_{\color{black}
t_2+N, \;\;\; t_1+N
}
\;\;\,
\underbrace{
    \mystrut{7pt}
    \color{black}
\tikzmarkin[subketstyle2]{ihateforcedvarnames}
    \choivec{\dots}\numsub{w}
\tikzmarkend{ihateforcedvarnames} \;\;\;
\tikzmarkin[subketstyle1]{theymakemecry}
\choivec{
    \dots 
}\numsub{N-w}
\tikzmarkend{theymakemecry}}_{\color{black} 
t_2, \;\;\; t_1
 }
}
\tag*{}
\end{align}
There are \textit{three} distinct ways that $t_1$ and $t_2$ can be found among these sub-registers.

\begin{enumerate}
\itemsep-.4em
\item 
\textbf{When} $t_2 < N-w$ (both 
 qubits \textcolor{subketborder1}{pink}), the principal bits $i_{[t_1]}$, $i_{[t_1+N]}$, $i_{[t_2]}$, $i_{[t_2+N]}$ are all determined by local index $j$.

\item 
\textbf{When} $t_1 \ge N-w$ (both qubits \textcolor{subketborder2}{blue}), bits $i_{[t_1]+N}$ and $i_{[t_2]+N}$ are fixed per-node and determined by the rank $r$, an the remaining principal bits by $j$.

\item \textbf{When} $t_1 < N-w$ (in \textcolor{subketborder1}{pink}) \textbf{and} $t_2 \ge N-w$ (in \textcolor{subketborder2}{blue}), then bit $i_{[t_2]+N}$ is determined by the rank, and all other principal bits by $j$.
\end{enumerate}

With bit interleaving, we \textit{could} devise non-branching loops which directly enumerate only local indices $j$ whose global indices $i$ satisfy $i_{[t_1]} \ne i_{[t_1+N]}$ or $i_{[t_2]} \ne i_{[t_2+N]}$.
 However, Eq.~\ref{eq:dephasing_2qb_update_rule} reveals $75\%$ of all amplitudes are to be modified. It is ergo worthwhile to enumerate \textit{all} indices and modify \textit{every} element, with a quarter multiplied by unity, accepting a $25\%$ increase in flops and memory writes in exchange for significantly simplified code. 
 We present Alg.~\ref{alg:distrib_dens_matr_two_qb_dephasing}.

\begin{algorithm}[h!]
\DontPrintSemicolon
\caption[blah blah]{
    \AlgTagDistributed \AlgTagDensityMatrix  \\
    Two-qubit dephasing of qubits $t_1$ and $t_2>t_1$ with probability $p$ of an $N$-qubit density matrix with Choi-vector distributed between arrays $\vec{\rho}$ of length $\Lambda$ among $2^w$ nodes.
    \protect{\begin{center} 
        $
\Qcircuit @C=.25em @R=.15em {
& \multigate{1}{\mathcal{E}_\phi} & \qw \\
& \ghost{\mathcal{E}_\phi} & \qw
}
        $
    \end{center}}
    \begin{center}
        \AlgTagBops{$\mathcal{O}(\Lambda)$}%
        \AlgTagFlops{$\Lambda$}%
        \AlgTagNumSerialRounds{$0$}\\
        \AlgTagAmpsTransferred{$0$}%
        \AlgTagMemOverhead{$\mathcal{O}(1)$}%
        \AlgTagMemWrites{$\Lambda$}
    \end{center}
}
\label{alg:distrib_dens_matr_two_qb_dephasing}

\textbf{distrib\_twoQubitDephasing}($\vec{\rho}$, $t_1$, $t_2$, $p$):

\Indp 

    $\Lambda$ = $\dim(\vec{\rho})$



    $r'$ = \textbf{getRank}() $\BitShiftLeft \; \log_2(\Lambda)$
    \tcp*{Alg.~\ref{alg:mpi_defs}}

    $c$ = $1-4\,p/3$


    \AlgThreadComment{multithread}

    \textbf{for} $j$ \textbf{in} \textbf{range}($0$, $\Lambda$):

    \Indp 

        $i$ = $r' \BitOr j$
        \tcp*{$\choivec{i}\numsub{2N} = \choivec{r}\numsub{w} \choivec{j}\numsub{\lambda}$}

        $b_1$ = \textbf{getBit}($i$, $t_1$)
        \tcp*{Alg.~\ref{alg:bit_twiddles}}

        $b_1'$ = \textbf{getBit}($i$, $t_1+N$)

        $b_2$ = \textbf{getBit}($i$, $t_2$)
        
        $b_2'$ = \textbf{getBit}($i$, $t_2+N$)


        $b$ = $(b_1 \BitXor b_1') \BitOr (b_2 \BitXor b_2')$
        \tcp*{$\in \{0,1\}$}

        $f$ = $b\,(c-1) + 1$
        \tcp*{$\in \{1,c\}$}

        $\vec{\rho}[j]$ \TimesEq{} $f$

    \Indm 

\Indm 

 \end{algorithm}

\clearpage

\subsection{Depolarising channel} %
\label{sec:depolarising} %
\noindent %
\begin{equation}
\Qcircuit @C=1em @R=.7em {
& \gate{\mathcal{E}_\Delta} & \multigate{1}{\mathcal{E}_\Delta} & \qw \\
& \qw & \ghost{\mathcal{E}_\Delta} & \qw \\
& \qw & \qw  & \qw
}
\end{equation}

The depolarising channel, 
also known as the uniform Pauli channel, 
transforms qubits towards the maximally mixed state, describing incoherent noise resulting from the erroneous application of \textit{any} Pauli qubit.
It is a ubiquitous noise model deployed in quantum error correction~\cite{RevModPhys.87.307}, is often a suitable description of the average noise in deep, generic circuits of many qubits~\cite{urbanek2021mitigating}, and is the effective noise produced by randomised compiling (also known as twirling) of circuits suffering coherent noise
~\cite{cai2020mitigating,wallman2016noise}.
Like the dephasing channel, we \textit{could} describe depolarising as a Kraus map and simulate it via Alg.~\ref{alg:distrib_dens_matr_kraus_map}, though this would not leverage the sparsity of the resulting superoperator.
In this section, we instead derive superior distributed algorithms to simulate both the one and two-qubit depolarising channels upon a density matrix distributed between $\Lambda$-length arrays, in $\mathcal{O}(\Lambda)$ operations and at most \textit{two} rounds of communication.
For simplicity, we study \textit{uniform} depolarising, though an extension to a general Pauli channel is straightforward.

\subsubsection{One-qubit}

\afterpage{%
\begin{algorithm}
\DontPrintSemicolon
\caption[blah blah]{
    \AlgTagDistributed \AlgTagDensityMatrix  \\
    One-qubit depolarising of qubit $t$ with probability $p$ of an $N$-qubit density matrix with Choi-vector distributed between arrays $\vec{\rho}$ of length $\Lambda$ among $2^w$ nodes.
    \protect{\begin{center} 
        $
\Qcircuit @C=1em @R=.7em {
& \gate{\mathcal{E}_\Delta} & \qw
}
        $
    \end{center}}
    \begin{center}
        \AlgTagBops{$\mathcal{O}(\Lambda)$}%
        \AlgTagFlops{$\mathcal{O}(\Lambda)$}%
        \AlgTagNumSerialRounds{$0$ or $1$}\\
        \AlgTagAmpsTransferred{$2^{2N}/2$}%
        \AlgTagMemOverhead{$\mathcal{O}(1)$}%
        \AlgTagMemWrites{$\mathcal{O}(\Lambda)$}
    \end{center}
}
\label{alg:distrib_dens_matr_one_qb_depolarising}

\textbf{distrib\_oneQubitDepolarising}($\vec{\rho}$, $\vec{\varphi}$, $t$, $p$):

\Indp 

    $\Lambda$ = $\dim(\vec{\rho})$

    $N$ = \textbf{getNumQubits}($\vec{\rho}$)
    \tcp*{Alg.~\ref{alg:distrib_dens_convenience_funcs}}

    $w$ = $\log_2(\textbf{getWorldSize()})$
    \tcp*{Alg.~\ref{alg:mpi_defs}}

    $c_1$ = $2\,p/3$

    $c_2$ = $1 - 2\,p/3$

    $c_3$ = $1-4\,p/3$

    \codegap 



    \tcp{embarrassingly parallel}

    \textbf{if} $t < N-w$:

    \Indp

        \tcp{loop $\choivec{k}\numsub{\lambda-2}$, interleave $\choivec{j}\numsub{\lambda}$}
        
        \AlgThreadComment{multithread}

        \textbf{for} $k$ \textbf{in} \textbf{range}($0$, $\Lambda/4$):

        \Indp 

            $j_{00}$ = \textbf{insertBits}($k$, $\{t,\,t+N\},\,0$)

            $j_{01}$ = \textbf{flipBit}($j_{00}$, $t$)
            \tcp*{Alg.~\ref{alg:bit_twiddles}}

            $j_{10}$ = \textbf{flipBit}($j_{00}$, $t+N$)
            \tcp*{Alg.~\ref{alg:bit_twiddles}}
            
            $j_{11}$ = \textbf{flipBit}($j_{01}$, $t+N$)
            \tcp*{Alg.~\ref{alg:bit_twiddles}}

            \codegap  

            $\gamma$ = $\vec{\rho}[j_{00}]$

            $\vec{\rho}[j_{00}]$ = $c_2\;\gamma + c_1 \; \vec{\rho}[j_{11}]$

            $\vec{\rho}[j_{01}]$ \TimesEq{} $c_3$

            $\vec{\rho}[j_{10}]$ \TimesEq{} $c_3$

            $\vec{\rho}[j_{11}]$ = $c_1 \; \gamma + c_2 \; \vec{\rho}[j_{11}]$

        \Indm 

    \Indm

    \codegap

    \tcp{exchange required}

    \textbf{else}:

    \Indp 

        \tcp{$b=i_{[t+N]}$ for all local $i$}

        $r$ = \textbf{getRank}()
        \tcp*{Alg.~\ref{alg:mpi_defs}}

        $b$ = \textbf{getBit}($r$, $t-(N+w)$)
        \tcp*{Alg.~\ref{alg:bit_twiddles}}

        \codegap 

        \tcp{pack half of local amps into buffer}

        \AlgThreadComment{multithread}

        \textbf{for} $k$ \textbf{in} \textbf{range}($0$, $\Lambda/2$):

        \Indp 

            $j$ = \textbf{insertBit}($k$, $t$, $b$)
            \tcp*{Alg.~\ref{alg:bit_twiddles}}
            
            $\vec{\varphi}[k]$ = $\vec{\rho}[j]$

        \Indm

        \codegap 

        \tcp{swap half buffer with pair node}

        $r'$ = \textbf{flipBit}($r$, $t-(N+w)$)
        \tcp*{Alg.~\ref{alg:bit_twiddles}}

        \textbf{exchangeArrays}($\vec{\varphi}$, $0$ $\vec{\varphi}$, $\Lambda/2$, $\Lambda/2$, $r'$)

        \codegap 

        \tcp{update $\beta_i$ where $i_{[t]} = j_{[t]}\ne i_{[t+N]} = b$}

        \AlgThreadComment{multithread}

        \textbf{for} $k$ \textbf{in} \textbf{range}($0$, $\Lambda/2$):

        \Indp 

            $j$ = \textbf{insertBit}($k$, $t$, $\LogicalNot \, b$)
            \tcp*{Alg.~\ref{alg:bit_twiddles}}

            $\vec{\rho}[j]$ \TimesEq{} $c_3$

        \Indm

        \codegap 

        \tcp{update $\beta_i$ where $i_{[t]} = j_{[t]} = i_{[t+N]} = b$}

        \AlgThreadComment{multithread}

        \textbf{for} $k$ \textbf{in} \textbf{range}($0$, $\Lambda/2$):

        \Indp 

            $j$ = \textbf{insertBit}($k$, $t$, $b$)
            \tcp*{Alg.~\ref{alg:bit_twiddles}}

            $\vec{\rho}[j]$ = $c_2\;\vec{\rho}[j] + c_1 \; \vec{\varphi}[k + \Lambda/2]$

        \Indm

    \Indm 
\Indm 

 \end{algorithm} %
} 

The one-qubit uniformly depolarising channel upon qubit $t$ of an $N$-qubit density matrix $\bm{\rho}$ produces state
\begin{align}
    \varepsilon(\bm\rho) &=
    (1 -p)\bm\rho + \frac{p}{3} \left(
        \hat{X}_t \bm\rho \hat{X}_t + 
        \hat{Y}_t \bm\rho \hat{Y}_t + 
        \hat{Z}_t \bm\rho \hat{Z}_t 
    \right),
\end{align} 
where $p$ is the probability of \textit{any} error occurring. Each Pauli operator upon a basis state $\ket{k}\bra{l}$ produces
\begin{align}
\hat{X}_t  \ket{k}\bra{l} \hat{X}_t 
    &=
\ket{k_{\neg t}}\bra{l_{\neg t}},
\label{eq:paulis_on_dens_matr_basis_x}
\\
\hat{Y}_t  \ket{k}\bra{l} \hat{Y}_t 
    &=
    (-1)^{k_{[t]}+l_{[t]}}
\ket{k_{\neg t}}\bra{l_{\neg t}},
\\
\hat{Z}_t \ket{k}\bra{l} \hat{Z}_t
&= (-1)^{k_{[t]}+l_{[t]}} \ket{k}\bra{l},
\label{eq:paulis_on_dens_matr_basis_z}
\end{align}
and ergo the depolarising channel maps a general state $\bm\rho = \sum_{kl}\alpha_{kl}\ket{k}\bra{l}$ to
\begin{align}
\mathcal{E}(\bm\rho) = 
\sum\limits_{kl} & \left( 
1 - p + \frac{p}{3}(-1)^{k_{[t]+l_{[t]}}} \right) \alpha_{kl} \ket{k}\bra{l}  \tag*{}
\\  + & 
\frac{p}{3} \left( 1 + (-1)^{k_{[t]}+l_{[t]}}\right)\alpha_{kl} \ket{k_{\neg t}}\bra{l_{\neg t}}.
\end{align} 
This prescribes a change of amplitudes 
\begin{align}
\alpha_{kl} \rightarrow \begin{cases}
\left(1 - \frac{2p}{3}\right)\alpha_{kl} + \frac{2p}{3} \alpha_{k_{\neg t}\, l_{\neg t}},
&
    k_{[t]} = l_{[t]},
    \\
\left(1 - \frac{4p}{3}\right) \alpha_{kl},
&
    k_{[t]} \ne l_{[t]},
\end{cases}
\end{align}
or the equivalent change to the equivalent Choi-vector $\choivec{\bm\rho}\numsub{2N} =  \sum_i \beta_i \choivec{i}\numsub{2N}$ of
\begin{align}
\beta_i \rightarrow 
\begin{cases}
\left(1 - \frac{2p}{3}\right)\beta_i + \frac{2p}{3} \beta_{i_{\neg \{t,t+N\} }},
&
i_{[t]} = i_{[t+N]},
\\
\left(1 - \frac{4p}{3}\right) \beta_i,
&
i_{[t]} \ne i_{[t+N]}.
\end{cases}
\label{eq:depolarising_1qb_amp_update}
\end{align}
Unlike the dephasing channel, we see already that the depolarising channel upon the Choi-vector is \textit{not} diagonal; it will linearly combine amplitudes and require communication.
Recall that the $2^{2N}$ amplitudes of $\choivec{\bm\rho}$ are uniformly distributed between arrays $\vec{\rho}$ among $2^w$ nodes (where $N \ge w$),
 such that the $j$-th local amplitude $\vec{\rho}[j]\equiv \beta_i$ of node $r$ corresponds to global basis state 
\begin{align}
\choivec{i}\numsub{2N}
&\equiv
\choivec{r}\numsub{w}\choivec{j}\numsub{2N-w}
\\
&\equiv \;
{
    \color{lightgray}
\underbrace{
    \mystrut{7pt}
    \color{black}
\hfsetbordercolor{subketborder2}
\tikzmarkin[subketstyle2]{flooploop}
\choivec{r}\numsub{w}
\tikzmarkend{flooploop} \;\;\;
\tikzmarkin[subketstyle1]{glompenshtein}
    \choivec{  \dots 
    }\numsub{N-w}
\tikzmarkend{glompenshtein}
}_{\color{black}
t+N
}
\;\;\,
\underbrace{
    \mystrut{7pt}
    \color{black}
\tikzmarkin[subketstyle2]{kablammybrothers}
    \choivec{\dots}\numsub{w}
\tikzmarkend{kablammybrothers} \;\;\;
\tikzmarkin[subketstyle1]{groopnscoop}
\choivec{
    \dots 
}\numsub{N-w}
\tikzmarkend{groopnscoop}}_{\color{black} 
t
 }
}
\tag*{}
\end{align}
Two communication scenarios emerge, informed by qubit $t$.
\begin{enumerate}
\item \textbf{When} $t < N-w$, the principal bits $i_{[t]}$ and $i_{[t+N]}$ are determined entirely by a local index $j$, and the pair amplitude $\beta_{i \neg \{t,t+N\}}$ resides within the same node as $\beta_i$. Simulation is embarrassingly parallel.
\item \textbf{When} $t \ge N-w$, bit $i_{[t]}$ is determined by local index $j$, but $i_{[t+N]}$ is fixed by the node rank $r$. Precisely,
$i_{[t+N]} = r_{[t - (N - w)]}$.
The pair amplitude $\beta_{i \neg \{t,t+N\}}$ resides within pair node
$
r' = r_{\neg (t-(N-w))},
$
requiring communication. Since only local amplitudes with indices $j$ 
 satisfying $j_{[t]}=i_{[t+N]}$ need to be exchanged, we first pack only this half into the node's buffer $\vec{\varphi}$ before exchange. This is communication paradigm \textbf{b)} of Fig.~\ref{fig:distrib_mem_exchange_patterns}. 
 
\end{enumerate}

Translating these schemes (and in effect, implementing Eq.~\ref{eq:depolarising_1qb_amp_update} upon distributed $\{\beta_i\}$) into efficient, non-branching, cache-friendly code is non-trivial.
We present such an implementation in Alg.~\ref{alg:distrib_dens_matr_one_qb_depolarising}. Interestingly, its performance is similar to a one-target gate (Alg.~\ref{alg:distrib_1qb_gate}) upon a statevector of $2N$ qubits, but exchanges only \textit{half} of all amplitudes when communication is necessary.

\subsubsection{Two-qubit}

We now consider the \textit{two}-qubit uniformly depolarising channel, inducing any Pauli error upon qubits $t_1$ and $t_2$ with probability $p$.
\begin{align}
    \mathcal{E}(\bm\rho) &=
    \left( 1 - \frac{16\,p}{15} \right) \bm\rho 
    + 
    \frac{p}{15} 
    \sum\limits_{\substack{\hat\sigma,\hat\sigma' \in \\ \{\mathbb{1}, X, Y, Z\}}}
    {\hat\sigma'}_{t_2} \hat\sigma_{t_1} \; \bm\rho \;
    {\hat\sigma'}_{t_2} \hat\sigma_{t_1} \, .
\end{align}
As per Eq.~\ref{eq:paulis_on_dens_matr_basis_x}~-~\ref{eq:paulis_on_dens_matr_basis_z}, this maps a general state $\bm\rho=\sum_{kl}\alpha_{kl}\ket{k}\bra{l}$ to
\begin{align}
 \mathcal{E}(\bm\rho) 
 = &
 \sum\limits_{k}^{2^N}\sum\limits_{l}^{2^N}
\left(1 - \frac{16\,p}{15} + \gamma_{kl} \right) \alpha_{kl} \ket{k}\bra{l} 
\\
& + \gamma_{kl} \, \alpha_{kl} 
\left( 
\begin{aligned} 
&\ket{k_{\neg t_1}}\bra{l_{\neg t_1}}
+
\ket{k_{\neg t_2}}\bra{l_{\neg t_2}} \\
+  &
\ket{k_{\neg \{t_1,t_2\}}}\bra{l_{\neg \{t_1,t_2\}}}
\end{aligned}
\right),
\tag*{}
\end{align} 
having defined (where $s_{kl}^{(t)} = (-1)^{k_{[t]} + l_{[t]}} = \pm 1$)
\begin{align}
\gamma_{kl} &= 
\frac{1}{15}p \left( 
    1 + s_{kl}^{(t_1)} + s_{kl}^{(t_2)}
    + s_{kl}^{(t_1)} s_{kl}^{(t_2)}
\right) 
\\
&= 
\begin{cases}
    \frac{4\,p}{15}, & 
        k_{[t_1]}=l_{[t_1]} \wedge k_{[t_2]}=l_{[t_2]}
    \\
    0, &
        \text{otherwise}.
\end{cases}
\end{align}
The channel therefore modifies amplitudes under
\begin{align}
\alpha_{kl} \rightarrow 
\begin{cases}
\left(1 - \frac{4\,p}{5}\right)\alpha_{kl}
+ \frac{4\,p}{15}
\left( 
\begin{gathered}
\alpha_{k_{\neg t_1}, l_{\neg t_1}} 
    \;+
    \\
\alpha_{k_{\neg t_2}, l_{\neg t_2}} 
    \;+
    \\
\alpha_{k_{\neg \{t_1,t_2\}}, l_{\neg \{t_1,t_2\}}} 
\end{gathered}
\right),
\\
\hphantom{\left(1 - \frac{4\,p}{5}\right)\alpha_{kl}
+ \frac{4\,p}{15}}
        k_{[t_1]}=l_{[t_1]} \wedge k_{[t_2]}=l_{[t_2]},
    \\
\left(1 - \frac{16\,p}{15}\right) \alpha_{kl},
\hphantom{\;\frac{4\,p}{15}}
         \text{otherwise}.
\end{cases}
\label{eq:2qb_depol_matrix_amp_change}
\end{align} 
This suggests a \textit{quarter} of all amplitudes are modified to become a combination of four previous amplitudes, and the remaining amplitudes are merely scaled. 

Assume $\bm\rho$ describes $N$ qubits.
Amplitudes of the equivalent Choi-vector $\choivec{\bm\rho}\numsub{2N} = \sum_i\beta_i \choivec{i}\numsub{2N}$
are modified as

\begin{align}
\beta_i \rightarrow \begin{cases}
\left(1 - \frac{4\,p}{5}\right)\beta_i
+ \frac{4\,p}{15}
\left( 
\begin{gathered}
\beta_{i_{\neg \{t_1, t_1+N\}}}
    \;+
    \\
\beta_{i_{\neg \{t_2, t_2+N\}}}
    \;+
    \\
\beta_{i_{\neg \{t_1, t_1+N, t_2,t_2+N\}}}
\end{gathered}
\right),
\\
\hspace{2.3cm} 
    i_{[t_1]}=i_{[t_1+N]} \wedge
    i_{[t_2]}=i_{[t_2+N]},
\\
\left(1 - \frac{16\,p}{15}\right) \beta_i,
\hspace{.35cm} \text{otherwise}.
\end{cases}
\label{eq:2qb_depolarising_choi_amp_update}
\end{align}

It is worth clarifying which indices $i$ satisfy $i_{[t_1]}=i_{[t_1+N]} \wedge i_{[t_2]}=i_{[t_2+N]}$, since the distributed simulation of Eq.~\ref{eq:2qb_depolarising_choi_amp_update} will be markedly more complicated than that required by the dephasing channel.
Consider a basis state of $4$ fewer qubits:
\begin{align}
\choivec{h}\numsub{2N-4} 
\equiv 
&
\choivec{e}\numsub{N-t_2-1}
\choivec{d}\numsub{t_2-t_1-1}
\choivec{c}\numsub{N-(t_2-t_1)-1} \; \otimes
\tag*{}
\\
&
\choivec{b}\numsub{t_2-t_1-1}
\choivec{a}\numsub{t_1},
\end{align}
and that produced by interleaving $4$ zero bits into $h$ at indices $t_1$, $t_2$, $t_1+N$ and $t_2+N$:
\begin{align}
\choivec{i_{0000}}\numsub{2N}
=
{
    \color{lightgray}
    \choivec{e}
}
\choivec{0}\numsub{1}
{
    \color{lightgray}
    \choivec{d}
}
\choivec{0}\numsub{1}
{
    \color{lightgray}
    \choivec{c}
}
\choivec{0}\numsub{1}
{
    \color{lightgray}
    \choivec{b}
}
\choivec{0}\numsub{1}
{
    \color{lightgray}
    \choivec{a}
}
\end{align}
Let $i_x$ be the index produced by changing the zero bits above to $x$.
Eq.~\ref{eq:2qb_depolarising_choi_amp_update} informs us that for every group of $16$ basis states sharing fixed substates $a,b,c,d,e$, the amplitudes of \textit{four} states are combined together (the \textcolor{Green}{green}-highlighted diagonals below) while the remaining amplitudes are scaled.
\begin{align*}
&{
    \color{Green}
    \choivec{i_{0000}}
}
    &
&\choivec{i_{0100}}
    &
&\choivec{i_{1000}}
    &
&\choivec{i_{1100}}
    \\
&\choivec{i_{0001}}
    &
& { \color{Green} 
    \choivec{i_{0101}} }
    &
&\choivec{i_{1001}}
    &
&\choivec{i_{1101}}
    \\
&\choivec{i_{0010}}
    &
&\choivec{i_{0110}}
    &
& {
    \color{Green}
    \choivec{i_{1010}}
    }
    &
&\choivec{i_{1110}}
    \\
&\choivec{i_{0011}}
    &
&\choivec{i_{0111}}
    &
&\choivec{i_{1011}}
    &
& {
    \color{Green}
    \choivec{i_{1111}}
}
\end{align*}

Recall that  the $2^{2N}$ amplitudes of $\choivec{\bm\rho}\numsub{2N}$ are uniformly distributed between $2^w$ nodes, such that
the $j$-th local amplitude $\vec{\rho}[j] = \beta_i$ of node $r$ again corresponds to global basis state
\begin{align}
\choivec{i}\numsub{2N}
&\equiv
\choivec{r}\numsub{w}\choivec{j}\numsub{2N-w}
\\
&\equiv \;
{
    \color{lightgray}
\underbrace{
    \mystrut{7pt}
    \color{black}
\hfsetbordercolor{subketborder2}
\tikzmarkin[subketstyle2]{grompy}
\choivec{r}\numsub{w}
\tikzmarkend{grompy} \;\;\;
\tikzmarkin[subketstyle1]{shlompy}
    \choivec{  \dots 
    }\numsub{N-w}
\tikzmarkend{shlompy}
}_{\color{black}
t_2+N, \;\;\; t_1+N
}
\;\;\,
\underbrace{
    \mystrut{7pt}
    \color{black}
\tikzmarkin[subketstyle2]{toottoot}
    \choivec{\dots}\numsub{w}
\tikzmarkend{toottoot} \;\;\;
\tikzmarkin[subketstyle1]{fuarkenblarkems}
\choivec{
    \dots 
}\numsub{N-w}
\tikzmarkend{fuarkenblarkems}}_{\color{black} 
t_2, \;\;\; t_1
 }
}
\tag*{}
\end{align}

\begin{figure*}[t]
    \centering 
    \includegraphics[width=\textwidth]{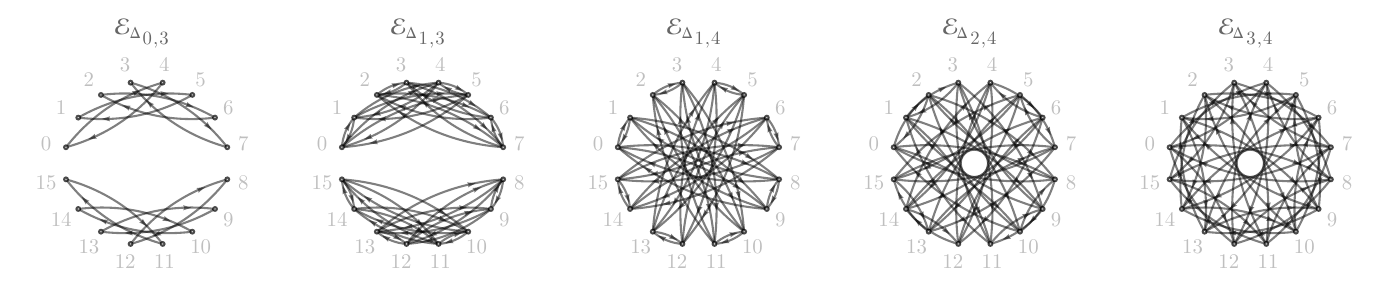}
    \caption{
        Some communication patterns of Alg.~\ref{alg:distrib_dens_matr_two_qb_depolarising}'s distributed simulation of two-qubit depolarising upon a $5$-qubit density matrix distributed between $16$ nodes. While ${\mathcal{E}_{\Delta}}_{0,3}$ is pairwise, the other channels are simulated through \textit{two} consecutive rounds of pairwise communication.
    }
    \label{fig:distrib_depol_2qb_comm}
\end{figure*}

We must consider the division of the $4$ principal bits of $i$ (informing the location of the above \textcolor{Green}{green} states) between the blue and pink regions. 
Like two-qubit dephasing of Sec.~\ref{sec:two_qubit_dephasing}, three distinct communication and simulation scenarios are evident.

\begin{enumerate}
\item
\textbf{When} $t_2 < N-w$ (both 
 qubits \textcolor{subketborder1}{pink}), all principal bits are determined by local index $j$, and all amplitudes to combine reside with the same node. Simulation is embarrassingly parallel.

\item 
\textbf{When} $t_2 \ge N-w$ (in \textcolor{subketborder2}{blue})  \textbf{and} $t_1 < N-w$ (in \textcolor{subketborder1}{pink}),
bit $i_{[t_2+N]}$ (the left-most subscripted bit in ${\color{Green} i_{0000} }$ above) is determined by the node's rank $r$ while all other principal bits by local index $j$. Ergo amplitudes of states ${
    \color{Green}
    \choivec{i_{0000}}
}$
and
${
    \color{Green}
    \choivec{i_{0101}}
}$
reside in a different node to that storing
${
    \color{Green}
    \choivec{i_{1010}}
}$
and
${
    \color{Green}
    \choivec{i_{1111}}
}$,
prescribing pairwise communication. 
Only a \textit{quarter} of a node's local amplitudes determine those in the pair node, suggesting we should pack outbound amplitudes into the buffer before communicating. However, since Eq.~\ref{eq:2qb_depolarising_choi_amp_update} does not distinguish remote amplitudes, we can cleverly pre-combine the outbound amplitudes and exchange only \textit{one eighth} of the buffer capacity. We visualise this below.
\begin{center}    \includegraphics[width=.5\columnwidth]{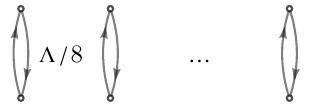}
    \end{center}

\item 
\textbf{When} $t_1 \ge N-w$ (both 
 qubits \textcolor{subketborder2}{blue}), bits $i_{[t_1+N]}$ and $i_{[t_2+N]}$ (the two left-most subscripted bits in ${\color{Green} i_{0000} }$ above) are determined entirely by the rank $r$, while $i_{[t_1]}$ and $i_{[t_2]}$ depend on local index $j$. Hence each of the three other amplitudes $\beta_{i_{\neg\{t_1,t_1+N\}}}$,
 $\beta_{i_{\neg\{t_2,t_2+N\}}}$ and $\beta_{i_{\neg\{t_1,t_1+N, t_2, t_2+N\}}}$ which inform the updated $\beta_i$ reside in distinct nodes with respective ranks
 \begin{align}
r' &= r_{\neg (t_1-(N-w))},
    \\
r'' &= r_{\neg (t_2-(N-w))},
    \\
r''' &= r_{\neg \{t_1-(N-w), \, t_2-(N-w)\}}.
 \end{align}
This naively suggests a communication pattern dividing all nodes into fully-connected $4$-node groups.
\begin{center}    \includegraphics[width=.5\columnwidth]{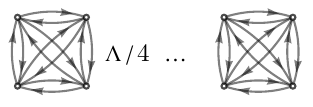}
\end{center}
 
\end{enumerate}

Scenario \textcolor{red}{\textbf{3}.} presents a new challenge; a non-pairwise communication pattern. This is not a major obstacle because each node sends only a quarter of its Choi-vector partition, so all remote amplitudes needed by a particular node ($3/4\,\Lambda$) can simultaneously fit within its communication buffer (of length $\Lambda$). Still, a direct implementation would necessitate \textit{three} rounds of communication and a total of $3/4 \times 2^{2N}$ amplitudes sent over the network. We can improve this.

We \textit{could} employ the SWAP gate of Alg.~\ref{alg:distrib_swap} to swap qubit $t_2$ to one with index $<N-w$, achieving pairwise scenario \textcolor{red}{\textbf{2}} above, much like our scheme to simulate the many-target general gate of Alg.~\ref{alg:distrib_manytarg_gate}. This would still require three rounds of communication and a total traffic of $3/4 \times 2^{2N}$.

A superior scheme is possible, prescribing only \textit{two} rounds of communication and $1/2\times 2^{2N}$ total traffic.
We substitute Eq.~\ref{eq:2qb_depolarising_choi_amp_update} for \textit{two} separate transformations, first performing
\begin{align}
\beta_i \rightarrow 
\begin{cases}
\left( 1 - \frac{4\,p}{5} \right) \beta_i 
+ \frac{4\,p}{15}\beta_{i_{\neg \{t_1,\,t_1+N\}}},
\\
\hspace{2.4cm}
i_{[t_1]}=i_{[t_1+N]} \wedge
    i_{[t_2]}=i_{[t_2+N]},
\\
\left(1 - \frac{16\,p}{15}\right) \beta_i,
\hspace{.35cm} \text{otherwise}.
\end{cases}
\end{align}
then successively performing
\begin{align}
\beta_i \rightarrow 
\begin{cases}
\beta_i 
 + 
 \frac{4\,p}{15 (1 - 4\,p/5)}\beta_{i_{\neg \{t_2,\,t_2+N\}}},
\\
\hspace{.8cm}
i_{[t_1]}=i_{[t_1+N]} \wedge
    i_{[t_2]}=i_{[t_2+N]},
\\
\beta_i,
\hspace{.35cm} \text{otherwise}.
\end{cases}
\end{align}
Together, these transformations update all amplitudes under two-qubit depolarising, while each prescribes simple pairwise communication. We visualise this below, where red edges indicate the second round of communication.
\begin{center}    \includegraphics[width=.5\columnwidth]{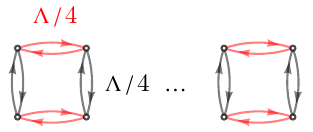}
\end{center}
Integrating these optimisations into a high-performance algorithm which avoids branching is non-trivial, and we present the result in Alg.~\ref{alg:distrib_dens_matr_two_qb_depolarising}. Some prescribed communication patterns are visualised in Fig.~\ref{fig:distrib_depol_2qb_comm}.

\begin{algorithm}[tb]
\DontPrintSemicolon
\caption[blah blah]{
    \AlgTagDistributed \AlgTagDensityMatrix  \\
    Two-qubit depolarising of qubits $t_1$ and $t_2>t_1$ with probability $p$ of an $N$-qubit density matrix with Choi-vector distributed between arrays $\vec{\rho}$ of length $\Lambda$ among $2^w$ nodes.
    \protect{\begin{center} 
        $
\Qcircuit @C=1em @R=.7em {
& \multigate{1}{\mathcal{E}_\Delta} & \qw \\
& \ghost{\mathcal{E}_\Delta} & \qw
}
        $
    \end{center}}
    \begin{center}
        \AlgTagBops{$\mathcal{O}(\Lambda)$}%
        \AlgTagFlops{$\mathcal{O}(\Lambda)$}%
        \AlgTagNumSerialRounds{$0$, $1$ or $2$}\\
        \AlgTagAmpsTransferred{$0$, $2^{2N}/8$ or $2^{2N}/2$}%
        \AlgTagMemOverhead{$\mathcal{O}(1)$}\\
        \AlgTagMemWrites{$17\Lambda/16$, $\,3\Lambda/2$ or $2\Lambda$}
    \end{center}
}
\label{alg:distrib_dens_matr_two_qb_depolarising}

\textbf{distrib\_twoQubitDepolarising}($\vec{\rho},\;\vec{\varphi},\;t_1,\;t_2,\;p$):

\Indp 

    $\Lambda$ = $\dim(\vec{\rho})$

    $N$ = \textbf{getNumQubits}($\vec{\rho}$)
    \tcp*{Alg.~\ref{alg:distrib_dens_convenience_funcs}}

    $w$ = $\log_2(\textbf{getWorldSize()})$
    \tcp*{Alg.~\ref{alg:mpi_defs}}

    $\lambda'$ = $N-w$

    $\vec{q}$ = $\{t_1, \,t_2, \, t_1+N,\, t_2+N\}$

    $c_1$ = $1 - 4\,p/5$

    $c_2$ = $4\,p/15$

    $c_3$ = $-16\,p/15$

    \codegap 

    \textbf{if} $t_2<\lambda'$:

    \Indp 

        \textbf{local\_twoQubitDepolarising}(
        \tcp{Alg.~\ref{alg:distrib_dens_matr_two_qb_depolarising_subroutines_1}}
        
        \Indp 
            $\vec{\rho},\;\vec{q},\;\Lambda,\;c_1,\;c_2,\;c_3$)

        \Indm

    \Indm

    \Indp
















    \Indm 

    \codegap  

    \textbf{else} \textbf{if} $t_2\ge \lambda'$ \textbf{and} $t_1 < \lambda'$:

    \Indp 

        \textbf{pair\_twoQubitDepolarising}(
        \tcp*{Alg.~\ref{alg:distrib_dens_matr_two_qb_depolarising_subroutines_1}}

        \Indp 
            $\vec{\rho},\;\vec{\varphi},\;\vec{q},\;\Lambda,\;\lambda',\;c_1,\;c_2,\;c_3$)

        \Indm 

    \Indm

    \codegap 

    \textbf{else}:

    \Indp 

        \textbf{quad\_twoQubitDepolarising}(
        \tcp*{Alg.~\ref{alg:distrib_dens_matr_two_qb_depolarising_subroutines_2}}

        \Indp 
            $\vec{\rho},\;\vec{\varphi},\;t_1,\,t_2,\;\Lambda,\;\lambda',\;c_1,\,c_2,\,c_3$)

        \Indm 
    
    \Indm 

\Indm 

 \end{algorithm}

\begin{algorithm}[tb]
\DontPrintSemicolon
\caption[blah blah]{
    Subroutines of Alg.~\ref{alg:distrib_dens_matr_two_qb_depolarising}
}
\label{alg:distrib_dens_matr_two_qb_depolarising_subroutines_1}

\textbf{local\_twoQubitDepolarising}($\vec{\rho},\;\vec{q},\;\Lambda,\;c_1,\;c_2,\;c_3$):

\Indp 

    \tcp{scale all amps by $1$ or $1+c_3=1-\frac{16\,p}{15}$}

    \AlgThreadComment{multithread}

    \textbf{for} $j$ \textbf{in} \textbf{range}($0$, $\Lambda$):

    \Indp 

        $f_1$ = \textbf{getBit}($j,\,
        \vec{q}[0]$) == \textbf{getBit}($j,\,\vec{q}[2]$)

        $f_2$ = \textbf{getBit}($j,\,\vec{q}[1]$) == \textbf{getBit}($j,\,\vec{q}[3]$)

        $f$ = $1 + (\LogicalNot (f_1 \BitAnd f_2)) \, c_3$

        $\vec{\rho}[j]$ \TimesEq{} $f$

    \Indm 

    \codegap

    \tcp{combine 4 amplitudes among every 16}

    \AlgThreadComment{multithread}

    \textbf{for} $h$ \textbf{in} \textbf{range}($0$, $\Lambda/16$):

    \Indp 

        $j_{0000}$ = \textbf{insertBits}($h$, $\vec{q}$, $0$)

        $j_{0101}$ = \textbf{flipBits}($j_{0000}$, $\{\vec{q}[0],\;\vec{q}[2]\}$)

        $j_{1010}$ = \textbf{flipBits}($j_{0000}$, $\{\vec{q}[1],\;\vec{q}[3]\}$)

        $j_{1111}$ = \textbf{flipBits}($j_{0101}$, $\{\vec{q}[1],\;\vec{q}[3]\}$)

        $\kappa$ = $\vec{\rho}[j_{0000}] + \vec{\rho}[j_{0101}] + \vec{\rho}[j_{1010}] + \vec{\rho}[j_{1111}]$

        \textbf{for} $j$ \textbf{in} $\{j_{0000},\; j_{0101}, \; j_{1010}, \; j_{1111} \}$:

        \Indp 

            $\vec{\rho}[j]$ = $c_1\, \vec{\rho}[j] + c_2 \, \kappa$

        \Indm 

    \Indm 

\Indm 

\codegap 

\codegap 

\textbf{pair\_twoQubitDepolarising}(%
$\vec{\rho},\;\vec{\varphi},\;\vec{q},\;\Lambda,\;\lambda',\;c_1,\,c_2,\,c_3$):

\Indp

       $r$ = \textbf{getRank}() 
        \tcp*{Alg.~\ref{alg:mpi_defs}}

        $b$ = \textbf{getBit}($r$, $\vec{q}[1]-\lambda'$)

        \codegap 

        \tcp{scale amplitudes by $1$ or $(1-\frac{16\,p}{15})$}

        \AlgThreadComment{multithread}

        \textbf{for} $j$ \textbf{in} \textbf{range}($0$, $\Lambda$):

        \Indp 

            $f_1$ = \textbf{getBit}($j,\,
            \vec{q}[0]$) == \textbf{getBit}($j,\,\vec{q}[2]$)

            $f_2$ = \textbf{getBit}($j,\,\vec{q}[1]$) == $b$

            $f$ = $1 + (\LogicalNot (f_1 \BitAnd f_2)) \, c_3$

            $\vec{\rho}[j]$ \TimesEq{} $f$

        \Indm 


        \codegap 

        \tcp{pack eighth of buffer}

        \AlgThreadComment{multithread}

        \textbf{for} $k$ \textbf{in} \textbf{range}($0$, $\Lambda/8$):

        \Indp 

            $j_{000}$ = \textbf{insertBits}($k$, $\vec{q}[:3]$, $0$)

            $j_{0b0}$ = \textbf{setBit}($j_{000}$, $\vec{q}[1]$, $b$)

            $j_{1b1}$ = \textbf{flipBits}($j_{0b0}$, $\{\vec{q}[0],\; \vec{q}[2]\}$)

            $\vec{\varphi}[k]$ = $\vec{\rho}[j_{0b0}] + \vec{\rho}[j_{1b1}]$

        \Indm 

        \codegap 

        \tcp{swap sub-buffers, receive at $\vec{\varphi}[\Lambda/8\dots]$}

        $r'$ = \textbf{flipBit}($r$, $\vec{q}[1]-\lambda'$)

        \textbf{exchangeArrays}($\vec{\varphi}$, $0$, $\vec{\varphi}$, $\Lambda/8$, $\Lambda/8$, $r'$)

        \codegap 

        \tcp{combine elements with remote}

        \AlgThreadComment{multithread}

        \textbf{for} $k$ \textbf{in} \textbf{range}($0$, $\Lambda/8$):

        \Indp 

            $j_{000}$ = \textbf{insertBits}($k$, $\vec{q}[:3]$, $0$)

            $j_{0b0}$ = \textbf{setBit}($j_{000}$, $\vec{q}[1]$, $b$)

            $j_{1b1}$ = \textbf{flipBits}($j_{0b0}$, $\{\vec{q}[0],\;\vec{q}[2]\}$)

            $\vec{\rho}[j_{0b0}]$ = $c_1 \; \vec{\rho}[j_{0b0}] + c_2\, ( \vec{\rho}[j_{1b1}] + \vec{\varphi}[k + \Lambda/8])$

            $\vec{\rho}[j_{1b1}]$ = $c_1 \; \vec{\rho}[j_{1b1}] + c_2\, ( \vec{\rho}[j_{0b0}] + \vec{\varphi}[k + \Lambda/8])$

        \Indm

    \Indm

\Indm

\end{algorithm}

\begin{algorithm}[tb]
\DontPrintSemicolon
\caption[blah blah]{
    Subroutines of Alg.~\ref{alg:distrib_dens_matr_two_qb_depolarising}
}
\label{alg:distrib_dens_matr_two_qb_depolarising_subroutines_2}

\textbf{quad\_twoQubitDepolarising}(%
$\vec{\rho},\,\vec{\varphi},\,t_1,\,t_2,\,\Lambda,\,\lambda',\,c_1,\,c_2,\,c_3$):

\Indp 

    $r$ = \textbf{getRank}()
    \tcp*{Alg.~\ref{alg:mpi_defs}}

    $b_1$ = \textbf{getBit}($r$, $t_1-\lambda'$)
    \tcp*{Alg.~\ref{alg:bit_twiddles}}

    $b_2$ = \textbf{getBit}($r$, $t_2-\lambda'$)

    \codegap

    \tcp{scale amplitudes by $1$ or $(1-\frac{16\,p}{15})$}

    \AlgThreadComment{multithread}

    \textbf{for} $j$ \textbf{in} \textbf{range}($0$, $\Lambda$):

    \Indp 

        $f_1$ = \textbf{getBit}($j,\,
        t_1$) == $b_1$

        $f_2$ = \textbf{getBit}($j,\,t_2$) == $b_2$

        $f$ = $1 + (\LogicalNot (f_1 \BitAnd f_2)) \, c_3$

        $\vec{\rho}[j]$ \TimesEq{} $f$

    \Indm 

    \codegap 

    \tcp{pack fourth of buffer}

    \AlgThreadComment{multithread}

    \textbf{for} $k$ \textbf{in} \textbf{range}($0$, $\Lambda/4$):

    \Indp 

        $j$ = \textbf{insertBit}($k$, $t_1$, $b_1$)

        $j$ = \textbf{insertBit}($j$, $t_2$, $b_2$)

        $\vec{\varphi}[k]$ = $\vec{\rho}[j]$

    \Indm 

    \codegap 

    \tcp{swap sub-buffer with first pair node}

    $r'$ = \textbf{flipBit}($r$, $t_1-\lambda'$)

    \textbf{exchangeArrays}($\vec{\varphi}$, $0$, $\vec{\varphi}$, $\Lambda/4$, $\Lambda/4$, $r'$)

    \codegap 

    \tcp{update amplitudes and buffer}

    \AlgThreadComment{multithread}

    \textbf{for} $k$ \textbf{in} \textbf{range}($0$, $\Lambda/4$):

    \Indp 

        $j$ = \textbf{insertBit}($k$, $t_1$, $b_1$)

        $j$ = \textbf{insertBit}($j$, $t_2$, $b_2$)

        $\vec{\rho}[j]$ = $c_1 \; \vec{\rho}[j] + c_2 \; \vec{\varphi}[k + \Lambda/4]$ 

        $\vec{\varphi}[k]$ = $\vec{\rho}[j]$

    \Indm 

    \codegap

    \tcp{swap sub-buffer with second pair node}

    $r''$ = \textbf{flipBit}($r$, $t_2-\lambda'$)

    \textbf{exchangeArrays}($\vec{\varphi}$, $0$, $\vec{\varphi}$, $\Lambda/4$, $\Lambda/4$, $r''$)

    \codegap 

    \tcp{update amplitudes}

    \AlgThreadComment{multithread}

    \textbf{for} $k$ \textbf{in} \textbf{range}($0$, $\Lambda/4$):

    \Indp 

        $j$ = \textbf{insertBit}($k$, $t_1$, $b_1$)

        $j$ = \textbf{insertBit}($j$, $t_2$, $b_2$)

        $\vec{\rho}[j]$ \PlusEq{} $(c_2 /c_1) \; \vec{\varphi}[k + \Lambda/4]$ 

    \Indm  

\Indm

\end{algorithm}

\clearpage

\subsection{Damping channel} %
\label{sec:damping} %
\noindent %
\begin{center} 
        $
\Qcircuit @C=1em @R=.7em {
& \gate{\mathcal{E}_\gamma} & \qw
}
        $
\end{center}

The amplitude damping channel is a widely considered realistic noise model in quantum information~\cite{khatri2020information},
describing dissipation of energy to the environment~\cite{nielsen2002quantum}, or (conventionally) a qubit to the $\ket{0}\numsub{1}$ state. Although damping of qubit $t$ with decay probability $p$ can be described through Kraus operators
\begin{align}
\hat{K}_t^{(1)} = \begin{pmatrix}
1 & 0
    \\
0 & \sqrt{1-p}
\end{pmatrix},
\;\;\;\;\;\;
\hat{K}_t^{(2)} = 
\begin{pmatrix}
0 & \sqrt{p}
    \\
0 & 0
\end{pmatrix},
\end{align}
its simulation as a Kraus map through Alg.~\ref{alg:distrib_dens_matr_kraus_map} is sub-optimal. We here instead derive an optimised simulation of the damping channel which communicates only a \textit{quarter} of all amplitudes across the network.
Interestingly, the prescribed communication is \textit{not} pairwise, but is fortunately cheaper.

Observe that the above Kraus operators map a basis projector $\ket{k}\bra{l}$ to
\begin{align}
\hat{K}_t^{(1)} \ket{k}\bra{l} \hat{K}_t^{(1)\dagger}
&= 
\ket{k}\bra{l} \cdot \begin{cases}
1, & k_{[t]} = l_{[t]} = 0,
\\ 
\sqrt{1-p}, & k_{[t]} \ne l_{[t]},
\\
1 - p, & k_{[t]} = l_{[t]} = 1,
\end{cases}
\\
\hat{K}_t^{(2)} \ket{k}\bra{l} \hat{K}_t^{(2)\dagger}
&=
\ket{k_{\neg t}}\bra{l_{\neg t}} 
\cdot 
\begin{cases}
p, & k_{[t]}=l_{[t]}=1, \\
0, & \text{otherwise}.
\end{cases}
\end{align}
The damping channel ergo modifies an amplitude of an $N$-qubit general state $\bm\rho = \sum_{kl}\alpha_{kl}\ket{k}\bra{l}$ via
\begin{align}
\alpha_{kl} \rightarrow 
\begin{cases}
\alpha_{kl} + p \; \alpha_{k_{\neg t},l_{\neg t}}
    &
    k_{[t]} = l_{[t]} = 0,
    \\
\sqrt{1-p} \; \alpha_{kl}
    &
    k_{[t]} \ne l_{[t]},
    \\ 
(1-p) \; \alpha_{kl},
    &
    k_{[t]} = l_{[t]} = 1,
\end{cases}
\end{align} 
and an amplitude of the equivalent Choi-vector $\choivec{\bm\rho}\numsub{2N} = \sum_i \beta_i \choivec{i}\numsub{2N}$ as
\begin{align}
\beta_i \rightarrow 
\begin{cases}
\beta_i + p \; \beta_{i_{\neg \{t,t+N\}}}
    &
    i_{[t]} = i_{[t+N]} = 0,
    \\
\sqrt{1-p} \; \beta_i
    &
    i_{[t]} \ne i_{[t+N]},
    \\
(1-p) \; \beta_i
    &
    i_{[t]} = i_{[t+N]} = 1.
\end{cases}
\label{eq:damping_amps_update}
\end{align}
We see every amplitude is scaled, and those of index $i$ with principle bits ($i_{[t]}$ and $i_{[t+N]}$) equal to zero are linearly combined with a pair amplitude of opposite bits.

\begin{figure}[b]
    \centering 
    \includegraphics[width=.47\textwidth]{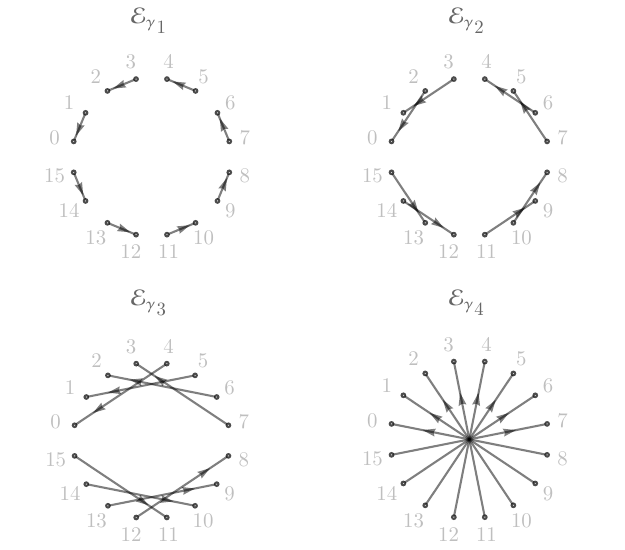}
    \caption{
        Communication patterns of Alg.~\ref{alg:distrib_dens_matr_damping} simulating the amplitude damping channel upon a $5$-qubit density matrix distributed between $16$ nodes. Each arrow indicates the one-way sending of $32$ amplitudes, via Fig.~\ref{fig:distrib_mem_exchange_patterns}~\textbf{c)}.
    }
    \label{fig:distrib_damping_comm}
\end{figure}

\afterpage{ %
\begin{algorithm}
\DontPrintSemicolon
\caption[blah blah]{
    \AlgTagDistributed \AlgTagDensityMatrix  \\
    Amplitude damping of qubit $t$ with decay probability $p$ of an $N$-qubit density matrix with Choi-vector distributed between arrays $\vec{\rho}$ of length $\Lambda$ among $2^w$ nodes.
    \protect{\begin{center} 
        $
\Qcircuit @C=1em @R=.7em {
& \gate{\mathcal{E}_\gamma} & \qw 
}
        $
    \end{center}}
    \begin{center}
        \AlgTagBops{$\mathcal{O}(\Lambda)$}%
        \AlgTagFlops{$\mathcal{O}(\Lambda)$}%
        \AlgTagNumSendRounds{$0$ or $1$ sends}\\
        \AlgTagAmpsTransferred{$2^{2N}/4$}%
        \AlgTagMemOverhead{$\mathcal{O}(1)$}%
        \AlgTagMemWrites{$\mathcal{O}(\Lambda)$}
    \end{center}
}
\label{alg:distrib_dens_matr_damping}

\textbf{distrib\_damping}($\vec{\rho}$, $t$, $p$):

\Indp 

    $\Lambda$ = $\dim(\vec{\rho})$

    $N$ = \textbf{getNumQubits}($\vec{\rho}$)
    \tcp*{Alg.~\ref{alg:distrib_dens_convenience_funcs}}

    $w$ = $\log_2(\textbf{getWorldSize()})$
    \tcp*{Alg.~\ref{alg:mpi_defs}}

    $c_1$ = $\sqrt{1-p}$

    $c_2$ = $1-p$

    \codegap 

    \textbf{if} $t < N-w$:

    \Indp 

        \codegap 

        \AlgThreadComment{multithread}

        \textbf{for} $k$ \textbf{in} \textbf{range}($0$, $\Lambda/4$):

        \Indp 

            $j_{00}$ = \textbf{insertBits}($k$, $\{t,t+N\}$, $0$)

            $j_{01}$ = \textbf{flipBit}($j_{00}$, $t$)
            \tcp*{Alg.~\ref{alg:bit_twiddles}}

            $j_{10}$ = \textbf{flipBit}($j_{00}$, $t+N$)

            $j_{11}$ = \textbf{flipBit}($j_{01}$, $t+N$)

            \codegap 

            $\vec{\rho}[j_{00}]$ \PlusEq{} $p \, \vec{\rho}[j_{11}]$

            $\vec{\rho}[j_{01}]$ \TimesEq{} $c_1$

            $\vec{\rho}[j_{10}]$ \TimesEq{} $c_1$

            $\vec{\rho}[j_{11}]$ \TimesEq{} $c_2$

        \Indm 

    \Indm 

    \codegap

    \textbf{else}:

    \Indp 

        $r$ = \textbf{getRank}()
        \tcp*{Alg.~\ref{alg:mpi_defs}}

        $r'$ = \textbf{flipBit}($r$, $t-(N-w)$)
        \tcp*{Alg.~\ref{alg:bit_twiddles}}

        $b$ = \textbf{getBit}($r$, $t-(N-w)$)
        \tcp*{Alg.~\ref{alg:bit_twiddles}}

        \codegap

        \tcp{if $i_{[t+N]}=1$, pack sub-buffer}

        \textbf{if} $b$ \EqEq{} $1$:

        \Indp 

            \AlgThreadComment{multithread}

            \textbf{for} $k$ \textbf{in} \textbf{range}($0$, $\Lambda/2$):

            \Indp 

                $j$ = \textbf{insertBit}($k$, $t$, $1$)
                \tcp{Alg.~\ref{alg:bit_twiddles}}

                $\vec{\varphi}[k]$ = $\vec{\rho}[j]$

                $\vec{\rho}[j]$ \TimesEq{} $c_2$

            \Indm 

            \codegap 

            \tcp{non-blocking send half of buffer}

            \textbf{MPI\_Isend}($\vec{\varphi}$, $\Lambda/2$, \textbf{MPI\_COMPLEX}, $r'$,
            
            \Indp 
            
                \textbf{MPI\_ANY\_TAG}, \textbf{MPI\_COMM\_WORLD})

            \Indm 

        \Indm 

        \codegap 

        \tcp{all nodes update $i$ where $i_{[t]}\ne i_{[t+N]}$}

        \AlgThreadComment{multithread}

        \textbf{for} $k$ \textbf{in} \textbf{range}($0$, $\Lambda/2$):

        \Indp 

            $j$ = \textbf{insertBit}($k$, $t$, $\LogicalNot b$)

            $\vec{\rho}[j]$ \TimesEq{} $c_1$

        \Indm

        \codegap 

        \tcp{if $i_{[t+N]}=0$, receive sub-buffer}

        \textbf{if} $b$ \EqEq{} $0$:

        \Indp 

            \tcp{receive half of buffer}

            \textbf{MPI\_Recv}($\vec{\varphi}$, $\Lambda/2$, \textbf{MPI\_COMPLEX}, $r'$, 
            
            \Indp 
            
                \textbf{MPI\_ANY\_TAG}, \textbf{MPI\_COMM\_WORLD})

            \Indm 

            \codegap 

            \AlgThreadComment{multithread}

            \textbf{for} $k$ \textbf{in} \textbf{range}($0$, $\Lambda/2$):

            \Indp 

                $j$ = \textbf{insertBit}($k$, $t$, $0$)

                $\vec{\rho}[j]$ \PlusEq{} $p\;\vec{\varphi}[k]$

            \Indm

        \Indm 

    \Indm 

\Indm 

 \end{algorithm} %
 } 

Next, we consider when these amplitudes are distributed.
Recall that the $2^{2N}$ amplitudes of $\choivec{\bm\rho}$ are uniformly distributed between arrays $\vec{\rho}$ among $2^w$ nodes (where $N \ge w$),
 such that the $j$-th local amplitude $\vec{\rho}[j]\equiv \beta_i$ of node $r$ corresponds to global basis state 
 $\choivec{i}\numsub{2N} \equiv \choivec{r}\numsub{w}\choivec{j}\numsub{2N-w}$. As we saw of $1$-qubit dephasing and depolarising, communication is only required when index $t+N$ falls within the leftmost $w$ qubits of $\choivec{\bm\rho}\numsub{2N}$.
 Eq.~\ref{eq:damping_amps_update} ergo distributes into two scenarios:

 \begin{itemize}
     \item \textbf{When} $t < N-w$, amplitude $\beta_{i_{\neg \{t,t+N\}}}$ is always contained within the same node containing $\beta_i$. Simulation is embarrassingly parallel.
     \item \textbf{When} $t \ge N-w$, amplitude $\beta_{i}$ (within rank $r$) is stored within a separate node to $\beta_{i_{\neg \{t,t+N\}}}$ (within rank $r'=r_{\neg t-(N-w)}$), and the bit $i_{[t+N]}$ is determined entirely by $r$. Ergo, every node (of rank $r$) satisfying $i_{[t+N]} \equiv r_{[t-(N-w)]} = 1$ (as do \textit{half} of all nodes) sends \textit{half} of its amplitudes (those satisfying $i_{[t]}=1$) to pair node $r'$, receiving none in return. 
     This is memory exchange pattern \textbf{c)} of Fig.~\ref{fig:distrib_mem_exchange_patterns}.
     In total, \textit{half} of all nodes send \textit{half} of their local amplitudes to the other non-sending half of all nodes.
    We illustrate this below.
     \begin{center}    \includegraphics[width=.5\columnwidth]{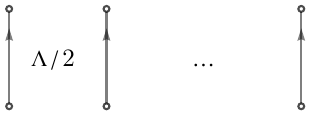}
    \end{center}
 \end{itemize}

We formalise our method in Alg.~\ref{alg:distrib_dens_matr_damping}, which overlaps concurrent buffer packing with local amplitude modification. Examples of the resulting communication patterns are shown in Fig.~\ref{fig:distrib_damping_comm}.


\subsection{Pauli string expectation value} %
\label{sec:pauli_string_expec_value} %
\noindent %
\begin{center} $ \boxed{ \text{Tr}\left( 
\bm\rho \; \sum\limits_n h_n \bigotimes\limits_t \hat{\sigma}_t  \right) } $ \end{center}

Expectation values of Hermitian operators are of fundamental importance in quantum computation. Beyond their modelling of experimental, probabilistic processes, expectation values appear extensively as primitives in quantum algorithms~\cite{buhrman2001quantum}, and are often themselves the desired output~\cite{masaya22aps}, such as they are in variational algorithms~\cite{yuan2019theory}, real-time simulators~\cite{georgescu2014quantum}, and condensed matter and chemistry calculations~\cite{mcardle2020quantum}.
Physically meaningful Hermitian operators are often naturally expressed in the Pauli basis~\cite{mcardle2020quantum}, and indeed Pauli strings have emerged as the canonical basis for realising operators like Hamiltonians on quantum computers.
It is essential that quantum simulators support calculation of expectation values, and prudent these calculations are highly optimised and bespoke, since even modest Pauli strings
of interest often contain many individual Pauli operators.
The need for bespoke treatment is especially evident when simulating realistic, noisy processes via density matrices, where naive simulation of a Pauli operator as a gate is already quadratically more expensive than if it were upon a statevector.
Alas, it is tempting to naively compute the expectation value using existing simulator primitives, like the previous algorithms of this manuscript, in a grossly inefficient manner.
In this section, we derive an embarrassingly parallel calculation (with a {single} scalar reduction) of a $T$-term $N$-qubit Pauli string expectation value under an $N$-qubit density matrix, distributed into $\Lambda$-length sub-Choi-vectors, requiring \textit{no} memory writes and only $\mathcal{O}(T \, \Lambda)$ flops per node.
We assume that the description of $\hat{H}$ (a list of real coefficients, and of Pauli-operator flags) is tractably small so that a local copy is present on every node; in essence that $T \, N \ll \Lambda$.
Sadly our method \textit{cannot} accelerate the evaluation of expected values of \textit{statevectors}, which we will elaborate upon.

The Pauli operators $\hat{X}$, $\hat{Y}$ and $\hat{Z}$ are a natural basis for Hermitian operators and have emerged as the canonical basis for realising Hamiltonians on quantum computers, encoded as an $N$-qubit $T$-term Pauli string of the form
\begin{gather}
\hat{H} = \sum\limits_n^T h_n \bigotimes\limits_t^N \hat{\sigma}_t^{(n)},
\\
h_n \in \mathbb{R}, 
\;\;\;\;\;
\hat{\sigma}\in \{\hat{\mathbb{1}}, \hat{X}, \hat{Y}, \hat{Z}\}.
\tag*{}
\end{gather}
This is a real-weighted sum of the Pauli tensors seen in Sec.~\ref{sec:pauli_tensor}, where we computed $\bm\rho \rightarrow \hat{\sigma}^{\otimes} \bm\rho$. Here, we instead seek scalar $\langle E \rangle = \text{Tr}(\hat{H} \bm\rho) \in \mathbb{R}$.
It is tempting to employ the Pauli tensor (upon density-matrix) simulation of Alg.~\ref{alg:distrib_dens_matr_specific_unitaries}, using clones of the density matrix which are summed (weighted by $h_n$) to produce $\hat{H}\bm\rho$, before a trace evaluation via
\begin{gather}
\choivec{\bm\rho}
= \sum\limits_i^{2^{2N}} \beta_i \choivec{i}
\;
\implies 
\;
\text{Tr}(\bm\rho) = \sum\limits_j^{2^N} \beta_{j(2^N+1)}.
\end{gather}
Such a scheme would require $\mathcal{O}(T)$ exchanges, a total of $\mathcal{O}(T)\,2^{2N}$ exchanged amplitudes and memory writes, and a $\mathcal{O}(2^{2N})$ memory overhead. 
An astonishingly more efficient scheme is possible.

Assume $\bm\rho =\sum_{kl} \alpha_{kl} \ket{k}\bra{l}$ and that $\matr{H}:\mathbb{C}^{2^N\times 2^N}$ is a $\hat{Z}$-basis matrix instantiating $\hat{H}$. Of course, we will not instantiate such an expensive object, but it permits us to express
\begin{align}
\langle E \rangle 
&=
\sum\limits_k^{2^N}
\sum\limits_l^{2^N} \,
\matr{H}_{\,lk} \; \alpha_{kl},
\end{align}
Through a natural 2D extension of our ket indexing, we may express 
\begin{align}
\matr{H}_{\,lk}
&=
\sum\limits_n^T h_n\left( \bigotimes_t^N \hat{\sigma}_t^{(n)}\right)_{lk}
\\
&=
\sum\limits_n^T h_n
\prod\limits_t^N \left( \hat{\sigma}_{N-t-1}^{(n)} \right)_{l_{[t]}, \, k_{[t]}}.
\end{align}
This reveals a $\mathcal{O}(N)$ bop calculation of a matrix element ($\in \{\pm 1, \pm \iu\}$) of the $N$-qubit Pauli tensor is possible,
informed by the elements of the Pauli matrices,
and hence that a single element $\matr{H}_{lk} \in \mathbb{C}$ is calculable in $\mathcal{O}(T)$ flops and $\mathcal{O}(T N)$ bops.
We can in-principle leverage Hermitivity of $\hat{H}$ and $\bm\rho$ to evaluate $\langle E \rangle$ in $\approx 2\times$ fewer operations, but this will not meaningfully accelerate distributed simulation.

Expressed in terms of the equivalent Choi-vector $\choivec{\bm\rho}\numsub{2N} = \sum_i \beta_i \choivec{i}\numsub{2N}$, we can write
\begin{gather}
\langle E \rangle 
= \sum\limits_i^{2^{2N}} 
\mathcal{H}_i \; \beta_i,
\label{eq:dens_expec_val_sum}
\\
\mathcal{H}_i = \sum\limits_n^T h_n 
\prod_t^N \left( \hat{\sigma}^{(n)}_{N-t-1} \right)_{i_{[t+N]},\,i_{[t]}},
\end{gather} 
where each $\mathcal{H}_i$ is concurrently and independently evaluable. 
We distribute the $2^{2N}$ amplitudes of $\choivec{\bm\rho}$ between $W=2^w$ nodes such that each contains a sub-vector $\vec{\rho}$ of length $\Lambda=2^{2N-w}$. 
Equ.~\ref{eq:dens_expec_val_sum} is then trivially divided between these nodes, each concurrently weighted-summing its amplitudes, before a final global reduction wherein each node contributes a single complex scalar. 

We formalise our strategy in Alg.~\ref{alg:distrib_dens_matr_pauli_string_expec}. 
It requires no exchanging of Choi-vectors between nodes, and no writing to heap memory. Each amplitude of $\matr{\rho}$ is read precisely once.

The subroutine \texttt{\textbf{pauliTensorElem}} (Line~\ref{algline:paulitensorelem_def}) performs exactly $N$ multiplications of real or imaginary integers ($0, \pm 1, \pm i$). One may notice that half of all elements among the Pauli matrices are \textit{zero}, and encountering any one will yield a zero tensor element. This might lead one to erroneously expect only a factor $1/2^N$ of invocations of \texttt{\textbf{pauliTensorElem}} will yield a non-zero result, and prompt an optimistion which avoids their calculation.
We caution against this; the input Pauli strings to Alg.~\ref{alg:distrib_dens_matr_pauli_string_expec} will \textit{not} be dense (i.e. $T \ll 4^N$), and so will not uniformly invoke the subroutine for all permutations of arguments $i$ and $\vec{\sigma}$. We should expect instead to perform exponentially fewer invocations, precluding us to reason about the expected number of zero elements as this depends on the user's input Pauli string structure. Finally, a naive optimisation like an attempt to return early from the loop of Line~\ref{algline:paulitensorelem_loop} whenever $v=0$ will cause needless branching and disrupt multithreaded performance.

We now lament the lack of an analogous statevector algorithm.
The speedup of Alg.~\ref{alg:distrib_dens_matr_pauli_string_expec} over a naive method invoking Pauli tensor simulation, results from our \textit{not} propagating any amplitude uninvolved in the trace. It hence will not accelerate the equivalent statevector calculation which necessarily involves all amplitudes. That is, for the general $\ket{\Psi}\numsub{N} = \sum_k \alpha_k\ket{k}\numsub{N}$, the expected value
\begin{align}
\langle E \rangle 
&=
\bra{\Psi} \hat{H} \ket{\Psi} 
= \sum\limits_k^{2^N}\sum\limits_l^{2^N} \matr{H}_{\,lk} \, \alpha_k^* \, \alpha_l,
\end{align}
includes products of all pairs of amplitudes $\{ \alpha_k, \, \alpha_l \}$. This necessitates the products include amplitudes residing in distinct nodes, and ergo that its distributed calculation involves multiple rounds of inter-node exchange.

\begin{algorithm}[tb]
\DontPrintSemicolon
\caption[blah blah]{
    \AlgTagDistributed \AlgTagDensityMatrix  \\
    Expected value under a $T$-term $N$-qubit Pauli string of an $N$-qubit density matrix with Choi-vector distributed between arrays $\vec{\rho}$ of length $\Lambda$ among $W$ nodes. List $\vec{\sigma}$ stores $N\,T$ flags identifying Pauli operators, ordered by terms (matching coefficients $\vec{h}$) then qubits (least to most significant).
    \protect{
    \begin{center} $\boxed{ \text{Tr}\left( 
\bm\rho \; \sum\limits_n h_n \bigotimes\limits_t \hat{\sigma}_t  \right) }$ \end{center}
    }
    \begin{center}
        \AlgTagBops{$\mathcal{O}(N \, T \, \Lambda)$}%
        \AlgTagFlops{$\mathcal{O}(T\,\Lambda)$}%
        \AlgTagNumSerialRounds{$0$}\\
        \AlgTagAmpsTransferred{$\mathcal{O}(W)$}%
        \AlgTagMemOverhead{$\mathcal{O}(1)$}%
        \AlgTagMemWrites{$0$}
    \end{center}
}
\label{alg:distrib_dens_matr_pauli_string_expec}

\textbf{distrib\_pauliStringExpec}($\vec{\rho}$, $\vec{h}$, $\vec{\sigma}$):

\Indp 

    $T$ = $\dim(\vec{h})$

    $N$ = $\dim(\vec{\sigma})$ / $T$

    $\Lambda$ = $\dim(\vec{\rho})$

    $\lambda$ = $\log_2(\Lambda)$

    $r$ = \textbf{getRank}()

    \codegap 

    $x$ = $0$

    \codegap

    \AlgThreadComment{multithread}

    \textbf{for} $j$ \textbf{in} \textbf{range}($0$, $\Lambda$):

    \Indp 

        $i$ = $(r \; \BitShiftLeft \; \lambda) \BitOr j$

        $v$ = $0$

        \textbf{for} $n$ \textbf{in} \textbf{range}($0$, $T$):

        \Indp 

            $\vec{\sigma}'$ = $\vec{\sigma}[n\,N \; : \; n\,N + N]$

            $\lambda$ = \textbf{pauliTensorElem}($\vec{\sigma}',\;i,\;N$)

            $v$ \PlusEq{} $\vec{h}[n] \; \lambda$

        \Indm 

        \codegap 

        \AlgThreadComment{thread reduce}
        
        $x$ \PlusEq{} $v$

    \Indm 

    \codegap 

    \AlgThreadComment{MPI reduce $x$}

    $x$ = $\dots$

    \codegap



    


    \codegap 

    \textbf{return} $x$

\Indm

\codegap 

\codegap 

\textbf{pauliTensorElem}($\vec{\sigma}$, $i$, $N$):
\label{algline:paulitensorelem_def}

\Indp 

    $\vec{M}$ = $\{
        \renewcommand*{\arraystretch}{.8}
        \begin{pmatrix} 1 & 0 \\ 0 & 1 \end{pmatrix}, 
        \begin{pmatrix} 0 & 1 \\ 1 & 0 \end{pmatrix}, 
        \begin{pmatrix} 0 & -\iu \\ \iu & 0 \end{pmatrix},
        \begin{pmatrix}
            1 & 0 \\ 0 & -1
        \end{pmatrix}
    \}$

    $v$ = $1$

    \textbf{for} $q$ \textbf{in} \textbf{range}($0$, $N$):
    \label{algline:paulitensorelem_loop}

    \Indp 

        $b_c$ = \textbf{getBit}($i$, $q$)

        $b_r$ = \textbf{getBit}($i$, $q$ + $N$)

        $k$ = $\vec{\sigma}[q]$

        $\matr{m}$ = $\vec{M}[ k ]$

        $v$ \TimesEq{} $\matr{m}[b_r, b_c]$ 
        \tcp*{integer multiply}

    \Indm

    \textbf{return} $v$

\Indm 

 \end{algorithm}

\pagebreak
\hphantom{.}
\pagebreak


\subsection{Partial trace} %
\label{sec:partial_trace} %
\noindent %
\begin{center} $ \boxed{ \text{Tr}_{\vec{t}}\left( 
\bm\rho \right) } $ \end{center}

The partial trace is an extremely useful operation in quantum information theory which reduces the dimension of a density matrix, and obtains a description of a sub-state from a composite state~\cite{peres1997quantum}. It is ubiquitous in the study of noise, entanglement and quantum control, and an essential tool in the general study of mixed states~\cite{paris2012modern,nielsen2002quantum}.

Several serial, local algorithms for computing the partial trace of a dense matrix exist in the literature, although their innovations are that explicit tensoring of identity matrices can be avoided~\cite{maziero2017computing}, and that amplitude indices can be calculated bitwise~\cite{barkataki2018set}; these are properties of all algorithms presented in this manuscript.
To the best of the author's knowledge, there are no reports of a distributed partial trace of dense matrices. Indeed the task is intimidating; the partial trace combines amplitudes of the input density matrix which are distributed between many distinct nodes of our network, suggesting non-pairwise communication.

We here derive a pair-wise distributed algorithm (between $2^w$ nodes) to compute the partial trace of any $N$-qubit multi-partite mixed state, tracing out $n$ qubits, in $\mathcal{O}(2^{2N}/2^n)$ total flops, 
$\mathcal{O}(w \, 2^{2N})$ 
exchanged amplitudes (in fewer than $2w$ rounds), and as many memory writes. These are smaller network costs than $w$ one-qubit gates, and asymptotically negligible flops. Our trick is similar to that used in Sec.~\ref{sec:distrib_manytarg_gate}'s simulation of the many-target gate, albeit with a more complicated post-processing step.
We can trace out a maximum of $n \le N-\lceil w/2 \rceil$ qubits, as elaborated upon below. However, for our output density matrix to satisfy the preconditions of this manuscript's other algorithms, we require the stricter condition that $n \le N - w$, which permits the tracing out of fewer target qubits.

Let us frame our general partial trace as the tracing out of $n$ qubits $\vec{t}$ constituting subsystem $B$ of $N$-qubit composite density matrix ${\bm\rho}^{AB}$, in order to form reduced matrix ${\bm\rho}^{A}$ of $m = N-n$ qubits. The ordering of $\vec{t}$ is inconsequential.
Because we permit $\vec{t}$ to be any subset of all qubits in $\intrange{0}{N}$, it is \textit{not} generally true that $\bm\rho^{AB} = \bm\rho^A \otimes \bm\rho^B$. We ergo label the constituent bits of the $k$-th basis ket as
\begin{align}
    \ket{k}\numsub{N} \equiv \ket{k^A, k^B}\numsub{N},
\end{align}
where $k^A \in \intrange{0}{2^m}$ and $k^B \in \intrange{0}{2^n}$. These are the $m$ and $n$-qubit basis kets of $\bm\rho^A$ and $\bm\rho^B$ respectively.
We define function $f$ to interweave the bits of $k^B$ into positions $\vec{t}$ of $k^A$, such that
\begin{align}
    k = f_{\vec{t}}(k^A, k^B).
\end{align}
We can trivially compute $f$ with bitwise algebra.
Our general composite density matrix, with amplitudes $\alpha_{kl}$, can then be written
\begin{gather}
    \bm\rho^{AB} = \sum\limits_k^{2^N} \sum\limits_l^{2^N} \alpha_{kl} \ket{k}\bra{l}\numsub{N}
    \\
     \equiv
    \sum\limits_{k^A}^{2^m}\sum\limits_{k^B}^{2^n}
    \sum\limits_{l^A}^{2^m}\sum\limits_{l^B}^{2^n}
    \alpha_{f_{\vec{t}}(k^A, k^B), \, f_{\vec{t}}(l^A,l^B)}
    \ket{k^A,k^B}\bra{l^A,l^B}\numsub{N}.
\end{gather}
Let $\bra{\mathbb{1}^{\otimes m},v}$ notate an interwoven tensor product of the $m$-qubit identity operator with the $n$ qubits of the $v$-th basis bra of $\bm\rho^B$.
The reduced density matrix can then be expressed as
\begin{align}
    \bm\rho^A 
    &= \Tr_B(\bm\rho^{AB}) 
        = \sum\limits_v^{2^n} 
        \bra{\mathbb{1}^{\otimes m},v}
        \bm\rho^{AB}
        \ket{\mathbb{1}^{\otimes m},v}
        \tag*{}
        \\
    &=
    \sum\limits_{k^A}^{2^m}
    \sum\limits_{l^A}^{2^m}
    \left( 
    \alpha_{f_{\vec{t}}(k^A, v), \, f_{\vec{t}}(l^A, v)}
    \right)
    \ket{k^A}\bra{l^A}\numsub{N}.
\end{align} 
This makes clear that the amplitudes $\alpha'$ of $\bm\rho^A$ are
\begin{align}
    \alpha_{kl}'
    &=
    \sum\limits_v^{2^n} 
    \alpha_{f_{\vec{t}}(k, v), \, f_{\vec{t}}(l, v)},
    \;\;\;
    k,l \in \intrange{0}{2^m}.
\end{align}
Observe that a fraction $1/2^n$ of all amplitudes of $\bm\rho^{AB}$ are involved in the determination of $\bm\rho^A$, and that $\bm\rho^A$ is determined by a total of $2^{2m+n}$ sum terms. 
Before proceeding, we make several more immediate observations to inform subsequent optimisation.
\begin{itemize}
    \item 
        In principle, evaluating a single $\alpha_{kl}'$ amplitude requires polynomially \textit{fewer} than the $2^{2m+n}-1$ floating-point additions suggested directly by the sum over $v$, because the sum may be performed by a sequence of hierarchical reductions on neighbouring pairs. This is compatible with numerical stability techniques like Kahan summation~\cite{kahan1965further},
though reduces the floating-point costs by a modest and shrinking factor $1-2^{-n}$, easily outweighed by its introduced caching overheads.

    \item 
    The pair of subscripted indices $(f_{\vec{t}}(k, v), \, f_{\vec{t}}(l, v))$ are unique for every unique assignment of $(k,l,v)$.
    Assuming no properties of $\bm\rho^{AB}$ (e.g. relaxing Hermitivity), 
    each amplitude $\alpha_{kl}'$ of $\bm\rho^A$ is therefore a sum of unique amplitudes $\alpha$ of $\bm\rho^{AB}$. There are no repeated partial sums between different $\alpha_{kl}'$ which we might otherwise seek to re-use in heirarchal reductions.
    \item As we should expect by the arbitrarity of the ordering of $\vec{t}$, the sum in $\alpha_{kl}'$ is uniformly weighted. We can therefore iterate $v$ in any order, and set its constituent bits $v_{[q]}$ with simplified unordered bitwise operations. 
    \item We must optimise our memory strides by choosing whether to contiguously iterate the ``output" amplitudes $\alpha'$ (for each, computing a full sum of $2^n$ scalars), or over the ``input" amplitudes $\alpha$ (adding each to one of $2^m$ partial sums). We choose the former, since the cache penalties of a suboptimal write stride outweigh the read penalties, and also since its multithreaded implementation avoids race conditions and minimises false sharing~\cite{bolosky1993false}. 
\end{itemize}
The equivalent reduced Choi-vector $\choivec{\bm\rho^A}\numsub{2m} = \sum_i \beta_i' \choivec{i}$, resulting from tracing out qubits $\vec{t}$ of $\choivec{\bm\rho^{AB}}\numsub{2N} = \sum_i \beta_i \choivec{i}$, has amplitudes
\begin{gather}
    \beta_i' = \sum\limits_v^{2^n} \beta_{ g(i,v) },
    \;\;\;\;\;
    \text{where}
    \label{eq:partial_trace_reduced_amp_def}
    \\
    g(i,v) = f_{\vec{t} \;\cup\; (\vec{t}+N)}(i,(v\BitShiftLeft n)\BitOr v).
    \label{eq:partial_trace_tau_index_func}
\end{gather}
The function $g$ merely takes index $i$ and interweaves the bits of $v$ into positions $\vec{t}$ \textit{and} $\vec{t}+N$, the latter being the same positions shifted left by $N$. Local, serial evaluation of this sum is trivial, requiring $\mathcal{O}(2^{2N}/2^n)$ flops, and local parallelisation is straightforward.

We now distribute these amplitudes between $W=2^w$ nodes. Because $\choivec{\bm\rho^{AB}}$ and $\choivec{\bm\rho^{A}}$ differ in dimension, their partitioned arrays on each node have different sizes.
Recall we assume $N \ge w$ such that each node contains at least one column's worth of $\choivec{\bm\rho^{AB}}$. 
For the output \textit{reduced} density matrix $\choivec{\bm\rho^{A}}$ to satisfy this precondition, and ergo be compatible input to this manuscript's other algorithms, it must similarly satisfy $m \ge w$, equivalently that $n \le N - w$. This is \textit{inessential} to our algorithm however, which imposes a looser condition elaborated upon later.

The $j$-th local amplitude $\vec{\rho}^{AB}[j]\equiv \beta_g$ of node $r$ corresponds to global basis state 
\begin{align}
\choivec{g}\numsub{2N}
&\equiv
\choivec{r}\numsub{w}\choivec{j}\numsub{2N-w}
\\
&\equiv \;
{
    \color{lightgray}
\underbrace{
    \mystrut{7pt}
    \color{black}
\hfsetbordercolor{subketborder2}
\tikzmarkin[subketstyle2]{uniquelabel}
\choivec{r}\numsub{w}
\tikzmarkend{uniquelabel} \;\;\;
\tikzmarkin[subketstyle1]{woweemister}
    \choivec{  \dots 
    }\numsub{N-w}
\tikzmarkend{woweemister}
}_{\color{black}
\vec{t}+N
}
\;\;\,
\underbrace{
    \mystrut{7pt}
    \color{black}
\tikzmarkin[subketstyle2]{grrreh}
    \choivec{\dots}\numsub{w}
\tikzmarkend{grrreh} \;\;\;
\tikzmarkin[subketstyle1]{youfoundthesecretmessage}
\choivec{
    \dots 
}\numsub{N-w}
\tikzmarkend{youfoundthesecretmessage}}_{\color{black} 
\vec{t}
 }
}
\tag*{}
\end{align}
Similarly, the $j$-th local \textit{reduced} amplitude $\vec{\rho}^A[j] = \beta_i'$ corresponds to
\begin{align}
    \choivec{i}\numsub{2m} = \choivec{r}\numsub{w}\choivec{j}\numsub{2m-w}.
\end{align}

%
 %
Therefore all to-be-traced qubits ${t_q} \in \vec{t}$ (satisfying $0 \le t_q < N$) target the \textit{suffix} substate $\choivec{j}\numsub{2N-w}$, and a \textit{subset} of the shifted qubits in $\vec{t}+N$ will target the \textit{prefix} substate $\choivec{r}\numsub{w}$. 
Two communication patterns emerge:

\begin{enumerate}
    \item \textbf{When} all $t_q < N-w$, then \textit{no} qubits in $\vec{t}+N$ target the prefix substate.
    This means
    every index $g(i,v)$ (Eq.~\ref{eq:partial_trace_tau_index_func}) within a given node of rank $r$ is determined entirely by ($r$, $v$, and) the suffix bits of $i$, equivalent to $j$.
    Precisely:
    \begin{align} 
    i = (r \BitShiftLeft (2N-w)) \BitOr j
    \end{align} 
    for all local $j$.
    Ergo all summed amplitudes $\{ \beta_{g(i,v)} : v\}$ reside within the same node.
    Furthermore, the index $i$ of the destination amplitude $\beta_i'$ shares the same $w$-bit prefix ($r$) as the source/summed amplitudes, and ergo also resides in the same node.
    As a result, this scenario is embarrassingly parallel.
    \item
        \textbf{When} $\exists \; t_q \ge N-w$, the amplitudes featured in Eq.~\ref{eq:partial_trace_reduced_amp_def} reside within distinct nodes.
        Computing a single $\beta_i'$ will require prior communication.
        
        Like we did in Sec.~\ref{sec:distrib_manytarg_gate} to simulate the many-target general unitary on a statevector, we can in-principle use SWAP gates to first obtain locality of these amplitudes.
        However, the procedure here is complicated by the reduced dimension of the output structure $\choivec{\bm\rho^A}\numsub{2m}$, meaning that we cannot simply ``swap back" our swapped qubits.
\end{enumerate}

Let us focus on scenario \textbf{\textcolor{red}{2.}}, which admits a several step procedure. 
In essence, we apply SWAP gates to move all targeted qubits into the $(2N-w)$-qubit suffix substate and then perform the subsequently embarrassingly parallel reduction; thereafter we perform additional SWAP gates on the reduced density matrix to restore the relative ordering of the non-targeted qubits. 
This \textit{a priori} requires that all target qubits $\vec{t}$, \textit{and} their paired Choi qubits $\vec{t}+N$, can \textit{fit} into the suffix substate (a similar precondition of the many-target gate upon a statevector of Sec.~\ref{sec:distrib_manytarg_gate}). This requires
\begin{align}
    n \le N - \lceil w/2 \rceil.
\end{align}
The subsequent SWAP gates on the reduced $m$-qubit density matrix, treated as an unnormalised $2m$-qubit statevector, assume the equivalent precondition $2m \ge w$.

\def\vColor{blue}
\def\iColor{gray}

\begin{figure*}[tb]
    \centering 
    \includegraphics[width=\textwidth]{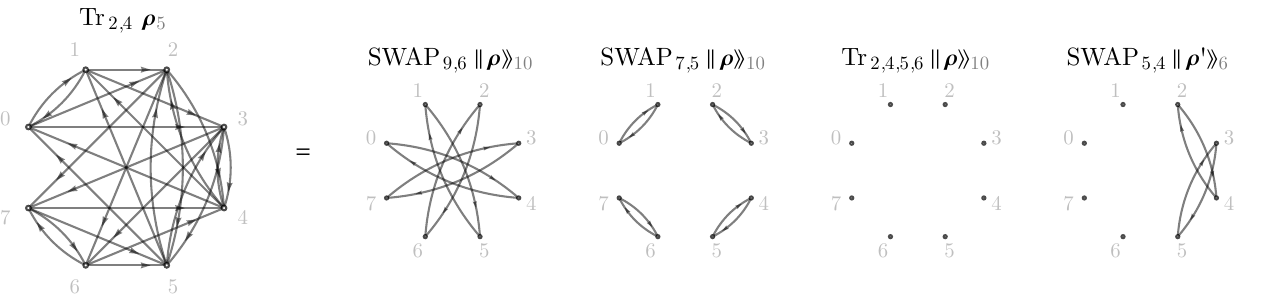}
    \caption{
        The communication pattern of Alg.~\ref{alg:distrib_partial_trace}'s distributed simulation of the
        partial trace of qubits $\vec{t}=\{2,4\}$ upon a $N=5$-qubit density matrix distributed between $W=8$ nodes.
        The left plot shows the necessary traffic of amplitudes to effect $\text{Tr}_{2,4}(\bm\rho)$ ``directly", in a way incompatible with our distributed partitioning. The right plots decompose this into the pairwise-communicating
        steps of our algorithm, which operates upon $\choivec{\bm\rho}\numsub{10}$ (and reduced state $\choivec{\bm\rho'}\numsub{6}$) treated as statevectors.
        It is interesting to monitor the movement of amplitudes from node $r=1$ to $2$, achieved via intermediate movements to nodes $5$, $4$, then $2$.
    }
    \label{fig:distrib_partial_trace_comm}
\end{figure*}

Even under these assumptions, the procedure is verbose in the general case. In lieu of its direct derivation, we opt to demonstrate it with an example.
Imagine that we distribute an $N=5$ qubit density matrix $\bm\rho$ between $W=8$ nodes (ergo $w = 3$).
Assume we wish to trace out qubits $\vec{t} = \{2,4\}$.
Let $\choivec{g}\numsub{2N} = \choivec{ {\color{\iColor}i} ,{\color{\vColor}v}}$ be a basis state of the Choi-vector $\choivec{\bm\rho}\numsub{2N}$, where ${\color{\vColor}{v}}$ is formed by the \textit{four} bits of $g$ at indices $\vec{t} \cup \vec{t}+N$, and ${\color{\iColor}i}$ is formed by the remaining \textit{six} untargeted bits of $g$.
The bits of $g$, constituted by the bits of ${\color{\iColor}i}$ and ${\color{\vColor}v}$, are arranged as:
\begin{align}
    \choivec{g}\numsub{2N} &= 
    \choivec{ g_{[9]} \, g_{[8]} \, g_{[7]} 
    \hphantom{\choivec{w}\,}
    g_{[6]} \, g_{[5]} 
    g_{[4]} \, g_{[3]} \, g_{[2]} \, g_{[1]} \, g_{[0]} }\numsub{2N}
    \tag*{}
    \\ 
    &=
    %
    %
    %
    \choivec{ {\color{\vColor}v_{[3]}} \, {\color{\iColor}i_{[5]}} \, {\color{\vColor}v_{[2]}} }\numsub{w} 
    %
    %
    \choivec{  
        {\color{\iColor}i_{[4]}} \,
        {\color{\iColor}i_{[3]}} \,
        {\color{\vColor}v_{[1]}} \,
        {\color{\iColor}i_{[2]}} \,
        {\color{\vColor}v_{[0]}} \,
        {\color{\iColor}i_{[1]}} \,
        {\color{\iColor}i_{[0]}} }\numsub{2N-w}.
\end{align}
Evaluating a reduced, output amplitude $\beta'_{\color{\iColor}i}$ requires summing all $\beta_g$ with indices satisfying ${\color{\vColor}v_{[0]}}={\color{\vColor}v_{[2]}}$ and ${\color{\vColor}v_{[1]}}={\color{\vColor}v_{[3]}}$ (with fixed ${\color{\iColor}i}$). Because bits ${\color{\vColor}v_{[2]}}$ and ${\color{\vColor}v_{[3]}}$ lie in the prefix state, the sum terms are distributed between multiple nodes. Precisely, between ranks 
%
$\{ r =  {\color{\vColor}v_{[3]}} 2^2 + \, {\color{\iColor}i_{[5]}} 2^1 + \,  {\color{\vColor}v_{[2]}} 2^0 \; : \; 0 \le v < 2^4 \}$.

So we first perform a series of SWAP gates upon $\choivec{\bm\rho}$ (treated as a statevector) via Sec.~\ref{sec:swap_gate}, in order to move all targeted prefix bits into the suffix state.
We heuristically swap the left-most prefix targets (starting with ${\color{\vColor}v_{[3]}}$) with the left-most non-targeted suffix qubits (initially ${\color{\iColor}i_{[4]}}$), minimising the relative displacement of the ${\color{\iColor}i}$ bits, which will accelerate a subsequent re-ordering step.
After effecting
\begin{align}
    \choivec{\bm\rho}\numsub{2N} \; \rightarrow \; 
    \text{SWAP}_{9,6}
    \;\;
    \text{SWAP}_{7,5}
    \,
    \choivec{\bm\rho}\numsub{2N},
\end{align}
the basis state $\choivec{g}$ has been mapped to $\choivec{g'}\numsub{2N} = $
\begin{align}
    %
    \choivec{ 
        {\color{\iColor}i_{[ \mathbf{4} ]}} \, {\color{\iColor}i_{[\mathbf{5}]}}
        \, {\color{\iColor}i_{[3]}}
        }\numsub{w} 
    \choivec{  
        {\color{\vColor}v_{[3]}} \,
        {\color{\vColor}v_{[2]}} \,
        {\color{\vColor}v_{[1]}} \,
        {\color{\iColor}i_{[2]}} \,
        {\color{\vColor}v_{[0]}} \,
        {\color{\iColor}i_{[1]}} \,
        {\color{\iColor}i_{[0]}} }\numsub{2N-w}.
        \tag*{}
\end{align}
All amplitudes $\beta_g$ across varying ${\color{\vColor}v}$ (fixing ${\color{\iColor}i}$) now reside within the same node,
permitting embarrassingly parallel evaluation of Eq.~\ref{eq:partial_trace_reduced_amp_def}.
This resembles a \textit{four}-qubit partial trace, of qubits $\vec{t}' = \{2,4,5,6\}$, upon a \textit{seven}-qubit statevector.
All $W$ nodes perform this reduction, producing a distributed $m=3$-qubit density matrix $\bm\rho'$.
Alas, this is \textit{not} yet the output state; the ${\color{\iColor}i}$-th global basis state of $\choivec{\bm\rho'}$ does \textit{not} correspond to the desired amplitude  $\beta'_i$, but is instead the state
\begin{align}
\choivec{i'}\numsub{2m} = 
    \choivec{ 
        {\color{\iColor}i_{[\mathbf{4}]}} \, {\color{\iColor}i_{[\mathbf{5}]}} \, {\color{\iColor}i_{[3]}}
        }\numsub{w} 
    \choivec{  
        {\color{\iColor}i_{[2]}} \,
        {\color{\iColor}i_{[1]}} \,
        {\color{\iColor}i_{[0]}} }\numsub{2m-w},
\end{align}
where the leftmost two prefix qubits are disordered.

Our final step is to restore the relative ordering of all bits of ${\color{\iColor}i}$ (mapping $\choivec{i'} \rightarrow \choivec{i}$) by performing additional SWAP gates upon the corresponding qubits of the reduced $3$-qubit density matrix.
Each qubit can be swapped directly to its known ordered location.
In this example, we simply perform
\begin{align}
    \choivec{\bm\rho'}\numsub{2m}
    \; \rightarrow \;
    \text{SWAP}_{5,4}\;
    \choivec{\bm\rho'}\numsub{2m} \, .
\end{align}
Choi-vector $\choivec{\bm\rho'}\numsub{2m}$ is now the correct, distributed, reduced density matrix. 
It is again beneficial to heuristically perform these ``post-processing" SWAPs upon the leftmost prefix qubits first, to avoid unnecessary displacement of subsequently swapped qubits between the suffix and prefix substates which causes wasteful communication.
We visualise the incurred communication pattern of this process in Fig.~\ref{fig:distrib_partial_trace_comm}.

We summarise the complexity of this method.
\begin{itemize}
    \item At most $w$ initial SWAP gates are required to remove all prefix targets. Each invokes Alg.~\ref{alg:distrib_swap} in communication scenario \textbf{\textcolor{red}{3.}} whereby \textit{half} a node's amplitudes are exchanged. A total of $\mathcal{O}(w)\,2^{2N}/2$ amplitudes are communicated in $\mathcal{O}(w)$ rounds, with as many memory writes, and \textit{zero} flops.
    \item The embarrassingly parallel evaluation of Eq.~\ref{eq:partial_trace_reduced_amp_def}, i.e. the ``local trace", involves $\mathcal{O}(2^{2N}/2^n)$ flops and bops, and a factor $1/2^n$ fewer memory writes.
    \item Fewer than $m$ final SWAPs are needed to reorder the reduced state, each invoking Alg.~\ref{alg:distrib_swap} in potentially \textit{any} of its three communication scenarios.
    However, 
    things simplify
    when we enforce $m \ge w$, i.e. the precondition assumed by this manuscript's other algorithms.
    In that case, all initially targeted prefix qubits get swapped into the \textit{leftmost} suffix positions, and ergo after reduction, only the prefix qubits are disordered (as per our example). The final SWAP costs ergo
    simplify to $\mathcal{O}(w)$ SWAPs in scenario \textcolor{red}{\textbf{3.}} of Alg.~\ref{alg:distrib_swap},
    exchanging a total of $\mathcal{O}(w)2^{2m}/2$ amplitudes.
    \textit{Without} this precondition, the final SWAP costs scale inversely with $2^{2n}$, so are anyway quickly occluded with increasing number of traced qubits $n$.
    Furthermore, our heuristic of initially swapping the leftmost qubits first reduces the necessary number of final SWAPs.
\end{itemize}

We formalise this algorithm in Alg.~\ref{alg:distrib_partial_trace}.
Below we discuss some potential optimisations, and some tempting but ultimately not worthwhile changes.
\begin{itemize}

    \item Our algorithm did not assume any properties (normalisation or otherwise) of $\bm\rho$. If we assume $\bm\rho$ is Hermitian, then we can reduce the network and floating-point costs by at most a factor $2$. This is because $\rho_{ij} = \rho_{ji}^*$ enables us to process only a factor $(1+1/2^N)/2$ of all $2^{2N}$ amplitudes, further reducing the fraction $1/2^n$ involved in the partial trace. Determining the exact reduction and consequential utility of the optimisation requires a careful treatment we do not here perform.

    \item The qubits of the reduced density matrix, before the post-processing SWAPs, are only ever out-of-order when an initial prefix qubit is swapped \textit{past} a non-targeted qubit. So it is tempting to swap only adjacent qubits, percolating prefix qubits toward the suffix substate one qubit at a time. This preserves the relative order of the non-targeted qubits, so no post-trace SWAPs are required.
    Alas, this is \textit{not} worthwhile; it necessitates more total SWAPs, and all of them will operate on the larger $N$-qubit density matrix, as opposed to the smaller reduced ($N-n$)-qubit density matrix, increasing communication and write costs.

\end{itemize}

\begin{itemize}

    \item Our initial SWAPs moved prefix targets into the \textit{leftmost} suffix qubits to reduce disordering of the untargeted qubits, and ergo reduce the number of subsequent SWAPs (and thus, the communication costs) on the reduced state. This means however that the local partial trace calculation (invocation of \textbf{local\_partialTraceSub} at Line~\ref{algline:partial_trace_local_invocation_inside_distrib} of Alg.~\ref{alg:distrib_partial_trace}) targets high-index qubits (in arrays $\vec{t}$ and $\vec{t}'$). This causes a large memory stride; the accessed amplitudes $\vec{\rho}[g]$ across $g$ at Line~\ref{algline:partialtrace_local_amp_get} are far apart (beyond cache lines), and their bounding addresses overlap across different $i$. This may lead to sub-optimal caching behaviour, especially in multithreaded settings (although we thankfully note it does \textit{not} induce false sharing, since we merely \textit{read} $\vec{\rho}$).
    It is worth considering to instead swap prefix targets into the \textit{rightmost} suffix qubits. This makes amplitudes $\vec{\rho}[g]$ at Line~\ref{algline:partialtrace_local_amp_get} \textit{contiguous} in memory, improving caching and multithreaded performance, but requiring more post-processing SWAPs and ergo modestly increased communication costs.
    Such a strategy may prove worthwhile when making use of so-called ``fused-swaps"~\cite{de2007massively,imamura2022mpiqulacs,stanwyck2022cuquantum}.

\end{itemize}

\begin{algorithm}[b]
\DontPrintSemicolon
\caption{
    $\mathcal{O}(N)$ subroutines of Alg.~\ref{alg:distrib_partial_trace}.
}
\label{alg:partial_trace_convenience_funcs}

\textbf{getNextLeftmostZeroBit}($b$, $i$):

\Indp 

    $i$ \MinusEq{} $1$

    \textbf{while} \textbf{getBit}($b$, $i$) \EqEq{} $1$:

    \Indp 

        $i$ \MinusEq{} $1$

    \Indm

    \textbf{return} $i$

\Indm

\codegap 

\codegap 

\textbf{getReorderedTargets}($\vec{s}$, $\lambda$):

\Indp


    \tcp{locate leftmost non-targeted suffix}

    $b$ = \textbf{getBitMask}($\vec{s}$)

    $\tau$ = \textbf{getNextLeftmostZeroBit}($b$, $\lambda$)


    \tcp{obtain new suffix-only targets}

    $\vec{s}'$ = $\{ \, \}$

    \textbf{for} $q$ \textbf{in} \textbf{range}($\dim(\vec{s})-1$, $-1$, $-1$):

    \Indp 

        \textbf{if} $\vec{s}[q] < \lambda$:

        \Indp 

            \textcolor{gray}{append} $\vec{s}[q]$ \textcolor{gray}{to} $\vec{s}'$

        \Indm 

        \textbf{else}:

        \Indp 

            \textcolor{gray}{append} $\tau$ \textcolor{gray}{to} $\vec{s}'$

            $\tau$ = \textbf{getNextLeftmostZeroBit}($b$, $\tau$)

        \Indm

    \Indm 


    \textbf{return} \textbf{reversed}($\vec{s}'$)

\Indm 

\codegap 

\codegap 

\textbf{getRemainingQubitOrder}($N$, $\vec{s}$, $\vec{s}'$, ):

\Indp 


    \tcp{determine post-swap qubit ordering}

    $\vec{q}$ = \textbf{range}($0$, $2\,N$)

    \textbf{for} $a$, $b$ in \textbf{zip}($\vec{s}$, $\vec{s}'$):

    \Indp 

        \textbf{if} $a$ \NotEqEq{} $b$:

        \Indp 

            $\vec{q}[a]$, $\vec{q}[b]$ = $\vec{q}[b]$, $\vec{q}[a]$

        \Indm

    \Indm 


    \tcp{remove traced-out qubits}

    $\vec{s}''$ = $\{ \, \}$

    $b'$ = \textbf{getBitMask}($\vec{s}'$)

    \textbf{for} $i$ \textbf{in} \textbf{range}($0$, $2N$):

    \Indp 

        \textbf{if} \textbf{getBit}($b'$, $i$) \EqEq{} $0$:

        \Indp 

            \textcolor{gray}{append} $\vec{q}[i]$ \textcolor{gray}{to} $\vec{s}''$

        \Indm

    \Indm 


    \tcp{make elements contiguous}

    $b''$ = \textbf{getBitMask}($\vec{s}''$)

    \textbf{for} $i$ \textbf{in} \textbf{range}($0$, $\dim(\vec{s}'')$):

    \Indp 

        \textbf{for} $j$ \textbf{in} \textbf{range}($0$, $\vec{s}''[i]$):

        \Indp 

            $\vec{s}''[i]$ \MinusEq{} $\LogicalNot$ \textbf{getBit}($b''$, $j$)

        \Indm 

    \Indm 


    \textbf{return} $\vec{s}''$

\Indm

\end{algorithm}


\begin{algorithm}[tb]
\DontPrintSemicolon
\caption[blah blah]{
    \AlgTagDistributed \AlgTagDensityMatrix  \\
    Partial tracing of $n$ qubits $\vec{t}$ from an $N$-qubit density matrix with Choi-vector distributed as $\Lambda$-length arrays between $2^w$ nodes.
    \protect{
    \begin{center} $ \boxed{ \text{Tr}_{\vec{t}}\left( \bm\rho \right) } $ \end{center}
    }
    \begin{center}
        \AlgTagBops{$\mathcal{O}(w \Lambda)$}%
        \AlgTagFlops{$\mathcal{O}(\Lambda/2^n)$}%
        \\
        \AlgTagNumSerialRounds{$\mathcal{O}(w+N-n)$} %
        \AlgTagAmpsTransferred{$\mathcal{O}(w)2^{2N}$} \\
        \AlgTagMemOverhead{$\mathcal{O}(1)$}%
        \AlgTagMemWrites{$\mathcal{O}(w)\Lambda$}
    \end{center}
}
\label{alg:distrib_partial_trace}

\textbf{distrib\_partialTrace}($\vec{\rho}$, $\vec{\varphi}$, $\vec{t}$):

\Indp 

    $N$ = \textbf{getNumQubits}($\vec{\rho}$)
    \tcp*{Alg.~\ref{alg:distrib_dens_convenience_funcs}}

    $\lambda$ = $\log_2( \dim(\vec{\rho}) )$

    \textbf{sort}($\vec{t}$)

    \codegap 

    \tcp{local if all targets are in suffix}

    \textbf{if} $\vec{t}[-1] + N < \lambda$:

    \Indp

        $\vec{\rho}'$ = \textbf{local\_partialTraceSub}($\vec{\rho}$, $\vec{t}$, $\vec{t} \, + \, N$)

        \textbf{return} $\vec{\rho}'$

    \Indm

    \codegap 

    \tcp{find where to swap prefix targets}

    $\vec{s}$ = $\vec{t} \; \cup \; (\vec{t}+N)$
    
    $\vec{s}'$ = \textbf{getReorderedTargets}($\vec{s}$, $\lambda$)
    \tcp*{Alg.~\ref{alg:partial_trace_convenience_funcs}}

    \codegap 

    \tcp{swap prefix targets into suffix}

    \textbf{for} $q$ \textbf{in} \textbf{range}($\dim(\vec{s})-1$, $-1$, $-1$):

    \Indp 

        \textbf{if} $\vec{s}'[q]$ $\ne$ $\vec{s}[q]$:

        \Indp 

            \textbf{distrib\_swapGate}($\vec{\rho}$, $\vec{\varphi}$, $\vec{s}[q]$, $\vec{s}[q']$)
            
            \tcp*{Alg.~\ref{alg:distrib_swap}}

        \Indm 

    \Indm 

    \tcp{perform embarrasingly parallel trace}

    $\vec{t}'$ = $\vec{s}'[ \dim(\vec{t}) : \; ]$

    $\vec{\rho}'$ = \textbf{local\_partialTraceSub}($\vec{\rho}$, $\vec{t}$, $\vec{t}'$)
    \label{algline:partial_trace_local_invocation_inside_distrib}

    \codegap

    \tcp{determine un-targeted qubit ordering}

    $\vec{s}''$ = \textbf{getRemainingQubitOrder}($N$, $\vec{s}$, $\vec{s}'$)

    \tcp*{Alg.~\ref{alg:partial_trace_convenience_funcs}}


    \tcp{reorder untargeted via swaps}

    \textbf{for} $q$ \textbf{in} \textbf{range}($\dim(\vec{s}'')-1$, $-1$, $-1$):

    \Indp 

        \textbf{if} $\vec{s}''[q]$ \NotEqEq{} $q$:

        \Indp 

            $p$ = \textcolor{gray}{index of} $q$ \textcolor{gray}{in} $\vec{s}''$

            \textbf{distrib\_swapGate}($\vec{\rho}'$, $\vec{\varphi}$, $q$, $p$)
            

            $\vec{s}''[q]$, $\vec{s}''[p]$ = $\vec{s}''[p]$, $\vec{s}''[q]$

        \Indm 

    \Indm

    \codegap 

    \textbf{return} $\vec{\rho}'$

\Indm

\codegap 

\codegap

\textbf{local\_partialTraceSub}($\vec{\rho}$, $\vec{t}$, $\vec{t}'$):

\Indp 

    $\Lambda$ = $\dim(\vec{\rho})$

    $N$ = \textbf{getNumQubits}($\vec{\rho}$)
    \tcp*{Alg.~\ref{alg:distrib_dens_convenience_funcs}}

    $w$ = $\log_2(\textbf{getWorldSize()})$
    \tcp*{Alg.~\ref{alg:mpi_defs}}

    $n$ = $\dim(\vec{t})$

    $\gamma$ = $2^{2(N-n)-w}$

    $\vec{\rho}'$ = $\vec{0}^{\gamma \times 1}$
    \tcp*{new Choi-vector}

    \codegap
    
    $\vec{s}$ = \textbf{sorted}($\vec{t}$ $\cup$ $\vec{t}'$)

    \codegap 

    \AlgThreadComment{multithread}

    \textbf{for} $i$ \textbf{in} \textbf{range}($0$,$\gamma$):

    \Indp 

        $g_0$ = \textbf{insertBits}($i$, $\vec{s}$, $0$)
        \tcp*{Alg.~\ref{alg:bit_twiddles}}

        \textbf{for} $v$ \textbf{in} \textbf{range}($0$, $2^n$):

        \Indp 

            $g$ = \textbf{setBits}($g_0$, $\vec{t}$, $v$) %
            \tcp*{Alg.~\ref{alg:bit_twiddles}}

            $g$ = \textbf{setBits}($g$, $\hphantom{_0}\vec{t}'$, $v$)

            $\vec{\rho}'$[$i$] $\PlusEq{}$ $\vec{\rho}$[$g$]
            \label{algline:partialtrace_local_amp_get}

        \Indm 

    \Indm

    \codegap 

    \textbf{return} $\vec{\rho}'$

\Indm

 \end{algorithm}


%

\clearpage

\section{\hphantom{o}Summary}
\label{sec:final_summary}

This work presented an overwhelming number of novel, distributed algorithms for the full-state simulation of statevectors and density matrices, under the evolution of gates, channels, Hermitian operators and partial traces.
We tabulate the algorithms and their costs in Table~\ref{tab:alg_summary_table}.
We derived all methods explicitly from first principles using the tools of quantum information theory, and avoided traditional formulations in terms of linear algebra primitives.

This is so that our work can serve as a first introduction to those interested in high-performance simulation of digital quantum computers. Our other intention is to advance the state of the art of full-state simulators, to accelerate researcher workflows.
To this end, we have implemented all algorithms in an open-source \texttt{C++} project, incorporating MPI~\cite{lusk2009mpi} and OpenMP~\cite{dagum1998openmp}, hosted on Github\textsuperscript{\ref{fn:link_to_repo}} with a permissive MIT license.
We invite re-implementations in other languages, integrations of the algorithms into other simulators, or the use of this code base as a backend in larger software stacks. 

For the reader's interest, we now summarise the core insights and treatments primarily leveraged by our algorithms.

\begin{enumerate}
    \item We invoked the correspondence between qubits of a quantum register, and bits of a classical register encoding a basis state, i.e.
    \begin{align}
        \ket{\Psi}\numsub{N} =
        \sum\limits_i^{2^N} \alpha_i
        \bigotimes\limits_j^N \ket{i_{[j]}}\numsub{1}.
    \end{align}
    Such a treatment has been reported in the literature~\cite{de2007massively,da2020qsystem,luo2020yao},
    and instantiates a classical register with a binary encoded unsigned integer.
    This allows otherwise necessary integer algebra to be replaced with bitwise operations, though this is of little performance benefit since memory access dominates the runtime at practical scales.
    Its main utility is enabling the next trick.
    
    \item We avoided abstracting operators as matrices to be multiplied (tensored with identities) upon a state abstracted as a vector. Such a treatment precludes optimised simulation of operators specified sparsely in the $\hat{Z}$ basis, and can make derivation of distributed simulation tedious. 
    Instead, we defined operators as bitwise maps between basis states. In this picture, we easily derive the change upon each individual amplitude, allowing bespoke communication scheduling.

    \item We partitioned qubits of a register into ``prefix" and ``suffix" substates, to reason about the prescribed communication of an operator. For instance, the $i$-th global amplitude of an $N$-qubit statevector distributed between $2^w$ nodes corresponds to basis state
    \begin{gather}
        \ket{i}\numsub{N} 
        \equiv \ket{r}\numsub{w} \ket{j}\numsub{N-w},
        \\
        r = \floor{i/2^{N-w}}, \;\;\;\; 
        j = i \;\text{mod}\; 2^{N-w},
        \tag*{}
    \end{gather}
    where $r$ is the rank of the node containing the amplitude and is encoded by the ``prefix bits", and $j$ is the amplitude's index in the node's local sub-state array, encoded by the ``suffix bits".
    This treatment makes it obvious whenever simulating a gate requires communication; when the gate upon a basis state modifies the prefix qubits. We used this treatment to systematically determine all communication edge-cases of our algorithms.

    \item We used SWAP gates to change the target qubits of many-target operators in order to make them compatible with our distribution constraints, and simulable by embarrassingly parallel subroutines.
    This technique, sometimes referred to as ``cache blocking", has been used extensively in the literature and in quantum simulators~\cite{de2007massively,jones2019quest,doi2020cache,imamura2022mpiqulacs,stanwyck2022cuquantum,adamski2023energy}, sometimes as the primary means to distribute simulation of operators.
    In this work, we only invoked it where necessary to circumvent limitations of our communication buffer,
    such as for evaluation of the partial trace of a density matrix.

    \item We invoked the Choi--Jamio\l{}kowski isomorphism~\cite{choi1975completely,jamiolkowski1972linear} to reuse our statevector distribution scheme, in order to represent and simulate density matrices. This often permits operations upon a density matrix to be decomposed into a series of simpler operations upon an unnormalised statevector.
    We first reported this strategy in Ref.~\cite{jones2019quest}.
    
\end{enumerate}

\begin{table*}[p]
    \centering

    \def\totalAmpSymb{\Sigma}

    \caption{
        The distributed algorithms presented in this manuscript, and their costs, expressed either in aggregate or per-node (pn).
        Assume each algorithm is invoked upon an $N$-qubit statevector or density matrix, constituted by a total of $\totalAmpSymb$ amplitudes, distributed between $2^w$ nodes, such that each node contains $\Lambda=\totalAmpSymb/2^w$ amplitudes and an equivalently sized communication buffer. All methods assume $N\ge w$.
        Where per-node costs differ between nodes (such as for the $s$-control one-target gate), the average cost is given.
    }

   \bigskip

   \textbf{Statevector algorithms ($\totalAmpSymb=2^N$, $\Lambda=2^{N-w}$)}

   \bigskip
    
    \makebox[\textwidth][c]{
    \begin{tabular}{ccccccccccc}
    Sec. 
    & 
    Alg. 
    & 
    Operation 
    &
    Symbol 
    &
    Bops (pn) 
    &
    Flops (pn) 
    &
    Writes (pn) 
    &
    Exchanges 
    &
    Exchanged 
    &
    Memory 
\\
\hline
    \ref{sec:distrib_1qb_gate}
    &
    \ref{alg:distrib_1qb_gate}
    &
    One-target gate
    &
    $\hat{M}_t$
    &
    %
    $\mathcal{O}(\Lambda)$
    &
    $\mathcal{O}(\Lambda)$
    &
    $\Lambda$
    &
    $0$ or $1$
    &
    $\totalAmpSymb$
    &
    $\mathcal{O}(1)$
\\
    \ref{sec:distrib_many_ctrl_one_targ_gate}
    &
    \ref{alg:distrib_manyctrl_gate}
    &
    $s$-control one-target gate
    &
    $C_{\vec{c}}(\hat{M}_t)$
    &
    %
    $\mathcal{O}(s\Lambda/2^s)$
    &
    $\mathcal{O}(\Lambda/2^s)$
    &
    $\Lambda/2^s$
    &
    $0$ or $1$
    &
    $\totalAmpSymb/2^s$
    &
    $\vdots$
\\
    \ref{sec:swap_gate}
    &
    \ref{alg:distrib_swap}
    &
    SWAP gate
    &
    $\text{SWAP}_{t_1,t_2}$
    &
    %
    $\mathcal{O}(\Lambda)$
    &
    $0$
    &
    $\Lambda/2$
    &
    $0$ or $1$
    &
    $\totalAmpSymb/2$
    &
    $\mathcal{O}(1)$
\\
    \ref{sec:distrib_manytarg_gate}
    &
    \ref{alg:distrib_manytarg_gate}
    &
    $n$-target gate\footnote{where $n \le N-w$}
    &
    $\hat{M}_{\vec{t}}$
    &
    $\mathcal{O}(2^n\Lambda)$
    &
    $\mathcal{O}(2^n\Lambda)$
    &
    $\mathcal{O}(2^n\Lambda)$
    &
    $\mathcal{O}(\min(n,w))$
    &
    $\mathcal{O}(\min(n,w))\totalAmpSymb$
    &
    $\mathcal{O}(2^n)$
\\
    \ref{sec:pauli_tensor}
    &
    \ref{alg:distrib_pauli_tensor}
    &
    $n$-qubit Pauli tensor
    &
    $\otimes^n \hat{\sigma}$
    &
    %
    $\mathcal{O}(n\Lambda)$
    &
    $\Lambda$
    &
    $\Lambda$
    &
    $0$ or $1$
    &
    $\totalAmpSymb$
    &
    $\mathcal{O}(1)$
\\
    \ref{sec:distrib_phase_gadget}
    & 
    \ref{alg:distrib_phase_gadget}
    & 
    $n$-qubit phase gadget
    & 
    $\exp(\iu \theta \hat{Z}^{\otimes})$
    & 
    %
    $\mathcal{O}(\Lambda)$
    &
    $\mathcal{O}(\Lambda)$
    & 
    $\Lambda$
    & 
    $0$
    & 
    $0$
    & 
    $\vdots$
\\
    \ref{sec:distrib_pauli_gadget} 
    & 
    \ref{alg:distrib_pauli_gadget}
    & 
    $n$-qubit Pauli gadget
    &
    $\exp(\iu \theta \otimes^n \hat{\sigma})$ 
    &
    %
    $\mathcal{O}(n\Lambda)$
    &
    $\mathcal{O}(\Lambda)$
    &
    $\Lambda$ 
    &
    $0$ or $1$
    &
    $\totalAmpSymb$ 
    &
\end{tabular}
} 

\bigskip 

\bigskip

   \textbf{Density matrix algorithms ($\totalAmpSymb=2^{2N}$, $\Lambda=2^{2N-w}$)}

\bigskip

\makebox[\textwidth][c]{
\begin{tabular}{ccccccccccc}
    Sec. 
    & 
    Alg. 
    & 
    Operation 
    &
    Symbol 
    &
    Bops (pn) 
    &
    Flops (pn) 
    &
    Writes (pn) 
    &
    Exchanges 
    &
    Exchanged 
    &
    Memory 
\\
\hline
    \ref{sec:dens_unitaries}
    & 
    \ref{alg:distrib_dens_matr_unitary}
    & 
    $n$-qubit unitary\footnote{\label{fn:dens_max_targ_cond}where $n \le N-\lceil w/2 \rceil$}
    &
    $\hat{U}_{\vec{t}}$ 
    &
    $\mathcal{O}(2^n \Lambda)$
    &
    $\mathcal{O}(2^n \Lambda)$
    &
    $\mathcal{O}(2^n \Lambda)$
    &
    $\mathcal{O}(\min(n,w))$
    &
    $\mathcal{O}(\min(n,w)) \totalAmpSymb/2$
    &
    $\mathcal{O}(2^n)$
\\
    \ref{sec:dens_unitaries}
    & 
    \ref{alg:distrib_dens_matr_specific_unitaries}
    & 
    SWAP gate 
    &
    $\text{SWAP}_{t_1, t_2}$ 
    &
    %
    $\mathcal{O}(\Lambda)$
    &
    $0$
    &
    $\Lambda$
    &
    $0$ or $1$
    &
    $\totalAmpSymb/2$
    &
    $\mathcal{O}(1)$ 
\\
    \ref{sec:dens_unitaries}
    & 
    \ref{alg:distrib_dens_matr_specific_unitaries}
    & 
    $n$-qubit Pauli tensor 
    &
    $\otimes^n \hat{\sigma}$ 
    &
    %
    $\mathcal{O}(n\Lambda)$
    &
    $2\Lambda$ or $3\Lambda$
    &
    $2\Lambda$ or $3\Lambda$
    &
    $0$ or $1$ 
    &
    $\totalAmpSymb$
    &
    $\vdots$
\\
    \ref{sec:dens_unitaries}
    & 
    \ref{alg:distrib_dens_matr_specific_unitaries}
    & 
    $n$-qubit phase gadget 
    &
    $\exp(\iu \theta \hat{Z}^{\otimes})$ 
    &
    %
    $\mathcal{O}(\Lambda)$
    &
    $\mathcal{O}(\Lambda)$
    &
    $2\Lambda$
    &
    $0$
    &
    $0$
    &
\\
    \ref{sec:dens_unitaries}
    & 
    \ref{alg:distrib_dens_matr_specific_unitaries}
    & 
    $n$-qubit Pauli gadget
    &
    $\exp(\iu \theta \otimes^n \hat{\sigma})$ 
    &
    %
    $\mathcal{O}(n\Lambda)$
    &
    $\mathcal{O}(\Lambda)$
    &
    $2\Lambda$
    &
    $0$ or $1$ 
    &
    $\totalAmpSymb$
    &
\\
    \ref{sec:kraus_map}
    & 
    \ref{alg:distrib_dens_matr_kraus_map}
    & 
    $n$-qubit Kraus map\textsuperscript{\ref{fn:dens_max_targ_cond}}
    &
    $\sum \hat{K}^\dagger \bm\rho \hat{K}$ 
    &
    $\mathcal{O}(4^n \Lambda)$
    &
    $\mathcal{O}(4^n \Lambda)$
    &
    $\mathcal{O}(4^n \Lambda)$
    &
    $\mathcal{O}(\min(2n,w))$ 
    &
    $\mathcal{O}(\min(2n,w)) \totalAmpSymb/2$
    &
    $\mathcal{O}(16^n)$
\\
    \ref{sec:dephasing}
    & 
    \ref{alg:distrib_dens_matr_one_qb_dephasing}
    & 
    one-qubit dephasing
    &
    ${\mathcal{E}_\phi}_t$ 
    &
    %
    $\mathcal{O}(\Lambda)$
    &
    $\Lambda/2$
    &
    $\Lambda/2$
    &
    $0$
    &
    $0$
    &
    $\mathcal{O}(1)$
\\
    \ref{sec:dephasing}
    & 
    \ref{alg:distrib_dens_matr_two_qb_dephasing}
    & 
    two-qubit dephasing
    &
    ${\mathcal{E}_\phi}_{t_1,t_2}$ 
    &
    %
    $\mathcal{O}(\Lambda)$
    &
    $\Lambda$
    &
    $\Lambda$
    &
    $0$
    &
    $0$ 
    &
    $\vdots$
\\
    \ref{sec:depolarising} 
    & 
    \ref{alg:distrib_dens_matr_one_qb_depolarising}
    & 
    one-qubit depolarising 
    &
    ${\mathcal{E}_{\Delta}}_t$ 
    &
    %
    $\mathcal{O}(\Lambda)$
    &
    $\mathcal{O}(\Lambda)$
    &
    $\mathcal{O}(\Lambda)$
    &
    $0$ or $1$ 
    &
    $\totalAmpSymb/2$
    &
\\
    \ref{sec:depolarising}
    & 
    \ref{alg:distrib_dens_matr_two_qb_depolarising}
    & 
    two-qubit depolarising 
    &
    ${\mathcal{E}_{\Delta}}_{t_1,t_2}$ 
    &
    %
    $\mathcal{O}(\Lambda)$
    &
    $\mathcal{O}(\Lambda)$
    &
    $\mathcal{O}(\Lambda)$
    &
    $0$, $1$ or $2$ 
    &
    $\totalAmpSymb/8$ or $\totalAmpSymb/2$ 
    &
\\
    \ref{sec:damping}
    & 
    \ref{alg:distrib_dens_matr_damping}
    & 
    amplitude damping 
    &
    ${\mathcal{E}_{\gamma}}_t$ 
    &
    %
    $\mathcal{O}(\Lambda)$
    &
    $\mathcal{O}(\Lambda)$
    &
    $\mathcal{O}(\Lambda)$
    &
    $0$ or $1$\footnote{only one node of a pair sends amplitudes}
    &
    $\totalAmpSymb/4$ 
    &
\\
    \ref{sec:pauli_string_expec_value}
    & 
    \ref{alg:distrib_dens_matr_pauli_string_expec}
    & 
    $T$-term Pauli string expectation 
    &
    $\Tr(\hat{H}\,\bm\rho)$ 
    &
    %
    $\mathcal{O}(N\,T\,\Lambda)$
    &
    $\mathcal{O}(T\,\Lambda)$
    &
    $0$
    &
    $0$
    &
    $0$\footnote{one scalar global reduction}
    &
\\
    \ref{sec:partial_trace}
    & 
    \ref{alg:distrib_partial_trace}
    & 
    $n$-qubit partial trace\textsuperscript{\ref{fn:dens_max_targ_cond}}
    &
    $\Tr_{\vec{t}}(\bm\rho)$ 
    &
    $\mathcal{O}(w\Lambda)$
    &
    $\mathcal{O}(\Lambda/2^n)$
    &
    $\mathcal{O}(w)\Lambda$
    &
    $\mathcal{O}(w+N-n)$
    &
    $\mathcal{O}(w)\totalAmpSymb$  
    &
    $\mathcal{O}(1)$\footnote{excluding the cost of the output density matrix} 
\end{tabular}
} 

\bigskip

\bigskip

    \label{tab:alg_summary_table}
\end{table*}


\begin{figure*}[t]
\centering
\makebox[\textwidth][c]{
\begin{overpic}[width=1.1\textwidth]{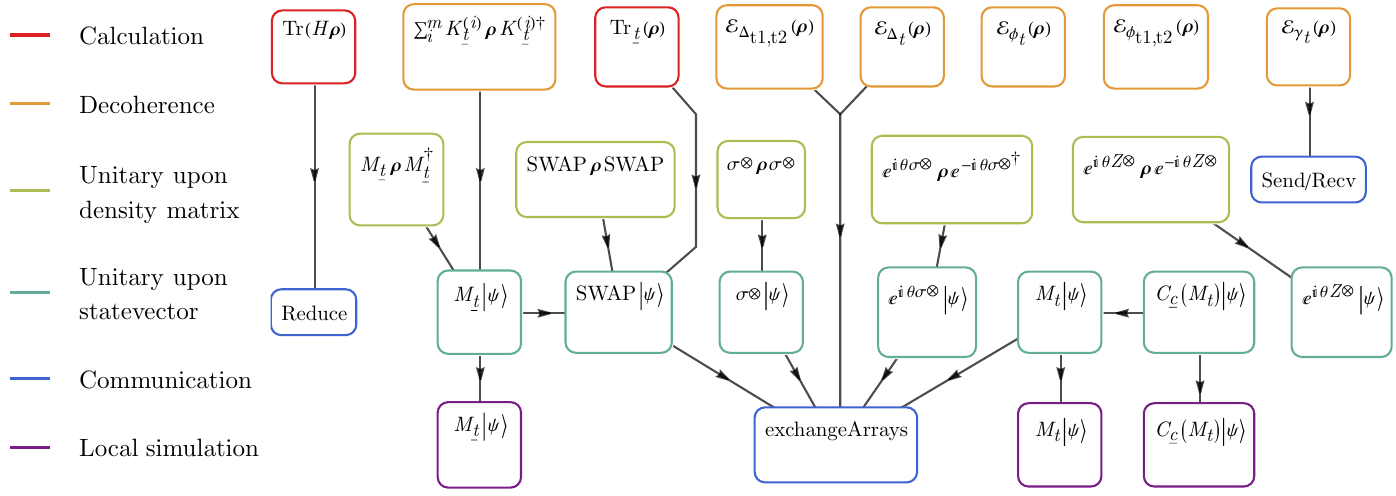}

\put(43.3,30.3){Alg.~\ref{alg:distrib_partial_trace}}

\put(26, 20.3){Alg.~\ref{alg:distrib_dens_matr_unitary}}

\put(40, 21){Alg.~\ref{alg:distrib_dens_matr_specific_unitaries}}
\put(52, 21){Alg.~\ref{alg:distrib_dens_matr_specific_unitaries}}
\put(65.24, 20.7){Alg.~\ref{alg:distrib_dens_matr_specific_unitaries}}
\put(80, 20.7){Alg.~\ref{alg:distrib_dens_matr_specific_unitaries}}

\put(80.5, 30.5) {Alg.~\ref{alg:distrib_dens_matr_two_qb_dephasing}}
\put(71, 30.5){Alg.~\ref{alg:distrib_dens_matr_one_qb_dephasing}}

\put(64, 11){Alg.~\ref{alg:distrib_pauli_gadget}}

\put(20, 30.6){Alg.~\ref{alg:distrib_dens_matr_pauli_string_expec}}

\put(91.5, 30.6){Alg.~\ref{alg:distrib_dens_matr_damping}}

\put(32, 30.4){Alg.~\ref{alg:distrib_dens_matr_kraus_map}}

\put(53, 11){Alg.~\ref{alg:distrib_pauli_tensor}}

\put(93.9, 11){Alg.~\ref{alg:distrib_phase_gadget}}

\put(32.1, 11){Alg.~\ref{alg:distrib_manytarg_gate}}

\put(42.5, 11.2){Alg.~\ref{alg:distrib_swap}}

\put(58.2, 1.9){Alg.~\ref{alg:mpi_defs}}

\put(84, 11){Alg.~\ref{alg:distrib_manyctrl_gate}}

\put(84, 1.9){Alg.~\ref{alg:serial_manyctrl_gate_alg}}

\put(32.4, 1.9){Alg.~\ref{alg:serial_manyqb_gate_alg}}

\put(62.4, 30.3){Alg.~\ref{alg:distrib_dens_matr_one_qb_depolarising}}

\put(53.1, 30.3){Alg.~\ref{alg:distrib_dens_matr_two_qb_depolarising}}

\put(74, 11){Alg.~\ref{alg:distrib_1qb_gate}}

\put(74, 1.9){Alg.~\ref{alg:serial_1qb_gate_alg}}

\end{overpic}
} 
\caption{
    The dependency tree of this manuscript's algorithms. The surprising connectivity becomes intuitive under the following observations:
    Distributed statevector simulation often includes edge-cases algorithmically identical to local statevector simulation; 
    Under the Choi--Jamio\l{}kowski isomorphism~\cite{choi1975completely,jamiolkowski1972linear},
    a unitary upon a density matrix resembles two unnormalised unitaries upon an unnormalised statevector (which we called a ``Choi-vector"); A Kraus channel upon a density matrix is equivalent to a superoperator upon a Choi-vector, itself equivalent to an unnormalised unitary upon an unnormalised statevector; SWAP gates permit transpiling communicating operators into embarrassingly parallel ones;
    The natural distribution of statevector amplitudes often admits simulation via pairwise communication and simple array exchanges.
}
\label{fig:all_alg_dependency_plot}
\end{figure*}


Our algorithms used the above properties whilst simultaneously satisfying several high-performance computing considerations. 
\begin{enumerate}

    \item We ensured exponentially large control-flow loops had independent iterations, and so could be locally parallelised, such as through multithreading. We labelled these loops, and where possible, organised memory access to be regular and of minimum stride to minimise caching conflicts.

    \item We avoided branching, circumventing the risks of failed branch prediction, by iterating directly the amplitudes requiring modification. The index algebra was performed bitwise, giving compilers the best chance of auto-vectorising.
    
    \item When we required to communicate a \textit{subset} of a node's local amplitudes, we first contiguously packed this subset into the communication buffer. This reduced the total network traffic.

    \item We endeavoured to make communication \textit{pairwise}, whereby nodes exchanged amplitudes in exclusive pairs. This simplifies the communication code, and makes network costs predictable and approximately uniform.

\end{enumerate}

The astute reader will have noticed that many of our algorithms invoked other algorithms as subroutines. For example, the distributed simulation of the one-qubit gate (Alg.~\ref{alg:distrib_1qb_gate}) invokes the \textit{local} simulation (Alg.~\ref{alg:serial_1qb_gate_alg}). The interdependence of our algorithms is visualised in Fig.~\ref{fig:all_alg_dependency_plot}.

We hope that functional, practical-scale quantum computers can one day repay the classical computational costs of their development. Happy simulating!


\section{\hphantom{o}Contributions}

TJ devised all algorithms not listed below, 
wrote the manuscript, and optimised and implemented all algorithms.
BK devised the algorithms of Section~\ref{sec:kraus_map},
and SCB devised those of Sections~\ref{sec:distrib_manytarg_gate}, \ref{sec:distrib_phase_gadget}, \ref{sec:dens_state}, and \ref{sec:dens_unitaries}.
All authors contributed to the development of QuEST~\cite{jones2019quest} which enabled the research of this manuscript.

\pagebreak

\section{\hphantom{o}Acknowledgements}

The authors thank Ania Brown for her pioneering role in the early development of the QuEST simulator.
TJ also thanks Keisuke Fujii and Kosuke Mitarai for hosting him at Osaka University where this manuscript was incidentally completed, and the UKRI NQCC for financially supporting QuEST development.
TJ additionally thanks Richard Meister for his contributions to QuEST and helpful input on matters HPC and software architecture; so too he thanks Cica Gustiani, Sam Jaques, Adrian Chapman and Arthur Rattew for useful discussions, the Clarendon fund for financial support, and peers Isabell Hamecher, Chris Whittle, Patrick Inns, Sinan Shi and Andrea Vitangeli for aesthetic advice.
This work was further supported by EPSRC grant EP/M013243/1. SCB acknowledges support from the EPSRC QCS Hub EP/T001062/1, from U.S. Army Research Office Grant No. W911NF-16-1-0070 (LOGIQ), and from EU H2020-FETFLAG-03-2018 under the grant agreement No 820495 (AQTION). The authors would also like to acknowledge the use of the University of Oxford Advanced Research Computing (ARC) facility to test the algorithms in this manuscript (\href{http://dx.doi.org/10.5281/zenodo.22558}{dx.doi.org/10.5281/zenodo.22558}).
Finally, TJ would like to offer no thanks at all to the many interruptions and delays to the writing of this manuscript which began amidst the 2020 pandemic.

\clearpage 
\pagebreak

\bibliographystyle{unsrt}
\bibliography{bibliography.bib}

\end{document}